\documentclass[10pt,draftcls,onecolumn]{IEEEtran}
\usepackage{acronym}
\usepackage{algpseudocode}
\usepackage{amsfonts}
\usepackage{amsmath}
\usepackage{amssymb}
\usepackage{amsthm}
\usepackage[ruled,vlined]{algorithm2e}
\usepackage{bm}

\usepackage{booktabs}

\usepackage{cite}

\usepackage{enumerate}
\usepackage{dsfont}

\usepackage{graphicx}

\usepackage{longtable}

\usepackage{mathtools}

\usepackage{nicefrac}

\usepackage[caption=false,font=footnotesize]{subfig}

\usepackage[colorinlistoftodos]{todonotes}

\usepackage{xspace}

\usepackage{xcolor}

\theoremstyle{plain}

\newtheorem{thm}{Theorem}
\newtheorem{cor}{Corollary}

\newtheorem{lem}{Lemma}

\theoremstyle{definition}
\newtheorem{defn}{Definition}
\newtheorem{exmp}{Example}

\theoremstyle{remark}
\newtheorem*{rem}{Remark}

\newcommand*{\fig}[1]{\figurename~\ref{fig:#1}\xspace}
\newcommand*{\tblref}[1]{Table~\ref{tbl:#1}\xspace}
\newcommand*{\eq}[1]{(\ref{eq:#1})\xspace}

\newcommand*{\thmref}[1]{Theorem~\ref{thm:#1}\xspace}
\newcommand*{\corref}[1]{Corollary~\ref{cor:#1}\xspace}

\newcommand*{\lemref}[1]{Lemma~\ref{lem:#1}\xspace}

\newcommand*{\secref}[1]{Section~\ref{sec:#1}\xspace}
\newcommand*{\appendixproof}[1]{See Appendix~\ref{sec:#1}\xspace}

\newcommand*{\iid}{i.i.d.\xspace}
\newcommand*{\wrt}{w.r.t.\xspace}
\newcommand*{\wlogen}{w.l.o.g.\xspace}

\newcommand*{\expected}[1]{\mathds{E}\left[#1\right]}
\newcommand*{\prob}[1]{\mathds{P}\left(#1\right)}

\DeclareMathOperator*{\argmax}{arg\,max}
\DeclareMathOperator*{\argmin}{arg\,min}
\DeclareMathOperator*{\sol}{sol}

\newcommand*{\Ica}{\mathcal{I}}

\newcommand*{\Lca}{\mathcal{L}}

\newcommand*{\Uca}{\mathcal{U}}

\newcommand*{\Xca}{\mathcal{X}}

\newcommand*{\Ubf}{\mathbf{U}}

\newcommand*{\Zbf}{\mathbf{Z}}

\acrodef{AMC}[AMC]{adaptive modulation and coding}
\acrodef{BA}[BA]{Blahut-Arimoto}
\acrodef{BS}[BS]{base station}
\acrodef{BSC}[BSC]{binary symmetric channel}
\acrodef{CEO}[CEO]{chief estimating officer}
\acrodef{CQI}[CQI]{channel quality indicator}
\acrodef{DFSQ}{distributed functional scalar quantization}
\acrodef{ECSQ}[ECSQ]{entropy-constrained scalar quantization}
\acrodef{HetSQ}[HetSQ]{heterogeneous scalar quantizer}
\acrodef{HomSQ}[HomSQ]{homogeneous scalar quantizer}
\acrodef{ID}[i.d.]{identically distributed}
\acrodef{MCS}{modulation and coding scheme}
\acrodef{MS}[MS]{mobile station}
\acrodef{MSE}[MSE]{mean squared error}
\acrodefplural{MS}[MSs]{mobile stations}
\acrodef{OFDMA}{orthogonal frequency-division multiple access}
\acrodef{PMF}[pmf]{probability mass function}
\acrodef{SQ}[SQ]{scalar quantizer}
\acrodefplural{SQ}[SQs]{scalar quantizers}
\acrodef{SW}[SW]{Slepian-Wolf}
\acrodef{MTIS}{Multi-Thresholds Interactive Scheme}
\acrodef{NBIS}{Non-Broadcasting Interactive Scheme}
\acrodef{RIS}{Relay Interactive Scheme}

\graphicspath{ {./img/} }

\title{Overhead Performance Tradeoffs---A Resource Allocation Perspective}
\author{%
	Jie~Ren, %
	Bradford~D.~Boyle,~\IEEEmembership{Student~Member,~IEEE,} %
	Gwanmo~Ku,~\IEEEmembership{Student~Member,~IEEE,} %
	Steven~Weber,~\IEEEmembership{Senior~Member,~IEEE,} and %
	John~MacLaren~Walsh,~\IEEEmembership{Member,~IEEE}%
	\thanks{This research has been supported by the Air Force Research Laboratory under agreement number FA9550-12-1-0086. The U.S. Government is authorized to reproduce and distribute reprints for Governmental purposes notwithstanding any copyright notation thereon.}
	\thanks{The authors are with the Department of Electrical and Computer Engineering, Drexel University, Philadelphia, PA USA. The contact author is John MacLaren Walsh (email: \textsf{jwalsh@ece.drexel.edu})}
	\thanks{Preliminary results were reported at CISS 2014 \cite{BoyWalWeb2014,RenWal2014}.}%
}
\showboxdepth=\maxdimen
\begin{document}
\maketitle
\begin{abstract}
	A key aspect of many resource allocation problems is the need for the resource controller to compute a function, such as the \(\max\) or \(\argmax\), of the competing users metrics.
	Information must be exchanged between the competing users and the resource controller in order for this function to be computed.
	In many practical resource controllers the competing users' metrics are communicated to the resource controller, which then computes the desired extremization function.
	However, in this paper it is shown that information rate savings can be obtained by recognizing that controller only needs to determine the result of this extremization function.
	If the extremization function is to be computed losslessly, the rate savings are shown in most cases to be at most \(2\) bits independent of the number of competing users.
	Motivated by the small savings in the lossless case, simple achievable schemes for both the lossy and interactive variants of this problem are considered.
	It is shown that both of these approaches have the potential to realize large rate savings, especially in the case where the number of competing users is large.
	For the lossy variant, it is shown that the proposed simple achievable schemes are in fact close to the fundamental limit given by the rate distortion function.
\end{abstract}

\section{Introduction}
\label{sec:introduction}
In this paper we consider a problem in which series of $N$ users have access to independent sequences $\boldsymbol{X}_i = \left[ X_{i,s}| s \in \{1,2,\ldots,\} \right], \ i\in\{1,\ldots,N\}$ of independent and identically distributed observations $X_{i,s}$ from a known distribution on a set $\mathcal{X}\subset \mathbb{R}^+$, a subset of the non-negative real numbers.  The users compress their observations for transmission to a \ac{CEO} that wishes to know for each element in the sequence:
\begin{enumerate}
	\item the largest observation, i.e. $\max_{i\in\{1,\ldots,N\}} X_{i,s}$ for each $s$;
	\item a source having the largest observation, i.e. a single member of $\arg \max_{i\in\{1,\ldots,N\}} X_{i,s}$ for each $s$, or;
	\item both the largest observation and the user that having the largest observation.
\end{enumerate}
We refer the three cases as the \(\max\) problem, the \(\argmax\) problem, and the \((\max, \argmax)\) problem respectively.
Although we present all our results terms of \(\max\), similar results will hold for the corresponding \emph{minimization} problems.
\begin{figure*}
	\centering
	\subfloat[]{\includegraphics{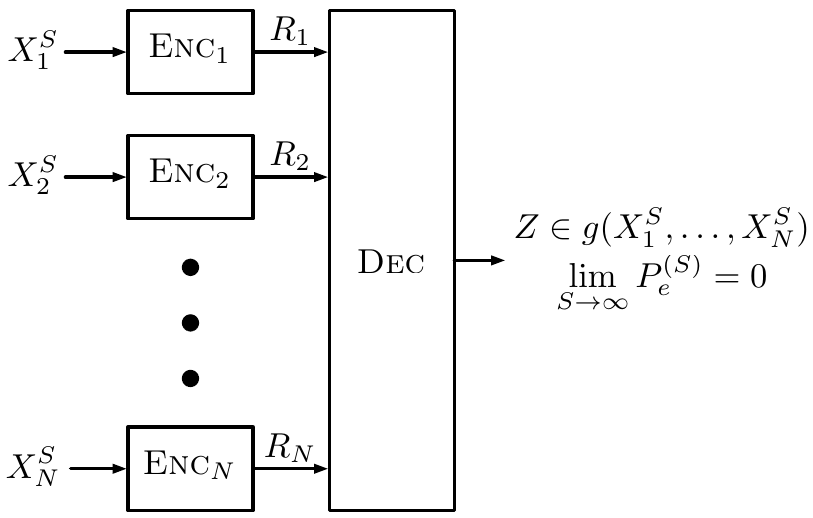}\label{fig:lossless-model-block-diagram}}\\
	\subfloat[]{\includegraphics{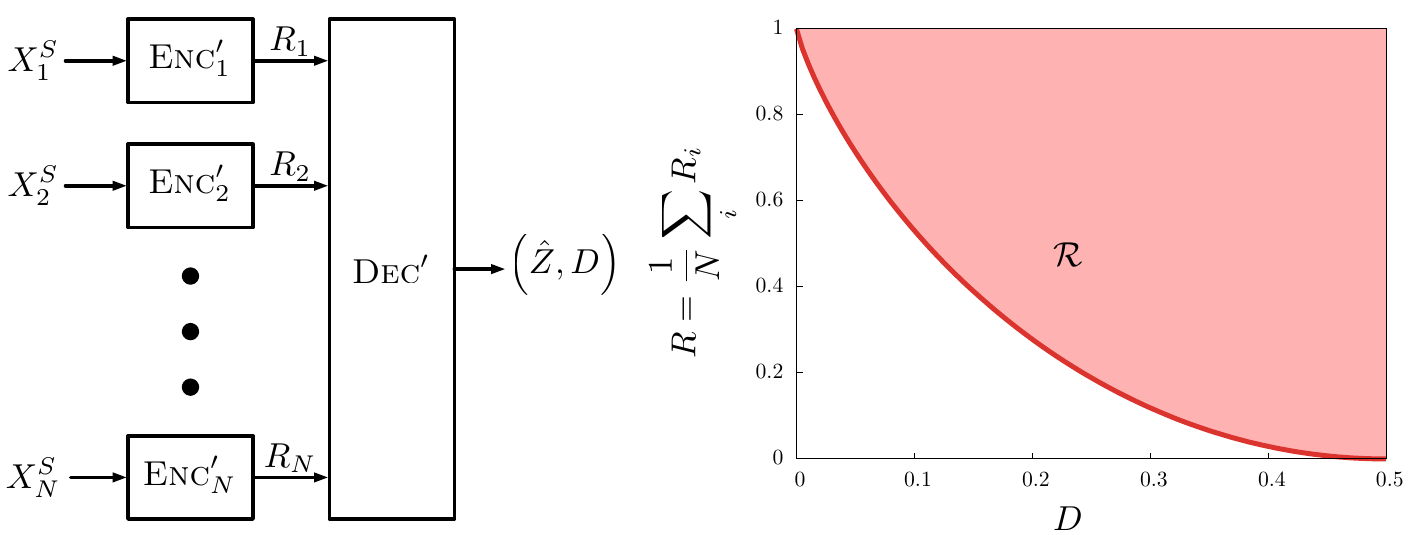}\label{fig:lossy-model-block-diagram}}
	\caption{Model block diagrams}
	\label{fig:model-block-diagrams}
\end{figure*}

\begin{table}
	\centering
	\begin{tabular}{ccccc}
		\toprule
		& & \(\max\) & \(\argmax\) & \((\max, \argmax)\) \\
		\cmidrule(r){3-5}
		\ref{exmp:ofdma} & \acs{OFDMA} resource allocation & anycasting & rateless PHY layer &  traditional \acs{AMC} PHY layer\\
		\ref{exmp:economics} & economics & asset pricing  & asset allocation & sealed-bid first-price auctions\\
		\ref{exmp:sensor-network} & sensor network/intrusion detection & is there an intruder & where is the intruder & is there an intruder \& where is the intruder\\
		\bottomrule
	\end{tabular}
	\caption{Example application of indirect extremal value computation}
	\label{tbl:examples}
\end{table}
This generic indirect extremal value computation problem finds examples in several fields; \tblref{examples} lists a few of these.
We consider three in more detail here:
\begin{exmp}[\ac{OFDMA} resource allocation]
	\label{exmp:ofdma}
	Rateless coding, also known as fixed-to-variable coding \cite{VariableRateCC}, can achieve performance close to the channel capacity without requiring the explicit feedback of channel state information and use of adaptive modulation and coding in a single user system \cite{LubyLT,RatelessAWGN}.
	These schemes operate by enabling the block length (in channel uses) for the modulation and coding to stretch or shrink based on the received channel quality in a manner that closely resembles H-ARQ.
	Rather than feeding back channel quality, the receiver only needs to indicate when it has successfully decoded the transmitted message, which it learns through an outer error detection code.
	In a multiuser OFDMA system, the \ac{BS} needs to assign \acp{MS} to subblocks of channles, even when a rateless code used (\fig{problem-model}).
	If the \ac{BS} wishes to maximize the sum-rate across users, the uplink feedback from the \acs{MS} only needs to enable the basestation to determine which \ac{MS} has the best channel.
	The \ac{BS} does not need to know the actual channel gain/capacity.
	\begin{figure}
		\centering
		\includegraphics[width=252.0pt]{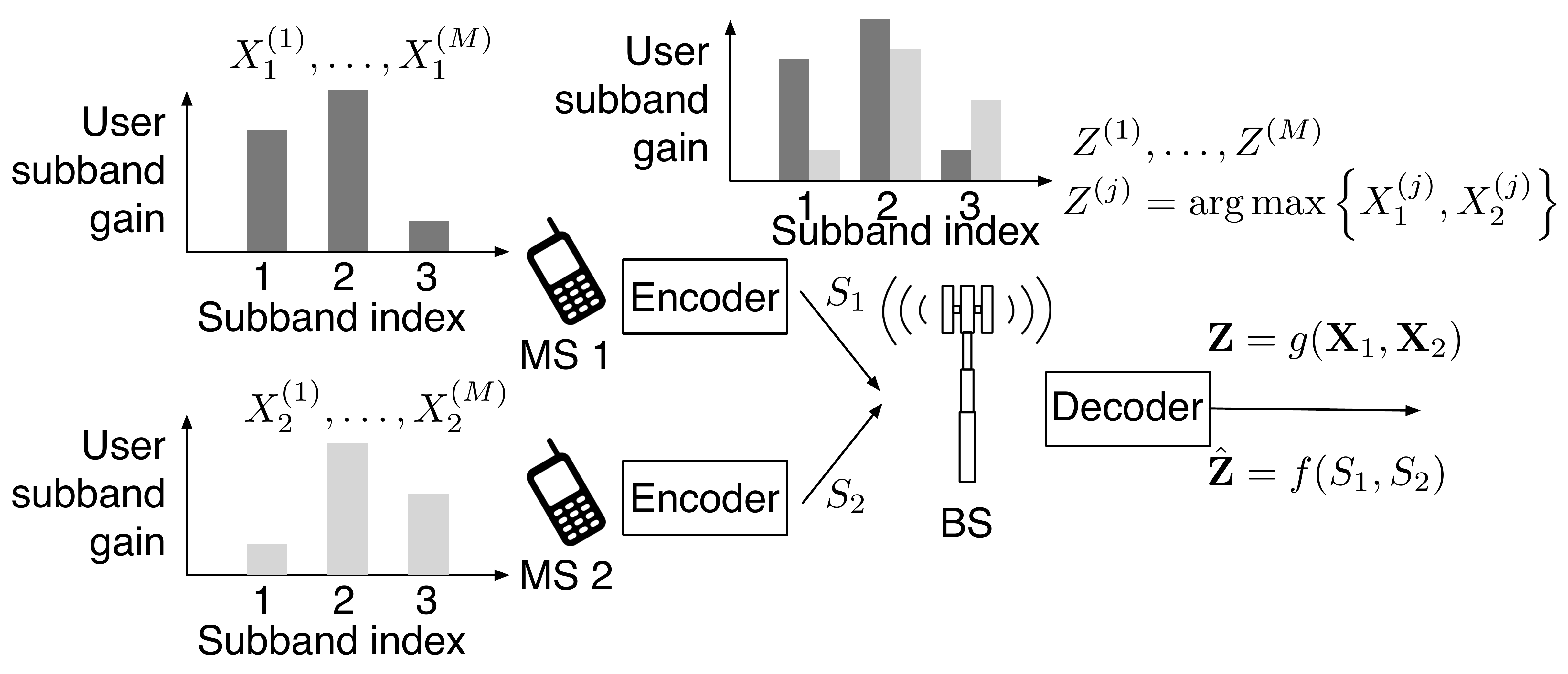}
		\caption{The \ac{BS} wishes to compute the index (arg max) of the user with the largest gain on each subband. The users encode their local gains across subbands using \acp{SQ}.}
		\label{fig:problem-model}
	\end{figure}
	Once the \ac{BS} has decided which user to schedule on a particular collection of subbands, it must signal this resource decision on the downlink as overhead control information in addition to the data to be transmitted to the user itself.
	These resource decisions control information, along with the \ac{MS}'s feedback, result in control overheads that are surprisingly large---the control overheads account for \(\approx 25\)--\(30\%\) of all downlink transmission in the LTE standard\cite{TS36211-8}.

	Anycasting is transmission scheme whereby a sender wishes to send a message to a group of potential receivers and ensure that it is received by at least one receiver \cite{KatWro2000}.
	This is contrasted with broadcasting, where \emph{every} receiver is required to receive the message.
	In this context, the \ac{CEO} is the \ac{BS} and the sourcess at the different receivers are the channel gains/capacities on the downlinks.
	The \ac{BS} needs to know the largest channel capacity in order to select an approriate transmission rate.
	Replacing \(\max\) with \(\min\), this setup becomes a broadcasting problem.
	By knowing the smallest channel capacity, the \ac{BS} can select a rate that ensures its message is received by all of the users.
	
	Traditional \ac{AMC} \cite{ProSal1994} proceeds by first defining a finite collection of codes and modulation schemes associated with different information rates \(r_k\) measured in bits per channel use.
	The index \(k \in \{1,\ldots, K\}\) indicating which scheme to use is called the \ac{MCS} index.
	The receiver measures the channel quality using reference or training signals, or pilots, and determines the information rate among this finite set corresponding to a modulation and coding scheme achieving a given target probability of error.
	The associated index \(k\), or some quantization of it, is then fed back to the transmitter under the label \ac{CQI}.
	The transmitter then takes into consideration factors such as the amount of data waiting to be sent to the various receivers associated with it and their necessary quality of service, then selects the modulation and coding scheme to use when transmitting to them.
\end{exmp}
\begin{exmp}[Economics]
	\label{exmp:economics}
	When a seller has a commodity that it wishes to sell, it sets the price with respect to the market \cite{Var1992}.
	If the seller wants to ensure that it does not price it self out of the market, it would want to compute the \(\max\) of the individual valuations of a representative sample of the market.
	Conversely, if the seller wants to undercut its competition it would need to compute the \(\min\) of the competitor's prices.

	In many situations, goods should be allocated/distributed to users based on their ``need'' or expected derived utility \cite{Raw1971}.
	For example, need-based financial aid for higher education.
	In this scenario, the entity distributing the goods would only need to calculate the individual with the largest expected derived utility (i.e., the \(\argmax\)).

	We think of the \ac{CEO} selling a good through an auction to a set of independent buyers.
	In a sealed-bid first-price auction, the buyers submit bids in ``sealed'' envelopes and the highest bidder is the winner, paying their bid amount\cite{Kri2010}.
	The auctioneer has no need for recovering this bids of the other users.
\end{exmp}
\begin{exmp}[Sensor network/intrusion detection \cite{ButMorSan2014}]
	\label{exmp:sensor-network}	
	A collection of sensor nodes are monitoring an area large enough that individual sensor readings are independent.
	As a very simple model, we can take the local sensor outputs to be binary: \(0\) if no intruder is present, \(1\) if an intruder is present.
	Computing the \(\argmax\) determines \emph{where} an intruder is (if in fact there is one); computing the\(\max\) sensor reading determines \emph{if} an intruder is present but not where,  and; computing both determines if and where an intruder is.
\end{exmp}

The remainder of the paper is organized as follows: In \secref{related-work}, we review the existing literature concerning fundamental limits and achievable schemes for the non-interactive lossless and lossy estimation. We also review literature for the interactive variant of this problem where the users and CEO are allowed to interactively communicate over multiple rounds. Next we formalize the mathematical model for this problem and propose natural distortion measures in \secref{model-specification}. In \secref{lossless-subband-allocation}, we derive the fundamental limit on rate to estimate losslessly (in the usual Shannon sense) and propose a scheme to achieve this limit. We observe that the rate savings compared with the source recovery (i.e. \ac{SW} \cite{SW}) are not large. In \secref{lossy-subband-allocation}, we consider the same problem, but allow the CEO to estimate the function in a lossy manner. We compute the rate-distortion function numerically (\secref{multi-source-blahut-arimoto}) and compare it with an achievable scheme based on scalar quantization (\secref{arg-max-hom-sq} \& \secref{two-user-het-sq}).
Finally in \secref{interactive-subband-allocation}, we consider interactive communications\cite{InteractiveFunctionComputation} between the CEO and the users. We propose interactive schemes in which the \ac{CEO} will losslessly determine the $\max$, $\argmax$ or the pair. For both the one-shot lossy and interactive lossless case, we show the rate saving can be substantial.

\section{Related Work}
\label{sec:related-work}
We first review some previous results of the CEO problem under both lossless and lossy setup to understand the fundamental limits of the rate region. We then review some results about quantization design which help us to give achievable schemes for the extremization problems we are interested in.We will also cover the lossless interactive communication results which allows multi-round communications.
In this paper, we consider aymptotically lossless, lossy (i.e., rate-distortion), and interactive limits for the problem of computing the functions of interest.
In this section we review the literature for the different approaches to the problem as well as the literature on quantization-based achievable schemes.

\subsection{Related Work---Lossless}
The two-terminal function computation with side information problem has been considered in \cite{Witsenhausen} and \cite{Orlitsky1} where two terminals (transmitter and receiver) each contain a source and the receiver wants to decode a function of the two sources. Earlier work by Witsenhausen considered the problem of minimizing the encoder's alphabet size with the constraint that the computation needs to be zero-error\cite{Witsenhausen}. He showed the minimum alphabet size is related to the chromatic number of the characteristic graph of the source. Orlitsky et al. considered a similar problem setup, but instead of zero-error they allowed an asymptotically small probability of error\cite{Orlitsky1}. With this assumption, they showed the fundamental limit on rate is the graph entropy of the characteristic graph.
 The distributed function computation problem has been considered in \cite{FunctionComputing} and \cite{FunctionCompression} where the problem is under the CEO setup\cite{CEO} where the CEO wants to compute a function of the sources from two or more users. Doshi et al. gave the rate region to the problem under a constraint that they term a ``zig-zag" condition \cite{FunctionCompression}. They showed that any achievable rate point can be realized  by graph coloring at each user and \ac{SW} encoding the colors. Sefidgaran et al. derived inner and outer bounds to the  rate region under for a class of tree structured networks, which includes the classical CEO problem formulation. They also showed that the two bounds coincide with each other if the sources are independent and hence obtained an expression for the rate region \cite{FunctionComputing}. The extremization functions that we are interested in are set-valued and the CEO only needs to know one value rather than the whole set of the function result. Under this setup, we give the fundamental limits of the minimum sum-rate to losslessly determine the extremizations. These results are in agreement with the results in \cite{FunctionComputing} and \cite{FunctionCompression} when the function is single-valued.

\subsection{Related Work---Lossy}
After Shannon introduced rate distortion function in source coding with a fidelity criterion \cite{Sha:Cod}, rate distortion theory was notably developed by Berger and Gallager \cite{Ber:Rat, Gal:Inf}. Recent work in rate distortion theory has been focused on lossy source coding by attempting to find efficient compression algorithms based on information theory.
Rate distortion theory was extended to multi-terminal systems in the 1970's by Berger and Tung \cite{Ber:Mul, Tun:Mul}. For point-to-point rate distortion problems, Arimoto and Blahut proposed numerical calculation algorithms based on alternating optimization to find the channel capacity and rate distortion function \cite{Ari:Alg, Bla:Com}. The convergence proof of Blahut and Arimoto' algorithms was developed by Csiszar \cite{Csi:Com} and Boukris \cite{Bou:Upp}. The generic CEO problem from multi-terminal source coding theory was introduced by Berger \cite{CEO} and the Quadratic Gaussian CEO rate region is known by Oohama \cite{Ooh:Rat} and Prabhakaran \cite{Pra:Rat}. A general outer bound to the CEO problem, not necessarily required Quadratic Gaussian CEO problem, was derived by Wagner \cite{Wag:Imp}. Lossy indirect function computation at the CEO was developed by Oohama \cite{Ooh:Rat} and Wyner \cite{Wyn:Rat}. An adaptation of the Blahut-Arimoto algorithm to the CEO model with independent sources is developed by the authors \cite{Ku:ABA}.

\subsection{Related Work---Scaler Quantization}
Recent work by Misra et al.\ considered the problem of \ac{DFSQ} \cite{MisGoyVar2011}.
By focusing on the high-rate regime and assuming a \ac{MSE} distortion, the authors are able to make several approximations to obtain distortion expressions that are optimal asymptotically (i.e., as the rate goes to infinity).
We assume a different distortion measure, derive an exact expression for the distortion as a function of the quantizer parameters, and derive necessary conditions for optimal parameters.
Moreover, our results hold for all rates.

Our focus on the use of \acp{SQ} as an achievable scheme is motivated by several results concerning the optimality of a layered architecture of quantization followed by entropy coding.
Zamir et al.\ considered the distributed encoding and centralized decoding of continuous valued sources and established that \emph{lattice} quantization followed by \ac{SW} encoding is optimal asymptotically in rate \cite{ZamBer1999}.
When the sources are Gaussian and the distortion is \ac{MSE}, local vector quantizers followed by \ac{SW} coding is optimal, not just asymptotically \cite{WagTavVis2008}.
For discrete valued random variables, \emph{scalar} quantization with block entropy encoding is optimal \cite{Ser2005}.
Each of the problem models considered in \cite{ZamBer1999,WagTavVis2008,Ser2005} can be understood as an instance of indirect distributed lossy source coding for the identity function.

\subsection{Related Work---Interaction}
\emph{Interactive communication} is the scheme that allows message passing forward and backward multiple times between two or more terminals. For the two terminals' interactive communication problem of lossy source reproduction, Kaspi first characterized the rate region in \cite{InteractiveTheory}. Followed by this,  Ishwar and Ma made some further contributions. They worked on both two and more than two terminals cases for computing any function of the sources in both lossy and lossless manner. They showed that interactive communication strictly improves the Wyner-Ziv rate-distortion function\cite{InteractionWynerZiv}. They also showed that in some distributed function computation problems, interactive communication can provide substantial benefits over non-interactive codes and infinite-many rounds of interaction may still improve the rate-region\cite{InteractiveFunctionComputation}.
In \secref{interactive-subband-allocation}, we consider resource allocation in the multiuser OFDMA systems that use rateless AWGN codes for downlink data transmission as a model of the distributed arg-max problem.  We propose an interactive communication scheme for this resource allocation problem. This scheme is inspired by the ideas of selective multiuser diversity \cite{SMUD} (SMUD) scheme as well as the multi-predefined thresholds \cite{MultiThresholds} scheme which is an extension of SMUD that set up multiple thresholds and allow the user nodes sending messages based on these thresholds.

\section{Model Specification}
\label{sec:model-specification}
As stated previously, we are considering the \(N\) user \ac{CEO} problem for estimating either \(\max\), \(\argmax\), or the pair \((\max, \argmax)\).
The \(i\textsuperscript{th}\) user observes the sequence \(\bm{X}_i \triangleq (X_{i,s} : s \in [S])\) of \emph{non-negative} random variables
\footnote{For any integer \(n\), let \([n] \triangleq \{1, \ldots, n\}\)}.
Let \(\bm{X} \triangleq (\bm{X}_i : i \in [N])\).
We assume that the sources are independent and identically distributed (\iid) across both users (\(i\)) and the sequence (\(s\)); that is
\begin{equation}
	f_{\bm{X}}(\bm{x}) = \prod_{i = 1}^{N}\prod_{s = 1}^{S}f_X(x_{n,s}).
\end{equation}
The quantities we are interested in for our problem are
\begin{equation}
	Z_A(s) \triangleq \{i| X_{i,s} = \max \{X_{i,s} : i \in [N]\}\}
\end{equation}
as the users with the maximum \(s\textsuperscript{th}\) source output and
\begin{equation}
	Z_M(s) \triangleq \max \{X_{i,s} : i \in [N]\}
\end{equation}
as the maximum \(s\textsuperscript{th}\) source output.
Specifically for the $\argmax$ case, we consider a class of problems where we need not estimate the set \(Z_A(s)\), but rather a representative user from this set.

\begin{longtable}{cl}
	\toprule \multicolumn{1}{c}{Symbol} & \multicolumn{1}{l}{Meaning} \\ \midrule
	\endfirsthead
	\multicolumn{2}{c}{\emph{\tablename\ \thetable---Continued from previous page}}\\ \toprule
	\multicolumn{1}{c}{Symbol} & \multicolumn{1}{l}{Meaning} \\ \midrule
	\endhead
	\bottomrule \multicolumn{2}{c}{\emph{Continued on next page}}\\
	\endfoot
	\bottomrule
	\endlastfoot
	\(\mathcal{C}(G(f))\) & set of all coloring method to color $G$ \\ 
	\(c(x_i)\) & coloring of source \(x_i\)\\
	\(\bm{\ell}\) & scalar quantizer decision boundaries\\ 
	\(\epsilon\) & fixed, but arbitrarily small value\\
	\(E_k\) & \(\expected{X | \ell_{k-1} \leq X \leq \ell_k}\)\\ 
	\(f\) & probability density function\\ 
	\(f_k\) & \(f_X(\ell_k)\)\\ 
	\(F\) & cumulative density function\\ 
	\(F_k\) & \(F_X(\ell_k)\)\\ 
	\(g\) & generic optimal resource allocation function\\
	\(G(V,E)\) & non directed graph with vertex set V and edge set E\\
	\(h\) & binary entropy function \(-p\log_2p -(1-p)\log_2(1-p)\)\\
	\(H(\cdot)\) & Shannon entropy\\
	\(H_G\) & graph entropy\\
	\(i\) & user indexing variable\\
	\(j\) & subcarrier indexing variable\\
	\(i,j\) & possible CQI levels/nodes in G usually in the proof in lossless limit section\\
	\(k\) & quantizer level indexing variable\\ 
	\(K\) & number of quantizer bins\\ 
	\(\kappa\) & rate region in Doshi's result \\ 
	\(L(\bm{\ell}, \bm{\mu})\) & Lagrangian\\ 
	\(L_f(X_1|X_2)\) & fundamental limit of the one-way Orlitsky's problem \\
	\(\mathcal{L}_k\) & \(k\)\textsuperscript{th} quantizer bin\\
	\(M_i\) & message index for \(i\)\textsuperscript{th} user\\
	\(N\) & number of users\\
	\(n\) & user indexing variable\\ 
	\(p\) & shorthand for probability\\
	\(p_k\) & \(F_X(\ell_k) - F_X(\ell_{k-1})\)\\ 
	\(R\) & total rate from all users\\
	\(R_i\) & rate from user \(i\)\\
	\(R_{HomSQ}\) & rate of \ac{HomSQ}\\ 
	\(R_{HetSQ}\) & rate of \ac{HetSQ}\\ 
	\(s\) & sequence indexing variable\\ 
	\(S\) & size of the source sequence\\
	\(t\) & discrete time index\\
	\(U_t\) & message sent from the BS to MSs at round t in interaction scheme\\
	\(\Gamma(G)\) & set of maximum independent sets of a graph\\
	\(W\) & maximum independent set\\
	\(X\) & random variable for channel capacity\\
	\(\mathcal{X}\) & support set for random variable \(X\)\\
	\(Z\) & generic resource allocation\\
	\(Z_A\) & arg-max of values\\
	\(Z_M\) & max of values\\
	\(Z_{M,A}\) & both max \& arg-max of values\\
	\(\lambda\) & possible threshold in interaction scheme\\
	\(\lambda^*_t\) & optimal threshold at round t in interaction scheme\\
	\(\mu\) & Lagrange multiplier (slope) for rate-distortion computation\\
	\(\nu\) & Lagrange multiplier associated with equality constraints\\
	\(\phi\) & encoding mapping\\
	\(\psi\) & decoding mapping\\
	\caption{Notation Guide}\label{tbl:notation}
\end{longtable}

\subsection{Distortion Measures}
Two typical distortion measures are Hamming and squared error, neither of which are appropriate for the problems of interest (\secref{introduction}, Examples~\ref{exmp:ofdma}--\ref{exmp:sensor-network}).
Hamming distortion is ``all-or-nothing'' in that all estimation errors are treated the same.
Squared error distortion more heavily penalizes larger estimation errors, but treats under- and over-estimation the same.
For the problems of interest, over-estimation needs to be more heavily penalized then under-estimation.
With that in mind, we propose the following distortion measures.
For estimating the \(\max\), the distortion measures linearly penalizes underestimation of the maximum value; we can think of this as the lost revenue (difference between what you could have gotten and what you got) when an asset is priced below market value.
It also captures the loss when the estimated max rate exceeds the actual max; continuing the analogy, this is the case where an asset does not sell because it is priced above market value.
\begin{equation}
\label{eq:D_max}
	d_M((X_{1,s},\ldots,X_{N,s}), \hat{Z}_M(s)) = \begin{dcases*}
		Z_M(s)- \hat{Z}_M(s) & if \(\hat{Z}_M(s) \leq Z_M(s)\)\\
		Z_M(s) & otherwise
	\end{dcases*}
\end{equation}
For estimating the \(\argmax\), a similar distortion measure is utilized.
The distortion measures the loss between source value of the user with the actual max and the source value for the user estimated to have the max.
Unlike the previous case, the \ac{CEO} cannot make an over-estimation error.
\begin{equation}
\label{eq:D_argmax}
	d_A((X_{1,s},\ldots,X_{N,s}), \hat{Z}_A(s)) = \begin{dcases*}
	0 & if $\hat{Z}_A \in Z_A$ \\
	Z_M(s) - X_{\hat{Z}_A(s),s} & otherwise
	\end{dcases*}
\end{equation}
Finally, for estimating the pair of values \((\max, \argmax)\) we propose a distortion measure that is a hybrid of the previous two.
The distortion is a combination of under-estimating the max value, provided the estimate does not exceed the value of the user estimated as having the max value.
It also captures the loss due to over-estimation, both exceeding the estimated \(\argmax\) user's value or exceeding the actual max value.
\begin{equation}
\label{eq:D_pair}
	d_{M,A}((X_{1,s},\ldots,X_{N,s}), (\hat{Z}_M(s), \hat{Z}_A(s))) = \begin{dcases*}
		Z_M - \hat{Z}_M(s) & if \(\hat{Z}_M(s) \leq X_{\hat{Z}_A(s),s}\)\\
		Z_M & otherwise
	\end{dcases*}
\end{equation}
Depending on the problem formulation being considered, let \(d(s)\) be
\begin{enumerate}
	\item \(d_{M,A}((X_{1,s},\ldots,X_{N,s}), (\hat{Z}_M(s), \hat{Z}_A(s)))\);
	\item \(d_A((X_{1,s},\ldots,X_{N,s}), \hat{Z}_A(s))\), or;
	\item \(d_M((X_{1,s},\ldots,X_{N,s}), \hat{Z}_M(s))\)
\end{enumerate}
and define
\begin{equation}
	d((\bm{X}_1, \ldots, \bm{X}_N), \hat{\bm{Z}}) = \frac{1}{S}\sum_{s = 1}^S d(s)
\end{equation}
as the distortion between sequences.
Finally, denote
\begin{equation}
	D = \expected{d((\bm{X}_1, \ldots, \bm{X}_N), \hat{\bm{Z}})}
\end{equation}
where the expectation is with respect to joint distribution on the sources.
In the next section, we consider the problem of finding the minimum sum rate necessary for computing the different extremization functions when the distortion is constrained \(D = 0\).
Later, we will consider the problem of the minimum sum rate necessary for computing the different extremization functions with a non-zero upper bound on the distortion \(D\).

\section{Lossless Extremization}
\label{sec:lossless-subband-allocation}
In this section, we determine the minimum amount of information necessary to remotely solve the extremization problems in a Shannon lossless sense. We begin by providing an achievable scheme for the $\argmax$ problem based on graph coloring in \secref{lossless_Nu_achievable}. We then prove in \secref{lossless_Nu_converse} that this scheme achieves a fundamental limit.  We also show via a computation of the fundamental limits, that no rate can be saved relative to simply forwarding the observations in the $\max$ and $(\max, \argmax)$ problems unless $\min \mathcal{X} = 0$.

In \cite{Orlitsky1}, a related problem is considered in which the node observing $\boldsymbol{X}_1$ sends a message to the node observing $\boldsymbol{X}_2$ in such a manner that the function $\boldsymbol{f}^S(\boldsymbol{X}_1^S,\boldsymbol{X}_2^S)=[f(X_{1,s},X_{2,s})| s \in [S]]$, taking values from the set $\mathcal{Z}^S$, can be computed losslessly.  In this problem, a rate $R$ is said to be achievable if for every $\epsilon > 0$ there exists a sufficiently large $S$ and $K$ with $R \geq \frac{K}{S}$, and an encoder
$
\varphi :  \mathcal{X}^S \rightarrow \{0,1\}^K
$ and a decoder
$
\psi : \{0,1\}^K \times \mathcal{X}^S \rightarrow \mathcal{Z}^S
$
such that
$\mathbb{P}(\psi(\varphi(\boldsymbol{X}_1^S),\boldsymbol{X}_2^S) \ne \boldsymbol{f}^S(\boldsymbol{X}_1^S,\boldsymbol{X}_2^S)) < \epsilon
$.
Orlitsky and Roche proved that for given $X_1$, $X_2$ and $f$, the infimum of the set of achievable rates is 
\begin{equation}
	\label{eq:Orlitsky}
	L_f(X_1|X_2) = H_G(X_1|X_2)
\end{equation}
where $H_G(X_1|X_2)$ is the conditional graph entropy of the characteristic graph of this problem in \cite{Orlitsky1}. The characteristic graph $G$ of $X_1$, $X_2$, and $f$ is a generalization of the definition given by Witsenhausen\cite{Witsenhausen}. Its vertex set is the support set $\mathcal{X}$ of $X_1$, and distinct vertices $x_1$, $x_1'$ are adjacent if there is a $x_2$ such that $p(x_1,x_2), p(x_1',x_2) > 0$ and $f(x_1,x_2) \ne f(x_1',x_2)$. The conditional graph entropy is
\begin{equation}
	\label{eq:condGentropy}
	H_G(X_1|X_2) \triangleq \min_{W - X_1 - X_2, X_1 \in W \in \Gamma(G)} I(W;X_1|X_2)
\end{equation}
where $\Gamma(G)$ is the set of all maximal independent sets in $G$, $W$ is a random variable that has $\Gamma(G)$ as its support set, and the minimization is over all conditional probabilities $p(w|x_1)$ which is supported on those maximal independent sets $w$ containing the vertex $x_1$, with the constraint that $W$, $X_1$ and $X_2$ form a Markov chain.

\indent Additionally, conditional graph entropy can be related to coloring a certain product graph.  In particular, the OR-product graph $G^S_1(V_S, E_S)$, based on the characteristic graph $G_1$ of $X_1$, $X_2$ and $f$, has a vertex set $V_S = \mathcal{X}^S$, and distinct vertices $(x_{1,1},\ldots,x_{1,S})$,$(x_{1,1}',\ldots,x_{1,S}')$ are connected if there exists an edge between $x_{1,s}$ and $x_{1,s}'$ in $G_1$ for any $s$. In \cite{FunctionCompressionSideInfo}, Doshi et al. showed that minimum-entropy coloring the OR-product graph, followed by lossless compression of the colors with \ac{SW} coding, yields a rate proportional to the conditional chromatic entropy, and can asymptotically reach the lower limit set out by the conditional graph entropy

 \begin{equation}
\lim_{S \rightarrow \infty} \min_{c \in \mathcal{C}_{\epsilon}(G_1^S(f))}\frac{1}{S}H(c(X_1)) = H_G(X_1|X_2)
\end{equation}
where $\mathcal{C}_{\epsilon}(G_1^S(f))$ is the set of all $\epsilon$-colorings of the product graph.

For the decentralized model where two users communicate with a CEO attempting to losslessly compute a function, Doshi et al. gave the rate region when the problem satisfies a given \emph{zig-zag condition}, which requires that for any ($x_1, x_2$) and ($x_1', x_2'$) in $\mathcal{X}_1\times\mathcal{X}_2$, $p(x_1,x_2)>0$ and $p(x_1',x_2')>0$ imply either $p(x_1,x_2')>0$ or $p(x_1',x_2)>0$ \cite{FunctionCompression}. The key idea is to let each user do an $\epsilon$-coloring \cite{FunctionCompression} of the OR-product graph of its own source and transmits the color by a \ac{SW} code. 

\cite{FunctionCompression} showed in Theorem 16 that the rate-region for the aforementioned distributed function computation problem  under the zig-zag condition is the set closure of $\kappa$, where $\kappa$ is the intersection of $\kappa^{\epsilon}$ for all $\epsilon>0$, and $\kappa^{\epsilon}$ is

\begin{equation} \label{eq:coloring_set}
\kappa^{\epsilon} = \displaystyle \bigcup_{n=1}^{\infty}\bigcup_{(c_{x_1}^n,c_{x_2}^n)}\mathcal{R}^n(c_{x_1}^n,c_{x_2}^n)
\end{equation}

\noindent where the regions $\mathcal{R}^n(c_{x_1}^n,c_{x_2}^n)$ are given by

\begin{equation} \label{eq:coloring_region}
\begin{aligned}
& R_{x_1} \ge \frac{1}{n}H(c_{x_1}^n({\bf X_1})|c_{x_2}^n({\bf X_2})) \\
& R_{x_2} \ge \frac{1}{n}H(c_{x_2}^n({\bf X_2})|c_{x_1}^n({\bf X_1})) \\
& R_{x_1}+R_{x_2} \ge \frac{1}{n}H(c_{x_1}^n({\bf X_1}),c_{x_2}^n({\bf X_2})).
\end{aligned}
\end{equation}
In Theorem 18, \cite{FunctionCompression} showed that the difference of the minimum sum-rate and $H_G(X_1|X_2) + H_G(X_2|X_1)$ is bounded by
\begin{equation} \label{eq:doshi_converse}
H_G(X_1|X_2) + H_G(X_2|X_1) - \left(R_{x_1} + R_{x_2}\right) \le \min \{I_{G_1} (X_1;X_2), I_{G_2} (X_1;X_2)\}
\end{equation}
where $I_{G_1} (X_1;X_2)$ is the graph information of $X_1$, and the right hand side is zero when $X_1$ and $X_2$ are independent. Note that $H_G(X_1|X_2) = H_G(X_1)$ when the sources are independent, where the graph entropy $H_G(X_1)$ is
\begin{equation}
	\label{eq:Gentropy}
	H_G(X_1) \triangleq \min_{W - X_1 - X_2, X_1 \in W \in \Gamma(G)} I(W;X_1).
\end{equation}
Hence when the sources are independent, the rate-region is
\begin{equation} \label{eq:rate_region_ind}
\begin{aligned}
& R_{x_1} \ge H_G(X_1) \\
& R_{x_2} \ge H_G(X_2) \\
& R_{x_1}+R_{x_2} \ge H_G(X_1)+H_G(X_2).
\end{aligned}
\end{equation}

Doshi et al. consider a very general class of problems, for which in general it is necessary to express the rate region in terms of the $\epsilon$-coloring, which essentially is an valid coloring on a high probability subset of the characteristic graph. We will now show how to apply these ideas and related ones to the extremization problems under investigation. In particular, we will show in \secref{lossless_Nu_converse} that we can achieve the fundamental limits of the sum-rate in the extremization problems by normally coloring the original characteristic graph as described in \secref{lossless_Nu_achievable}, thereby removing the need for both OR-product graph and $\epsilon$-coloring.

\subsection{Achievable Schemes of Determining the $\argmax$ Function} \label{sec:lossless_Nu_achievable}
In this subsection we present an achievable scheme for determining the $\argmax$ as we will show in \secref{lossless_Nu_converse}, there is no need for sophisticated coding schemes for the max and both functions, as simple Huffman coding achieves the fundamental limits of the sum-rates for these functions. 

We first consider N users, each observing $\boldsymbol{X}_n = (X_{n,s}| s \in \{1,\ldots,S\}, X_{n,s} \in \mathcal{X})$ and assume that $\mathcal{X} = \{\alpha_1,\alpha_2,\ldots,\alpha_L\}$ s.t. $0 \le \alpha_1 < \alpha_2 < \ldots < \alpha_L$ and $\mathbb{P}(X=\alpha_i) > 0$ for all $i \in [L]$ w.l.o.g.. For each element $X_{1,s},\ldots,X_{N,s}$ of these sequences, we are interested in the aggregate rate required to enable the CEO to learn a $\hat{Z}_A(s)$ in the $\argmax$ such that
\begin{equation}
	\mathbb{E}[d_A((X_{1,s},\ldots,X_{N,s}), \hat{Z}_A(s))] = 0.
\end{equation}
\begin{defn}
 A rate $R$ will be said to be achievable if for every $\epsilon$ there exists $S,R_1,\ldots,R_N$ with $R=\sum_{n=1}^{N}R_n$, N encoder maps $\phi_n:\mathcal{X}^S \rightarrow \{0,1\}^{S\cdot R_n}$, $n\in\{1,\ldots,N\}$, and a decoder map $\psi:\{0,1\}^{S \cdot R_1} \times \{0,1\}^{S \cdot R_2} \ldots \times\{0,1\}^{S \cdot R_N} \rightarrow \{1,\ldots,N\}^S$ such that 
 $
 d_A((X_{1,s},\ldots,X_{N,s}), \psi(\phi_1(\boldsymbol{X}_1^S),\phi_2(\boldsymbol{X}_2^S),\ldots,\phi_N(\boldsymbol{X}_N^S))) < \epsilon.
 $
\end{defn}
We say a tie happens in the $\argmax$ of the $s^{th}$ sources if two or more users attain the maximum value. Note that the $\argmax$ is not unique in such a case, because when a tie happens, the CEO can choose any user that achieves the maximum and will attain zero distortion. In other words, the extremization function is not uniquely determined everywhere. This will be useful when minimizing the amount of information necessary to determine this function.

\begin{defn}
	A response from the \(N\) users is \(j\)-ambigous if there are \(j\) maximizers.
\end{defn}
\begin{lem}
	The number of \(j\)-ambigous responses is
	\begin{equation}
		A_{j, N}(K) = \binom{N}{j} \sum_{i = 1}^{K - 1} i^{N - j}.
	\end{equation}
	The number of possible deterministic tie-breaking \(\argmax\) functions is
	\begin{equation}
		C_N(K) = N \prod_{j = 1}^{N} j^{A_{j,N}(K)}
	\end{equation}
\end{lem}
\begin{IEEEproof}
	There are
	\begin{equation}
		\binom{N}{j} (i - 1)^{N - j}
	\end{equation}
	possible responses from the \(N\) users such that \(i\) is the index of the maximum value and \(j\) is the number of maximizers.
	Summing over \(i\) gives the total number of \(j\)-ambgious responses.
	For each of these responses, we have \(j\) possible values for the candidate \(\argmax\) function; if \(i = 1\), then we have \(N\) possible values for the candidate function.
	Taking the product over \(j\) and making suitable changes of variable gives the result of the lemma.
\end{IEEEproof}
\begin{rem}
	We have
	\begin{equation}
		\begin{aligned}
			C_2(K) &= 2^K \\
			C_3(K) &= 3^K 2^{\frac{3 K (K - 1)}{2}}\\
			C_4(K) &= 4^K 3^{2 K (K - 1)} 2^{(K - 1) K (2K - 1)}\\
			C_5(K) &= 5^K 4^{\frac{5 K (K - 1)}{2}} 3^{\frac{10 (K - 1) K (2K - 1)}{6}} 2^{\frac{10 K^2 (K - 1)^2}{4}}
		\end{aligned}
	\end{equation}
	These functions are plotted as a function of \(K\) in \fig{argmax-candidates}.
	We see that the number of functions \(\mathcal{X}^N \mapsto [N]\) that returns the unique maximizer when there is a single maximizer and deterministically breaks ties when there are more than one maximizer is extremely large, even for small values of \(N\) and \(K\).
	\begin{figure}
		\centering
		\includegraphics[width=252.0pt]{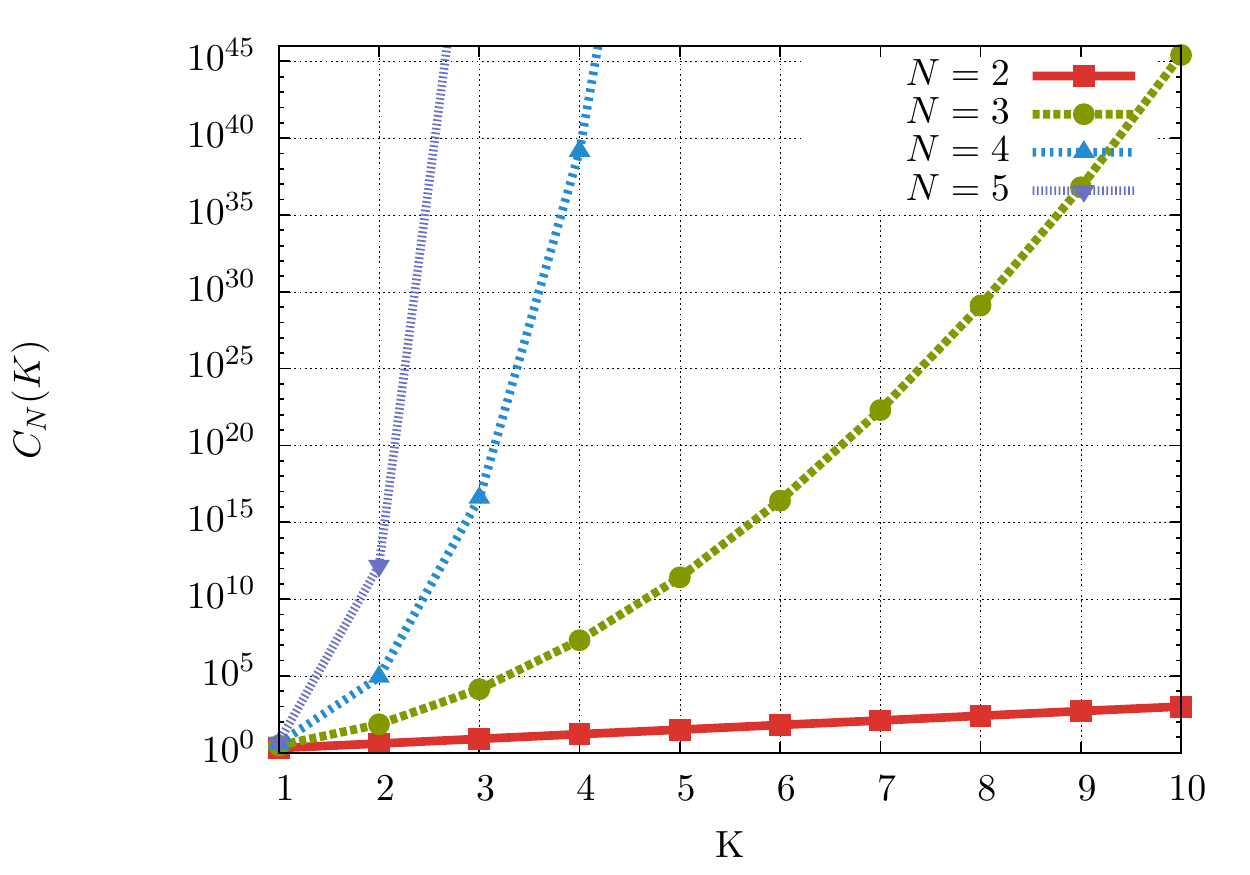}
		\caption{The number of candidate \(\argmax\) functions as a function of \(K\) the number of values the users can take parameterized by \(N\) the number of users.}
		\label{fig:argmax-candidates}
	\end{figure}
\end{rem}
We are able to realize a reduction in rate over the na\"{\i}ve approach of each user independently Huffman encoding their observations by searching over the space of functions that are consistent with \(\argmax\) (defined formally \eq{candidate_f_argmax}) in \thmref{N-user-achievability}.
Despite the incredibly large search space, we develop a characterization of a subset of these candidate \(\argmax\) functions and provide an expression for the rate acheived by these functions in \thmref{optimal_function}.
In \secref{lossless_Nu_converse}, we establish that this rate is in fact the best possible sum rate that can be attained.

\begin{defn}
A function $f: \mathcal{X}^N \rightarrow \{1,\ldots,N\}$ is a candidate \(\argmax\) function if and only if
\begin{equation} \label{eq:candidate_f_argmax}
\mathbb{E}[d_A((X_{1},\ldots,X_{N}),f(X_{1},\ldots,X_{N}))]=0.
\end{equation}
Let $\mathcal{F}_{A,N}$ be the set of all such candidate $\argmax$ functions with $N$ inputs. For any $f\in \mathcal{F}_{A,N}$, it indicates the index of a user attaining the $\max$.
\end{defn}
\begin{thm} \label{thm:N-user-achievability}
An achievable sum-rate for losslessly determining the $argmax$ among a set of $N$ users is
\begin{equation}\label{eq:musers_def}
R_A =  \min_{f_N \in \mathcal{F}_{A,N}} \sum_{n=1}^{N} \min_{c_n \in \mathcal{C}(G_n(f_N))} H(c_{n}(X_n))
\end{equation}
where the first minimization is over all candidate \(\argmax\) functions, and $\mathcal{C}(G_n(f))$ is the set of all colorings of the characteristic graph of user $n$ w.r.t. the function $f_N$.
\end{thm}
\begin{IEEEproof}
This achievable scheme directly follows the result from \cite{FunctionCompression} with a block size $S=1$ and by observing that an ordinary coloring is also an $\epsilon$-coloring.  Following  \cite{FunctionCompression}, we color the characteristic graph for each $\argmax$ function and transmit the colors by a \ac{SW} code.  \eq{musers_def} is the minimum sum-rate over all such schemes w.r.t. all candidate $\argmax$ functions and all possible coloring schemes on the OR-product graph of size $S=1$.
\end{IEEEproof}

In order to solve the optimizations in \eq{musers_def}, the following two lemmas will be useful. Throughout the discussion below, we will use $\{\alpha_i,\alpha_j\} \in G$ to denote the existence of an edge between node $\alpha_i$ and $\alpha_j$ in the characteristic graph $G$, and use $\{\alpha_i,\alpha_j\} \not \in G$ to denote that there is no such edge.

\begin{lem}\label{lem:diffColors}
 For any function $f_N \in \mathcal{F}_{A,N}$ that determines the $\argmax$, no 3 vertices can form an independent set in its characteristic graph $G_i(f_N)$ for any user $i$.
\end{lem}

\begin{IEEEproof}
	For any 3 vertices, there must exist two of them that their indices are not adjacent in number, say vertex $\alpha$ and vertex $\beta$, hence $\exists$ vertex $\gamma$, $\alpha<\gamma<\beta$ such that $f_N(x_1=\gamma,\ldots, x_{i-1}=\gamma,x_i=\alpha,x_{i+1}=\gamma,\ldots,x_N = \gamma) \ne f_N(x_1=\gamma,\ldots, x_{i-1}=\gamma,x_i=\beta,x_{i+1}=\gamma,\ldots, x_N =\gamma)$. Therefore, an edge must exist between $\alpha$ and $\beta$ in $G_i(f_N)$, and they can not be in the same independent set.
\end{IEEEproof}

\begin{lem}\label{lem:2}
 For any function $f_N \in \mathcal{F}_{A,N}$ that determines the $\argmax$, if $\{\alpha,\beta\} \not \in G_i(f_N)$, then $\{\alpha,\beta\} \in G_n(f_N)\ \forall n \in [N]\setminus\{i\} $.
\end{lem}

\begin{IEEEproof}
	Without loss of generality, we suppose $\alpha<\beta$. From the condition that $\{\alpha,\beta\} \not \in G_i(f_N)$, we know that $\forall \boldsymbol{x}_{\setminus \{i\}}\in \mathcal{X}^{N-1}$, $f_N(x_i=\alpha,\boldsymbol{x}_{\setminus \{i\}})=f_N(x_i=\beta,\boldsymbol{x}_{\setminus \{i\}})$.
In particular, we consider the following input sequences
\begin{equation}
\begin{aligned}
& \boldsymbol{x}^1 = (x_1^1,\ldots,x_N^1)\ s.t.\ x_n^1 = \alpha\ \forall n \in [N], \\
& \boldsymbol{x}^2 = (x_1^2,\ldots,x_N^2)\ s.t.\ x_i^2 = \beta, x_n^2=\alpha\ \forall n \in [N]\setminus \{i\}, \text{and} \\
& \boldsymbol{x}^3 = (x_1^3,\ldots,x_N^3)\ s.t.\ x_j^3 = \beta, x_n^3=\alpha\ \forall n \in [N]\setminus \{j\}. \\
\end{aligned}
\end{equation}
Begin by observing that $f_N(\boldsymbol{x}^2) = i$ and $f_N(\boldsymbol{x}^3) = j$ since $\beta > \alpha$ and the positions associated with other users are all $\alpha$. Next, we observe that $f_N(\boldsymbol{x}^1)=f_N(\boldsymbol{x}^2)$ because $\{\alpha,\beta\} \not \in G_i(f_N)$ by assumption. This then implies
$
f_N(\boldsymbol{x}^1) \ne f_N(\boldsymbol{x}^3) 
$, and hence there exists $\boldsymbol{x}_{\setminus \{j\}} = (\alpha,\ldots,\alpha)$ such that the function result differs for $x_j=\alpha$ and $x_j=\beta$, and there is an edge between $x_j=\alpha$ and $x_j=\beta$.
\end{IEEEproof}

As we mentioned above, the minimum achievable sum-rate $R_A$ depends on how we break the ties (i.e. how we choose the candidate $\argmax$ function). Denote $\mathcal{F}_{A,N}^*$ as the set of all candidate $\argmax$ functions that achieve $R_A$, the following theorem specifies the solution to the optimization problem introduced in \thmref{N-user-achievability}.

\begin{thm} \label{thm:optimal_function}
There exists a series of functions $\left\{f^*_n\left| n \in [N]\right.\right\}$ where $f^*_n$ is a candidate $\argmax$ function for $n$ users satisfying the properties that 
\begin{enumerate}
\item $f^*_{1} (x) = 1$ for any $x \in \{\alpha_1,\ldots,\alpha_L\}$,

\item $\forall \boldsymbol{x} \in \mathcal{S}^{-}_{n}(\alpha_i)$, where $\alpha_i \in \mathcal{X}$ and $\mathcal{S}^{-}_{n}(\alpha_i)= \{(x_1,\ldots,x_n)| x_1 = \alpha_i, \max \{\boldsymbol{x}_{\setminus\{1\}}\} < \alpha_i\}$,
\begin{equation} \label{eq:NDisAmb1}
f^*_n(\boldsymbol{x}) = 1,
\end{equation}

\item $\forall \boldsymbol{x} \in \mathcal{S}^{=}_{n}(\alpha_i)$, where $\alpha_i \in \mathcal{X}$ and $\mathcal{S}^{=}_{n}(\alpha_i)= \{(x_1,\ldots,x_n)| x_1 = \alpha_i, \max \{\boldsymbol{x}_{\setminus\{1\}}\} = \alpha_i\}$,
\begin{equation} \label{eq:NDisAmb2}
f^*_n(\boldsymbol{x}) =
\begin{dcases}
1 & \mod (n,2)=\mod(i,2)\\
f^*_{n-1}(\boldsymbol{x}_{\setminus\{1\}}) + 1 & \text{otherwise},
\end{dcases}
\end{equation}

\item $\forall \boldsymbol{x} \in \mathcal{S}^{+}_{n}(\alpha_i)$, where $\alpha_i \in \mathcal{X}$ and $\mathcal{S}^{+}_{n}(\alpha_i)= \{(x_1,\ldots,x_n)| x_1 = \alpha_i, \max \{\boldsymbol{x}_{\setminus\{1\}}\} > \alpha_i\}$,

\begin{equation} \label{eq:NDisAmb3}
f^*_n(\boldsymbol{x}) = f^*_{n-1}(\boldsymbol{x}_{\setminus \{1\}})+1,
\end{equation}
\end{enumerate}
such that
\begin{enumerate}
\item The minimum sum-rate achieved by graph coloring w.r.t. $f^*_N$ is
\begin{equation} \label{eq:musers_rate}
		R(f^*_N) = - (N-2) \sum_{i=1}^{L} p_i\log_2 p_i - \sum_{i=1}^{L-1}p_{i,i+1}\log_2 p_{i,i+1} - p_1\log_2 p_1 - p_L\log_2 p_L
\end{equation}
where $p_{i} = \mathbb{P}(\mathbb{X}=\alpha_i)$ and $p_{i,i+1}=p_i+p_{i+1}$,
\item $f^*_N \in \mathcal{F}_{A,N}^*$, i.e. $R_A$ can be achieved by $f^*_N$.
\end{enumerate}
\end{thm}

\begin{figure}
	\centering
	\includegraphics[width=252.0pt]{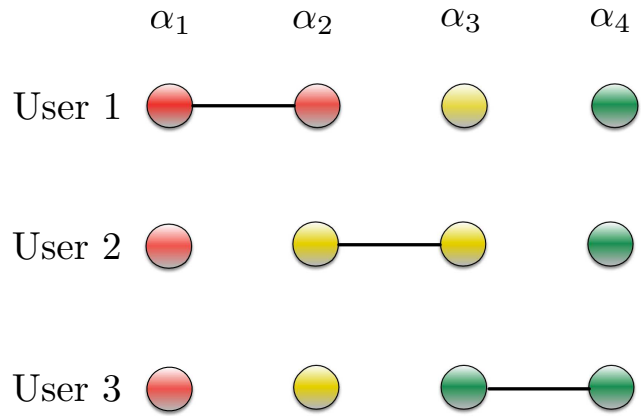}
	\caption{Coloring the characteristic graph under the optimal function with $N=3$ and $L=4$}
	\label{fig:N3L4Graph}
\end{figure}

\begin{IEEEproof}
\appendixproof{proof_optimal_function}
\end{IEEEproof}
\begin{exmp}[$N=3$ $L=4$ case]
The properties that $f_3^*$ must obey become
\begin{enumerate}
\item $\forall \boldsymbol{x} \in \mathcal{S}^{-}_{3}(\alpha_i)$, where $\alpha_i \in \mathcal{X}$ and $\mathcal{S}^{-}_{3}(\alpha_i)= \{(x_1,x_2,x_3)| x_1 = \alpha_i, \max \{x_2,x_3\}\} < \alpha_i\}$,
\begin{equation}
f^*_3(\boldsymbol{x}) = 1,
\end{equation}

\item $\forall \boldsymbol{x} \in \mathcal{S}^{=}_{3}(\alpha_i)$, where $\alpha_i \in \mathcal{X}$ and $\mathcal{S}^{=}_{3}(\alpha_i)= \{(x_1,x_2,x_3)| x_1 = \alpha_i, \max \{x_2,x_3\} = \alpha_i\}$,
\begin{equation}
f^*_3(\boldsymbol{x}) =
\begin{dcases}
1 & i\; \text{odd}\\
f^*_{2}(x_2,x_3) + 1 & \text{otherwise},
\end{dcases}
\end{equation}
where
\begin{equation} \label{eq:N2function}
f^*_{2}(x_2,x_3)=
\begin{dcases}
1 & x_2 = \alpha_i > x_3 \\
2 & x_2 < x_3 = \alpha_i \\
1 & x_2 = x_3 = \alpha_i , i\; \text{even} \\
2 & x_2 = x_3 = \alpha_i, i\; \text{odd} \\
\end{dcases}
\end{equation}
\item $\forall \boldsymbol{x} \in \mathcal{S}^{+}_{3}(\alpha_i)$, where $\alpha_i \in \mathcal{X}$ and $\mathcal{S}^{+}_{3}(\alpha_i)= \{(x_1,x_2,x_3)| x_1 = \alpha_i, \max \{x_2,x_3\} > \alpha_i\}$,
\begin{equation}
f^*_3(\boldsymbol{x}) = f^*_{2}(x_2,x_3)+1,
\end{equation}
\end{enumerate}

For convenience, we illustrate the complement characteristic graph as well as the coloring method in \fig{N3L4Graph}. Note that an edge connects two nodes in the complement graph represents that the two nodes forms an independent set in the original graph.
\end{exmp}

\begin{figure}
	\centering
	\includegraphics[width=252.0pt]{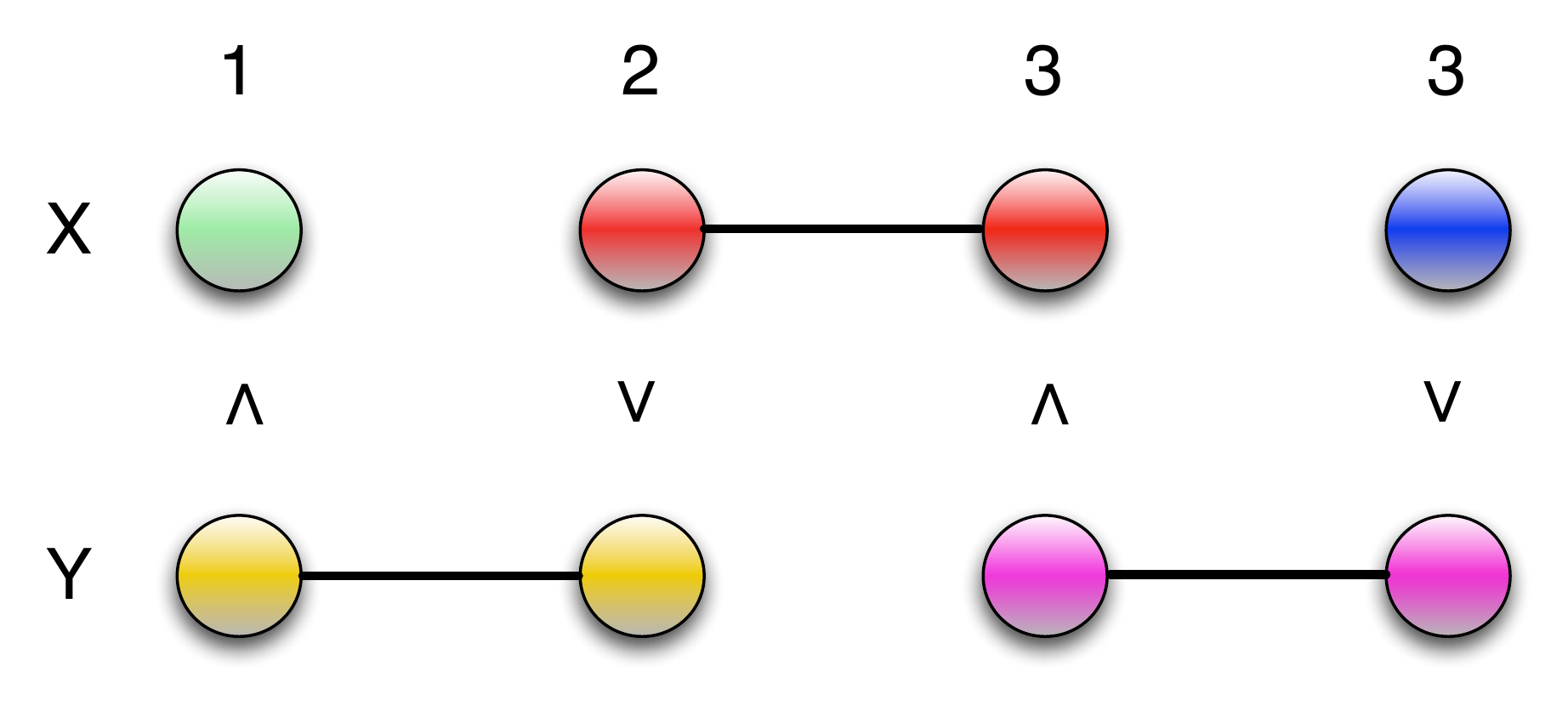}
	\caption{Coloring the characteristic graph under the optimal function with $N=2$ and $L=4$}
	\label{fig:OptcharGraph}
\end{figure}
For the case that $N=2$, since there will be $L$ different ties that need to be distinguished, and $2^L$ different candidate functions that need to be considered, we have:
\begin{cor}[$N=2$ case]\label{cor:1}
Among all $2^L$ $\argmax$ functions, the one that achieves the lowest sum-rate under minimum entropy graph coloring satisfies the property that for all $\alpha_i \in \mathcal{X}$,
\begin{equation} \label{eq:twoDisAmb}
f^*_2(\alpha_i,\alpha_i) = 
\begin{dcases*}
1, & $i$ odd\\
2, & $i$ even.\\
\end{dcases*}
\end{equation}
\end{cor}
\begin{figure}
	\centering
	\includegraphics[width=252.0pt]{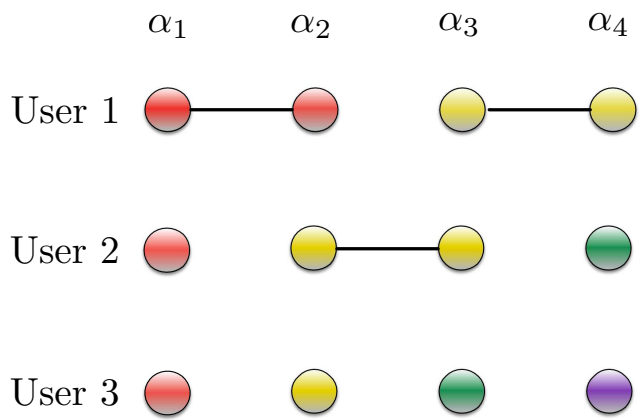}
	\caption{Coloring the characteristic graph under the non-recursive optimal function with $N=3$ and $L=4$}
	\label{fig:AlternativeN3L4}
\end{figure}
\begin{rem}
Another type of candidate $\argmax$ function which leads to the complement characteristic graph as shown in \fig{AlternativeN3L4}
with rate
\begin{equation}
R(f_N^*)= (N-2)H(X) + \min_{c_{1} \in \mathcal{C}(G_1(f^*_2))} H(c_{1}(X_1))+\min_{c_{2} \in \mathcal{C}(G_2(f^*_2))} H(c_{2}(X_2))
\end{equation}
although do not have the recursive property, can still achieve \eq{musers_rate}, and this structure can be interpreted as $2$ of the $N$ users do graph coloring by \corref{1}, and the rest of the $N-2$ users Huffman encode their own sources.
\end{rem}

Having introduced this scheme, we will show in the next section that no scheme can have a higher rate-savings than this one.


\subsection{Converse of Determining the Extremization Functions} \label{sec:lossless_Nu_converse}
The following lemma is necessary to aid in drawing the conclusion that joint graph coloring achieves the fundamental limit and there is no benefit from the OR-product graph for $\argmax$.
\begin{lem}\label{lem:3}
For any given candidate $\argmax$ function $f_N \in \mathcal{F}_{A,N}$, the conditional probability $p(w|x)$, which is supported on the maximal independent sets $w$ containing the vertices $x$, to achieve the minimum mutual information in the graph entropy expression \eq{Gentropy} must be either $1$ or $0$ for all $n \in [N]$.
\end{lem}
\noindent \emph{Proof:} We prove this by showing that no vertex exists in two different maximal independent sets. Without loss of generality, we consider vertex $\alpha_i$ in $X_1$'s characteristic graph. By \lemref{diffColors}, the two maximal independent sets that $\alpha_i$ may belong to are $w_1=\{\alpha_{i-1},\alpha_i\}$ and $w_2=\{\alpha_i,\alpha_{i+1}\}$. If vertex $\alpha_i \in w_1$ under the $\argmax$ function $f_N$ (which means there is no edge between $\alpha_{i-1}$ and $\alpha_i$), then  $\forall \boldsymbol{x}_{\setminus \{1\}} \in \mathcal{X}^{N-1}$, we have
\begin{equation}
f_N(\alpha_i, \boldsymbol{x}_{\setminus \{1\}}) = f_N(\alpha_{i-1}, \boldsymbol{x}_{\setminus \{1\}}).
\end{equation}
In particular, there exists $ \boldsymbol{x}_{\setminus \{1\}} = (\alpha_i, \alpha_{i-1},\ldots,\alpha_{i-1})$ such that
\begin{equation}
f_N(\alpha_i,  \boldsymbol{x}_{\setminus \{1\}})=f_N(\alpha_{i-1}, \boldsymbol{x}_{\setminus \{1\}}) = 2,
\end{equation}
and obviously
\begin{equation}
f_N(\alpha_{i+1}, \boldsymbol{x}_{\setminus \{1\}}) = 1.
\end{equation}
Therefore, $x_1=\alpha_i$ is connected to $x_1=\alpha_{i+1}$, and the set $w_2=\{\alpha_i,\alpha_{i+1}\}$ is not an independent set in $X_1$'s characteristic graph, and we have
\begin{equation}
p(w|x_1=\alpha_i)=
	\begin{dcases*}
	1, & $w=\{\alpha_{i-1},\alpha_{i}\}$ \\
	0, & otherwise. \\
	\end{dcases*}
\end{equation}
\hfill $\blacksquare$
\begin{thm} \label{thm:converse_argmax}
To losslessly determine the $\argmax$, the fundamental limit of the sum-rate can be achieved by coloring the characteristic graph of each user, hence the OR-product graph is not necessary.
\end{thm}
\begin{IEEEproof}
As reviewed at \eq{rate_region_ind}, the fundamental limit of the sum-rate with independent sources problems is the sum of the graph entropy, i.e. $R_A^* = \sum_{n=1}^N R_n$ with
$
 R_n = H_{G}(X_n)
$.
By \lemref{3}, for any given candidate $\argmax$ function,
\begin{equation} \label{eq:Gentropy_Hw}
\begin{aligned}
H_{G}(X_n) & =  \min_{p(w_{n}|x_n) \in \{0,1\}, w_n \in \Gamma(G_n)} I(W_{n};X_n) \\
                         & = \min_{p(w_{n}|x_n) \in \{0,1\}, w_n \in \Gamma(G_n)} H(W_{n}) - H(W_{n}|X_n)  \\
                         & = \min_{p(w_{n}|x_n) \in \{0,1\}, w_n \in \Gamma(G_n)} H(W_{n}).
\end{aligned}
\end{equation}
Note that the proof of \lemref{3} implies that the maximal independent sets are disjoint, and the fact that one can always Huffman encode the maximal independent sets with any given distribution. Consider using colors to represent the maximal independent sets, then Huffman encode the sets is the same as Huffman encode these colors. This color representation is a normal coloring method w.r.t. the characteristic graph since any two vertices connected by an edge will be in two different independent sets and no two maximal independent sets share the same color. Also note that when the maximal independents are disjoint, distinguish the vertices in the same independent set will result in a higher mutual information in the graph entropy optimization, since for any probabilities $p_i$, $p_{i+1}$ of the the nodes in a pairwise maximal independent set $\{\alpha_i, \alpha_{i+1}\}$, the difference of the mutual information will be
\begin{equation}
\begin{aligned}
- p_i \log_2 p_i - p_{i+1} \log_2 p_{i+1} + (p_i+p_{i+1})\log_2 (p_i+p_{i+1}) &= - p_i \log_2 \left(\frac{p_i}{p_i+p_{i+1}}\right) - p_{i+1} \log_2 \left(\frac{p_{i+1}}{p_i+p_{i+1}}\right) \\
& =\left(p_i+p_{i+1}\right) h_2 (\frac{p_i}{p_i+p_{i+1}}) \ge 0 \\
\end{aligned}
\end{equation}
where $h_2()$ is the binary entropy function.
Therefore \eq{Gentropy_Hw} can be achieved by graph coloring. Since the scheme we present in \thmref{optimal_function} is the optimal coloring method w.r.t. the non-product characteristic graph, it must achieve the minimum in \eq{Gentropy_Hw}. Therefore we have the following relationship for all $n \in [N]$ and the fundamental limit of the sum-rate can be achieved by \thmref{optimal_function}.
\begin{equation}
H_G(X_n) \overset{(a)}{\leq} \lim_{S\rightarrow \infty}\frac{1}{S} \min_{c_n \in \mathcal{C_{\epsilon}}(G_n^S(f))} H(c_n(X_n)) \overset{(b)}{\leq} \min_{c_n \in \mathcal{C}(G_n(f))} H(c_n(X_n)) \overset{(c)}{=} H_G(X_n)
\end{equation}
where (a) holds by \cite{FunctionCompressionSideInfo}; (b) holds by achievability: an ordinary coloring is also an $\epsilon$-coloring, and a valid ordinary coloring on the characteristic graph can be used in replication to achieve a valid coloring on the OR-product graph; and we have proved (c) above.  
\end{IEEEproof}

We now give the fundamental limit of the sum-rate in the problem that the CEO needs to determine $Z_M$ and $Z_{A,M}$ respectively.

\begin{defn}
A function $f: \mathcal{X}^N \rightarrow \mathcal{X}$ is a candidate \(\max\) function if and only if
\begin{equation} \label{eq:candidate_f_max}
\mathbb{E}[d_M((X_{1},\ldots,X_{N}),f(X_{1},\ldots,X_{N}))]=0.
\end{equation}
Let $\mathcal{F}_{M,N}$ be the set of all such candidate $\max$ functions with $N$ inputs. For any $f \in \mathcal{F}_{M,N}$, it indicates the $\max$. 
\end{defn}

\begin{defn}
A function $f: \mathcal{X}^N \rightarrow \{1,\ldots,N\}\times \mathcal{X} $ is a candidate \((\argmax,\max)\) function if and only if
\begin{equation}
\label{eq:candidate_f_pair}
\mathbb{E}[d_{A,M}((X_{1},\ldots,X_{N}),f(X_{1},\ldots,X_{N}))]=0.
\end{equation}
Let $\mathcal{F}_{P,N}$ be the set of all such candidate $(\argmax,\max)$ functions with $N$ inputs. For any $f\in \mathcal{F}_{P,N}$, it indicates both the index of a user attaining the $\max$ and the $\max$. 
\end{defn}

\begin{thm}\label{thm:converse_max}
In the problem that the CEO needs to decide $\hat{Z}_M$, if $\min \mathcal{X} > 0$, then the minimum sum-rate will be
\begin{equation}
R_M^* = \sum_{n=1}^N H(X_n).
\end{equation}
\end{thm}
\begin{IEEEproof}
 The distortion measure $d_M$ is
\begin{equation}
	d_M((X_{1,s},\ldots,X_{N,s}), \hat{Z}_M(s)) = \begin{dcases*}
		Z_M(s)- \hat{Z}_M(s) & if \(\hat{Z}_M(s) \leq Z_M(s)\)\\
		Z_M(s) & otherwise.
	\end{dcases*}
\end{equation}
Given $\min \mathcal{X} >0$, $Z_M(s)$ can never be $0$, the only way to make \eqref{eq:candidate_f_max} happen is to let $\hat{Z}_M(s)$ exactly estimate $Z_M(s)$, in other words, \eq{candidate_f_max} is satisfied if and only if
\begin{equation}
f(X_{1,s},\ldots,X_{N,s}) = \max (X_{1,s},\ldots,X_{N,s}).
\end{equation}
For any node pair ($\alpha_i$, $\alpha_{j}$) in user $n$'s characteristic graph, assume $i<j$ w.l.o.g., we will have $(\alpha_i,\alpha_j) \in G_n$ since there exists $\boldsymbol{x}_{\setminus \{n\}} = (\alpha_i,\ldots,\alpha_i)$ such that
\begin{equation}
f(\alpha_i, \boldsymbol{x}_{\setminus \{n\}}) \ne f(\alpha_j,\boldsymbol{x}_{\setminus \{n\}}).
\end{equation}
Therefore, the characteristic graph of user $n$ w.r.t. $f$ is complete, and $\Gamma(G_n) = \{\{\alpha_i\}: \alpha_i \in \mathcal{X}\}$, and the graph entropy is the same as the entropy of each source.
\end{IEEEproof}
\begin{rem}
To achieve this limit, we simply need each user to Huffman encode its source.
\end{rem}
\begin{cor} \label{cor:pair_complete}
In the problem that the CEO needs to decide $(\hat{Z}_{A},\hat{Z}_M)$, if $\min \mathcal{X} > 0$, then the minimum sum-rate will be
\begin{equation}
R_{A,M}^* = \sum_{n=1}^N H(X_n).
\end{equation}
\end{cor}
\begin{IEEEproof}
This directly follows the proof of \thmref{converse_max}, the characteristic graph is also complete if $\min \mathcal{X} >0$.
\end{IEEEproof}
\begin{thm} \label{thm:max_with_zero}
In the problem that the CEO needs to decide $\hat{Z}_M$, if $\min \mathcal{X} = 0$, then the minimum sum-rate satisfies
\begin{equation}
\begin{aligned}
R_M^* &= NH(X)+N\left(p_1\log_2 p_1 + p_{2}\log_2 p_{2}-\left(p_{1}+p_{2}\right)\log_2 \left(p_{1}+p_{2}\right)\right) \\
	&= NH(X) - N \left(p_1+p_{2}\right)h_2\left(\frac{p_1}{p_1+p_{2}}\right)
\end{aligned}
\end{equation}
where $p_1 = \mathbb{P}(X=\alpha_1=0)$ and $p_{2} = \mathbb{P}(X=\alpha_2)$. 

\end{thm}
\begin{IEEEproof}
Let $f \in \mathcal{F}_{M,N}$ satisfies that $f(\alpha_1,\ldots,\alpha_1) = \alpha_2$, then $(\alpha_1, \alpha_2) \not \in G$ for the characteristic graph of  each source $X_i$ w.r.t. $f$. The graph is not complete and the set of independent sets is $\Gamma(G_n) = \{\{\alpha_1,\alpha_2\},\{\alpha_3\},\ldots,\{\alpha_L\}\}$ for all $n \in [N]$. Hence by a similar proof as in \thmref{optimal_function} and \thmref{converse_argmax}, we have
\begin{equation}
R_M^* = \sum_{n=1}^{N} H_G(X_n)
\end{equation}
with
\begin{equation}
H_G (X_n) = -\left(p_1+p_2\right) \log_2 \left(p_1+p_2\right) - \sum_{i=3}^{L} p_i \log_2 p_i
\end{equation}
for all $n \in [N]$.
\end{IEEEproof}

\begin{thm} \label{thm:pair_with_zero}
In the problem that the CEO needs to decide $(\hat{Z}_A,\hat{Z}_M)$, if $\min \mathcal{X} = 0$, then the minimum sum-rate satisfies
\begin{equation}
\begin{aligned}
R_{A,M}^* &= NH(X)+\left(p_1\log_2 p_1 + p_{2}\log_2 p_{2}-\left(p_{1}+p_{2}\right)\log_2 \left(p_{1}+p_{2}\right)\right) \\
	&= NH(X) -  \left(p_1+p_{2}\right)h_2\left(\frac{p_1}{p_1+p_{2}}\right)
\end{aligned}
\end{equation}
where $p_1 = \mathbb{P}(X=\alpha_1) = \mathbb{P}(X = 0)$ and $p_{2} = \mathbb{P}(X=\alpha_2)$.
\end{thm}
\begin{IEEEproof}
Let $f = (f_A,f_M) \in \mathcal{F}_{P,N}$. Note that \eq{candidate_f_pair} is satisfied if and only if both \eq{candidate_f_argmax} and \eq{candidate_f_max} are satisfied, and hence $\{\alpha_i, \alpha_j\} \not \in G_n$ in user $n$'s characteristic graph if and only if $f_A(\alpha_i,\boldsymbol{x}_{\setminus \{n\}}) = f_A(\alpha_j,\boldsymbol{x}_{\setminus \{n\}})$ and $f_M(\alpha_i,\boldsymbol{x}_{\setminus \{n\}}) = f_M(\alpha_j,\boldsymbol{x}_{\setminus \{n\}})$ for all $\boldsymbol{x}_{\setminus \{n\}} = (x_{k}| k \in [N]\setminus\{n\}, x_{k} \in \mathcal{X})$.
Let $f_A \in \mathcal{F}_{A,N}$ be defined as in \thmref{optimal_function} and $f_M \in \mathcal{F}_{M,N}$ as in \thmref{max_with_zero}, then $\{\alpha_1, \alpha_2\} \not \in G_1$, and all characteristic graphs other than $G_1$ are complete. The sets of independent sets are $\Gamma(G_1) = \{\{\alpha_1,\alpha_2\},\{\alpha_3\},\ldots,\{\alpha_L\}\}$, and  $\Gamma(G_n) = \{\{\alpha_1\},\{\alpha_2\}, \ldots,\{\alpha_L\}\}$ for all $n \in [N]\setminus \{1\}$. By a similar proof as in \thmref{optimal_function} and \thmref{converse_argmax}, we have
\begin{equation}
R_{A,M}^* = \sum_{n=1}^{N} H_G(X_n)
\end{equation}
with
\begin{equation}
H_G (X_1) = -\left(p_1+p_2\right) \log_2 \left(p_1+p_2\right) - \sum_{i=3}^{L} p_i \log_2 p_i
\end{equation}
and
\begin{equation}
H_G(X_n) = H(X)
\end{equation}
for all $n \in [N]\setminus\{1\}$.
\end{IEEEproof}

\subsection{Scaling in Number of Users}
In this subsection, we consider the rate saving performance of graph coloring in a large scale of $N$.
Define the rate savings as the difference between the scheme that each user Huffman encode its source and the scheme by \thmref{optimal_function}, i.e. 
\begin{equation}
\Delta_A \triangleq \sum_{n=1}^{N}H(X_n) -R_A^* .
\end{equation}
\begin{thm}\label{thm:scale_users_argmax}
To losslessly determine the $\argmax$, the savings $\Delta_A$ is bounded by
\begin{equation}
\max_{p_1, \ldots, p_L} \Delta_A = \max_{p_1,\ldots,p_L} \sum_{i=1}^{L-1} \left(p_i+p_{i+1}\right)h_2\left(\frac{p_i}{p_i+p_{i+1}}\right) \le 2
\end{equation}
where $p_i=\mathbb{P}(X=\alpha_i)$, and hence the per user saving satisfies that
\begin{equation}
\lim_{N \rightarrow \infty} \frac{\Delta_A}{N} = 0.
\end{equation}
\end{thm}
\begin{IEEEproof}
\appendixproof{proof_scale_users_argmax}
\end{IEEEproof}
\begin{cor} \label{cor:scale_users_max}
In the problem that the CEO needs to decide the $max$, the per user saving $\nicefrac{\Delta_M}{N}$ satisfies
\begin{equation}
\lim_{N \rightarrow \infty} \frac{\Delta_M}{N} = \begin{dcases*}
0 & if $\min \mathcal{X} > 0$ \\
-\left(p_1+p_{2}\right) h_2\left(\frac{p_1}{p_1+p_{2}}\right) & if $\min \mathcal{X} = 0$
\end{dcases*}
\end{equation}
where $p_1 = \mathbb{P}(x= \alpha_1) =\mathbb{P}(x=0)$ and $p_{2} = \mathbb{P}(x=\alpha_2)$.
\end{cor}
\begin{IEEEproof}
\appendixproof{proof_scale_users_max}
\end{IEEEproof}
\begin{cor} \label{cor:scale_users_pair}
In the problem that the CEO needs to decide the pair $(\argmax,\max)$, the per user saving $\nicefrac{\Delta_{A,M}}{N}$ goes to $0$ as $N$ goes to infinity.
\end{cor}
\begin{IEEEproof}
Observe that
\begin{equation}
R_A^* \le R_{A,M}^* \le \sum_{n=1}^{N}H(X_n).
\end{equation}
\end{IEEEproof}
As we shall see in \secref{interactive-subband-allocation}, the lack of savings in this lossless non-interactive problem structure stands in stark contrast to an interactive setup in which it can be shown that, by allowing the CEO to communicate with the users over multiple rounds, a substantial saving in sum rate relative to the the non-interactive scheme can be achieved \cite{InteractionDP} while still obtaining the answer losslessly. Additionally, as we will see in \secref{lossy-subband-allocation}, substantial rate savings can be obtained if we are willing to tolerate a small amount of loss.

\section{Lossy Extremization}
\label{sec:lossy-subband-allocation}
In the previous section, it was shown that having a \ac{CEO} losslessly compute the \(\max\), \(\argmax\), or \((\max, \argmax)\) of a set of distributed sources does \emph{not} result in a significant rate savings relative to simply recovering all of the sources.
For applications where reducing the rate is critical, tolerating bounded estimation error may be necessary.
In this section, we consider the \emph{lossy} variant of the function computation problem where the \ac{CEO} need not determine the function output exactly.
In particular, we first bound the best achievable rate-distortion tradeoff by computing the rate-distortion curves for each of the three functions with an adapted version of the Blahut-Arimoto algorithm in \secref{multi-source-blahut-arimoto}.
Achievable schemes for each of the three functions based on scalar quantization followed by entropy encoding are then presented in \secref{arg-max-hom-sq} and \secref{two-user-het-sq}.
For certain problem instances, this scheme closely approximates the rate-distortion function as shown in \secref{sq-examples}.

\subsection{Fundamental Limits---Multi-Source Blahut-Arimoto}
\label{sec:multi-source-blahut-arimoto}
In this subsection, we first utilize a generalized \ac{BA} algorithm to compute the sum-rate distortion function for the independent \ac{CEO} extremization problems with discrete sources.
We then show by \thmref{continuous_RD_vector} and \corref{continuous_RD_vector_CEO} that the sum-rate distortion function for the case of continuous sources can be well approximated by discretizing the sources and applying the generalized \ac{BA} algorithm from \cite{Ku:ABA} to compute the rate distortion function for the discretized sources.
In the limit as the discretization becomes arbitrarily fine, the discretized rate distortion function provides an \(\epsilon\)-lower bound for the continuous one.
This calculated lower bound is used in \secref{sq-examples} to measure the performance of the continuous quantizations we propose in \secref{arg-max-hom-sq} and \secref{two-user-het-sq}.

We first prove our discretization result for the classical single-source rate distortion function for continuous sources with bounded support.
\begin{thm}
	\label{thm:continuous_RD_vector}
	For any continuous source $X$ with bounded support set $\mathcal{X} = (x_{min}, x_{max})$ and bounded PDF $f(x)$ if there exists a continuous quantizer
	\begin{equation}
		\bm{Q} : \mathcal{X}^S \rightarrow \{1,\ldots,2^{SR}\}
	\end{equation}
	a reconstruction function
	\begin{equation}
		g : \{1,\ldots, 2^{SR}\} \rightarrow \hat{\mathcal{X}} = \{\bm{\hat{x}}_{1},\ldots,\bm{\hat{x}}_{L}\}
	\end{equation}
	and a distortion metric 
	\begin{equation}
		d: \mathcal{X} \times \hat{\mathcal{X}} \rightarrow R^+
	\end{equation}
	that attain rate distortion pair $(R,D)$, where 
	\begin{equation}
		L = 2^{SR}
	\end{equation}
	\begin{equation}
		\bm{\hat{x}}_{\ell} = \left(\hat{x}_{\ell,1},\ldots,\hat{x}_{\ell,S}\right)
	\end{equation}
	\begin{equation}
		D =\mathbb{E}[\bm{d}(\bm{X},\bm{\hat{X}}_{\ell})]
	\end{equation}
	\begin{equation}
		\bm{d} (\bm{x}, \bm{\hat{x}}_{\ell}) = \frac{1}{S} \sum_{s=1}^{S} d(x_s,\hat{x}_{\ell,s}) 
	\end{equation}
	with $d(\cdot)$ bounded above by $d_{m}$, and, when regarded as a function of $\bm{x}$ for a fixed $\bm{\hat{x}_\ell}$, $\bm{d}(\bm{x}, \bm{\hat{x}}_{\ell})$ has at most a finite number of discontinuous points,
	then there must exist a discrete source $Y_K$, a discrete quantizer
	\begin{equation}
		\bm{Q}^d_K : \mathcal{Y}_K^S \rightarrow \{1,\ldots,2^{SR}\}
	\end{equation}
	along with the same distortion metric $d$ and reconstruction mapping $g$ that attains rate $R_K=R$ and distortion $D_K  \ge D$,
	where $Y_K=Q_K(X)$ is built by uniformly quantizing $X$ into $K$ intervals with the reconstruction levels $\mathcal{Y}_K = \{u_k : k \in [K]\}$. Further, $D_K$ can be arbitrarily close to $D$ for a large enough $K$, i.e.
	\begin{equation}
		\lim_{K \rightarrow \infty} D_K = D
	\end{equation}
\end{thm}
\begin{IEEEproof}
	Given a continuous source $X$ with $\mathcal{X} = [x_{min}, x_{max}]$ and PDF $f(x)$, the expected distortion is
	\begin{equation}
		\label{eq:continuousD}
		\begin{aligned}
			D
			&= \mathbb{E} [\bm{d} (\bm{X}, \bm{\hat{X}})] \\
			& = \int \bm{d} (\bm{x},\bm{Q}(\bm{x}))f(\bm{x}) d \bm{x}
		\end{aligned}
	\end{equation}
	where we denote $\bm{\hat{x}}_{\ell} = g(\bm{Q}(\bm{x}))$ by $\bm{Q}(\bm{x})$
	for convenience, and
	\begin{equation}
		\label{eq:vec_distortion}
		\bm{d} (\bm{x}, \bm{Q}(\bm{x})) = \frac{1}{S} \sum_{s=1}^{S} d\left(x_s,\left(\bm{Q}(\bm{x})\right)_s\right) 
	\end{equation}

	Now let $Q_K$ that uniformly quantizes $X$ with $K$ intervals, i.e. 
	\begin{equation}
		Q_K : \mathcal{X} \rightarrow \{\Ica_k : k \in[K]\}
	\end{equation}
	where 
	\begin{equation}
		\Ica_k = (x_{min} + \frac{k-1}{K} \left(x_{max}-x_{min}\right),\  x_{min} + \frac{k}{K} \left(x_{max}-x_{min}\right)).
	\end{equation}
	Let $g_K$ that maps the intervals to the reconstruction levels, i.e.
	\begin{equation}
		g_K : \{\Ica_k : k \in K\} \rightarrow \{u_k : k \in [K]\}
	\end{equation}
	with
	\begin{equation} \label{eq:reconsYK}
		u_k = \underset{x \in \Ica_k}{arg\min}\; d(x,Q(x)).
	\end{equation}
	The discrete random variable $Y_K$ by discretizing $X$ is then defined on the support set
	\begin{equation}
		\mathcal{Y}_K = \{u_{k} : k \in [K]\}
	\end{equation}
	with PMF
	\begin{equation}
		\mathbb{P}[Y_K = u_k] = \int_{x \in \Ica_k} f(x) dx.
	\end{equation}
	Let $\bm{y}_K = \left(y_{K,s} : s \in [S]\right)$, the discrete quantizer $\bm{Q}^d_K$ satisfies
	\begin{equation}
		\bm{Q}^d_K(\bm{y}_K) = \bm{Q}(\bm{y}_K).
	\end{equation}
	The distortion $D_K$ for quantization $Q^d_K$ will be 
	\begin{equation}
		\label{eq:discreteD}
		D_K = \mathbb{E}[\bm{d}(\bm{Y}_K, \bm{Q}^d_K(\bm{Y}_K)] = \sum_{\bm{y}_K \in \mathcal{Y}_K^S} \bm{d} (\bm{y}_K,\bm{Q}^d_K(\bm{y}_K))\mathbb{P}[\bm{Y}_K = \bm{y}_K]
	\end{equation}
	where
	\begin{equation}
		\bm{d}(\bm{y}_K,\bm{Q}^d_K(\bm{y}_K)) = \frac{1}{S} \sum_{s=1}^{S} d\left(y_{K,s},\left(\bm{Q}^d_K(\bm{y}_K)\right)_s\right).
	\end{equation}
	Let $\bm{Q}_K(\cdot)$ quantize $\bm{x}$ element wise as $Q_K$ does. In addition, let
	\begin{equation}
		\Ica(\bm{y}_K) = \{\bm{x} \in \mathcal{X}^S| \bm{Q}_K(\bm{x}) = \bm{y}_K\} = \Ica_1(\bm{y}_K) \cup \Ica_2(\bm{y}_K) \cup \Ica_3(\bm{y}_K) 
	\end{equation}
	be a subset of $\mathcal{X}^S$ that maps to $\bm{y}_K$ by $\bm{Q}_K(\cdot)$, where
	\begin{equation}
		\Ica_1(\bm{y}_K) = \begin{dcases*}
			\Ica(\bm{y}_K) & if $\bm{Q}(\bm{x}) = \bm{Q}(\bm{y}_K)$ for all $\bm{x} \in \Ica(\bm{y}_K)$ and $\bm{d}(\bm{x},\bm{\hat{x}})$ is continuous on $\Ica(\bm{y}_K)$\\
			\emptyset & otherwise
		\end{dcases*}
	\end{equation}

	\begin{equation}
		\Ica_2(\bm{y}_K) = \begin{dcases*}
			\Ica(\bm{y}_K) & if $\bm{Q}(\bm{x}) = \bm{Q}(\bm{y}_K)$ for all $\bm{x} \in \Ica(\bm{y}_K)$ and $\bm{d}(\bm{x}, \bm{\hat{x}})$ is not continuous on $\Ica(\bm{y}_K)$\\
			\emptyset & otherwise
		\end{dcases*}
	\end{equation}
	and
	\begin{equation}
		\Ica_3(\bm{y}_K) = \begin{dcases*}
			\Ica(\bm{y}_K) & if $\exists \bm{x}_1, \bm{x}_2\in \Ica(\bm{y}_K)$ such that $\bm{Q}(\bm{x}_1) \ne \bm{Q}(\bm{x}_2)$\\
			\emptyset & otherwise
		\end{dcases*}
	\end{equation}
	Clearly, $\Ica_i(\bm{y}_K) \cap \Ica_j(\bm{y}_K) = \emptyset$ for any $i \ne j,\ i,j \in \{1,2,3\}$. By comparing \eq{continuousD} and \eq{discreteD}, we have
	\begin{equation}
		\label{eq:D_Analy1}
		\begin{aligned}
			D
			&= \int \bm{d} (\bm{x},\bm{Q}(\bm{x}))f(\bm{x})d\bm{x} \\
			&= \sum_{\bm{y}_K \in \mathcal{Y}_K^S} \int_{\bm{x}\in \Ica(\bm{y}_K)} \bm{d} (\bm{x},\bm{Q}(\bm{x}))f(\bm{x})d\bm{x} \\
			&= \sum_{\bm{y}_K \in \mathcal{Y}_K^S} \int_{\bm{x}\in \Ica(\bm{y}_K)} \left[\bm{d} (\bm{x},\bm{Q}(\bm{x})) + \bm{d} (\bm{y}_K,\bm{Q}(\bm{y}_K))-\bm{d} (\bm{y}_K,\bm{Q}(\bm{y}_K))\right]f(\bm{x})d\bm{x} \\ 
			&= D_K+ \sum_{\bm{y}_K \in \mathcal{Y}_K^S} \int_{\bm{x}\in \Ica(\bm{y}_K)} \left[\bm{d} (\bm{x},\bm{Q}(\bm{x})) -\bm{d} (\bm{y}_K,\bm{Q}(\bm{y}_K))\right]f(\bm{x})d\bm{x} \\
		\end{aligned}
	\end{equation}
	First observe by \eq{vec_distortion} and \eq{reconsYK} that 
	\begin{equation}
		D = D_K + \sum_{\bm{y}_K \in \mathcal{Y}_K^S} \int_{\bm{x}\in \Ica(\bm{y}_K)} \left[\bm{d} (\bm{x},\bm{Q}(\bm{x})) -\bm{d} (\bm{y}_K,\bm{Q}(\bm{y}_K))\right]f(\bm{x})d\bm{x} \ge D_K
	\end{equation}
	then we can further express \eq{D_Analy1} as
	\begin{equation}
		\begin{aligned}
			D
			&= D_K+ \sum_{\bm{y}_K \in \mathcal{Y}_K^S} \int_{\bm{x}\in \Ica(\bm{y}_K)} \left[\bm{d} (\bm{x},\bm{Q}(\bm{x})) -\bm{d} (\bm{y}_K,\bm{Q}(\bm{y}_K))\right]f(\bm{x})d\bm{x} \\
			&= D_K + \sum_{\bm{y}_K \in \mathcal{Y}_K^S} \int_{\bm{x}\in \Ica_1(\bm{y}_K) \cup \Ica_2(\bm{y}_K) \cup \Ica_3(\bm{y}_K)} \left[\bm{d} (\bm{x},\bm{Q}(\bm{x})) -\bm{d} (\bm{y}_K,\bm{Q}(\bm{y}_K))\right]f(\bm{x})d\bm{x} \\
			&= D_K + A(\Ica_1)+A(\Ica_2)+A(\Ica_3)
		\end{aligned}
	\end{equation}
	where 
	\begin{equation}
		A(\Ica_i) = \sum_{\bm{y}_K \in \mathcal{Y}_K^S} \int_{\bm{x}\in \Ica_i(\bm{y}_K)} \left[\bm{d} (\bm{x},\bm{Q}(\bm{x})) -\bm{d} (\bm{y}_K,\bm{Q}(\bm{y}_K))\right]f(\bm{x})d\bm{x}.
	\end{equation}
	For any S-fold vector quantization $\bm{Q}(\cdot)$, we then have the following statements:
	\begin{enumerate}[(a)]
		\item $\forall$ $\epsilon_1 >0$, $\exists$ a large enough $K_1$ and a uniform quantization $Q_{K_1}$ such that $\forall$ $\bm{x} \in \Ica_1(\bm{y}_{K_1})$
			\begin{equation} \label{eq:state1}
				\bm{d} (\bm{x},\bm{Q}(\bm{x})) -\bm{d} (\bm{y}_{K_1},\bm{Q}(\bm{y}_{K_1})) \le \epsilon_1
			\end{equation}

		\item $\forall$ $\epsilon_2 >0$, $\exists$ a large enough $K_2$ and a uniform quantization $Q_{K_2}$ such that 
			\begin{equation}  \label{eq:state2}
				\sum_{\bm{y}_{K_2} \in \mathcal{Y}_{K_2}^S} \mathbb{P}[\bm{X} \in \Ica_2(\bm{y}_{K_2})] < \epsilon_2.
			\end{equation}
		\item $\forall$ $\epsilon_3 >0$, $\exists$ a large enough $K_3$ and a uniform quantization $Q_{K_3}$ such that 
			\begin{equation}  \label{eq:state3}
				\sum_{\bm{y}_{K_3} \in \mathcal{Y}_{K_3}^S} \mathbb{P}[\bm{X} \in \Ica_3(\bm{y}_{K_3})] < \epsilon_3.
			\end{equation}
	\end{enumerate}
	where \eq{state1} holds since $d(\cdot)$ is continuous on $\Ica_1(\bm{y}_{K})$; \eq{state2} holds since $\bm{d}(\bm{x}, \bm{\hat{x}})$ has a finite number (i.e. $m$) of discontinuous points on $\mathcal{X}^S \times \mathcal{\hat{X}}$, 
	and for given $\epsilon = \nicefrac{\epsilon_2}{m} >0$, there must exist a large enough $K_2$ such that when $\Ica_2(\bm{y}_{K_2}) \ne \emptyset$,
	\begin{equation}
		\mathbb{P}[\bm{X} \in \Ica_2(\bm{y}_{K_2})]  < \epsilon
	\end{equation}
	hence 
	\begin{equation}
		\begin{aligned}
			\sum_{\bm{y}_{K_2} \in \mathcal{Y}_{K_2}^S} \mathbb{P}[\bm{X} \in \Ica_2(\bm{y}_{K_2})]
			&= \sum_{\bm{y}_{K_2} \in \mathcal{Y}_{K_2}^S} \left(\mathbb{P}[\bm{X} \in \Ica_2(\bm{y}_{K_2}), \Ica_2(\bm{y}_{K_2})\ne \emptyset] + \mathbb{P}[\bm{X} \in \Ica_2(\bm{y}_{K_2}), \Ica_2(\bm{y}_{K_2}) = \emptyset] \right)\\
			&< m\epsilon + 0\\
			& =\epsilon_2.
		\end{aligned}
	\end{equation}
	Now we prove \eq{state3} also holds. Given a continuous quantizer
	\begin{equation}
		\bm{Q} : \mathcal{X}^S \rightarrow \{1,\ldots,2^{SR}\},
	\end{equation}
	a point $\bm{x} \in \mathcal{X}^S$ is a boundary point w.r.t. $\bm{Q}$ if for any $\epsilon>0$ there exist $\bm{x}_1 \in \mathcal{X}^S, \bm{x}_1 \ne \bm{x}$ such that
	\begin{equation}
		\|\bm{x}_1-\bm{x}\|_2 < \epsilon
	\end{equation}
	and
	\begin{equation}
		\bm{Q}(\bm{x}_1) \ne \bm{Q}(\bm{x}).
	\end{equation}

	Let $f(\bm{x})$ be a bounded PDF which is defined on $\mathcal{X}^S \subset \mathbb{R}^S$ with $\mathcal{X} = (x_{min},x_{max})$. For a continuous quantizer
	\begin{equation}
		\bm{Q} : \mathcal{X}^S \rightarrow \{1,\ldots,2^{SR}\},
	\end{equation}
	let $B(\bm{Q})$ be the set of all boundary points w.r.t. $\bm{Q}(\cdot)$.
	Since every k-dimensional subspace of $\mathbb{R}^S$ must have measure zero if $k < S$, by the definition of measure zero, we have that for any $\epsilon > 0$, there exist open cubes $\Uca_1,\Uca_2,\ldots$ such that $B(\bm{Q}) \subseteq \cup_{i=1}^{\infty} \Uca_i$, and
	\begin{equation}
		\sum_{i=1}^{\infty} \mathrm{vol}(\Uca_i) < \frac{\epsilon}{f_{max}},
	\end{equation}
	where
	\begin{equation}
		\Uca_i = (a_{i,1},b_{i,1}) \times (a_{i,2},b_{i,2})\times \cdots \times (a_{i,S},b_{i,S}) 
	\end{equation}

	\begin{equation}
		\mathrm{vol}(\Uca_i) = (b_{i,1}-a_{i,1})(b_{i,2}-a_{i,2})\cdots(b_{i,S}-a_{i,S})	
	\end{equation}
	and
	\begin{equation}
		f_{max} = \max_{\bm{x} \in \mathcal{X}^S} f(\bm{x}).
	\end{equation} 
	Hence
	\begin{equation}
		\begin{aligned}
			\mathbb{P}[\bm{X} \in \cup_{i=1}^{\infty} \Uca_i]
			&\le \sum_{i=1}^{\infty} \int_{\bm{x} \in \Uca_i} f(\bm{x}) d\bm{x} \\
			&\le f_{max} \left(\sum_{i=1}^{\infty}  \int_{\bm{x} \in \Uca_i} d\bm{x}\right)\\
			&=f_{max}\sum_{i=1}^{\infty} \mathrm{vol}(\Uca_i) \\
			&=\epsilon.
		\end{aligned}
	\end{equation}
	In other words, the boundaries of the quantization levels have probability measure $0$.

	Therefore, for any $\epsilon > 0$ and any $S$, there exists $K \ge \max \{K_1,K_2,K_3\}$ such that
	\begin{equation}
		\begin{aligned}
			D
			&= D_K + A(\Ica_1)+A(\Ica_2)+A(\Ica_3) \\
			&\le D_K +  \sum_{\bm{y}_K \in \bm{\mathcal{Y}}_K} \int_{\bm{x}\in \Ica_1(\bm{y}_K)} \epsilon_1f(\bm{x})d\bm{x} + \sum_{\bm{y}_K \in \bm{\mathcal{Y}}_K} \int_{\bm{x}\in \Ica_2(\bm{y}_K)} d_{m} f(\bm{x})d\bm{x} + \sum_{\bm{y}_K \in \bm{\mathcal{Y}}_K} \int_{\bm{x}\in \Ica_3(\bm{y}_K)} d_{m} f(\bm{x})d\bm{x} \\
			&= D_K + \epsilon_1 \sum_{\bm{y}_K \in \bm{\mathcal{Y}}_K} \mathbb{P}[\bm{X} \in \Ica_1(\bm{Y}_K)] + d_{m} \sum_{\bm{y}_K \in \bm{\mathcal{Y}}_K} \mathbb{P}[\bm{X} \in \Ica_2(\bm{Y}_K)]+d_{m} \sum_{\bm{y}_K \in \bm{\mathcal{Y}}_K} \mathbb{P}[\bm{X} \in \Ica_3(\bm{Y}_K)] \\
			&\le D_K + \epsilon_1 + \epsilon_2d_{m} +\epsilon_3d_{m}
		\end{aligned}
	\end{equation}
	and
	\begin{equation}
		\lim_{K \rightarrow \infty} D_K = D
	\end{equation}
\end{IEEEproof}

\begin{cor}
	\label{cor:continuous_RD_vector_CEO}
	In the \ac{CEO} problem for estimating a function $f(\cdot)$ of $N$ independent observations, user $n$ observe $\bm{X}_n = \left(X_{n,s} : s \in [S]\right)$, where $X_{n,s}$ is a continuous random variable drawn from a bounded support set $\mathcal{X} = (x_{min}^n,x_{max}^n)$ with a bounded PDF $f_n(x)$.
	If for each user there exists a continuous quantizer $\bm{Q}_n$
	\begin{equation}
		\bm{Q}_n : \mathcal{X}^S_n \rightarrow \{1,\ldots,2^{SR_n}\}
	\end{equation}
	a joint reconstruction function
	\begin{equation}
		g : \{1,\ldots,2^{SR_1}\} \times \{1,\ldots,2^{SR_2}\} \times  \cdots \times \{1,\ldots,2^{SR_N}\} \rightarrow \hat{\mathcal{X}} = \{\bm{\hat{x}}_1,\ldots,\bm{\hat{x}}_L\}
	\end{equation}
	and a distortion metric 
	\begin{equation}
		d: \mathcal{X}_1 \times \cdots \times \mathcal{X}_N \times \hat{\mathcal{X}}  \rightarrow R^+
	\end{equation}
	that attain rate distortion pair $(\bm{R},D)$, where 
	\begin{equation}
		\bm{R} = \left(R_1,\ldots,R_N\right)
	\end{equation}
	\begin{equation}
		\bm{\hat{x}}_{\ell} = \left(\hat{x}_{\ell,1},\ldots,\hat{x}_{\ell,S}\right)
	\end{equation}
	\begin{equation}
		D = \mathbb{E}[\bm{d} (\bm{X}_1, \ldots,\bm{X}_N,\bm{\hat{X}})]
	\end{equation}
	\begin{equation}
		\bm{d} (\bm{x}_1, \ldots,\bm{x}_N,\bm{\hat{x}}_{\ell}) = \frac{1}{S} \sum_{s=1}^{S} d(x_{1,s},\ldots,x_{N,s},\hat{x}_{\ell,s}) 
	\end{equation}
	with $d(\cdot)$ bounded above by $d_{m}$, and, when regarded as a function of $\bm{x}_1, \ldots,\bm{x}_N$ for a fixed $\bm{\hat{x}}_{\ell}$, $\bm{d} (\bm{x}_1, \ldots,\bm{x}_N,\bm{\hat{x}}_{\ell})$ has at most a finite number of discontinuous points, 
	then there must exist discrete sources $\{Y_{K_n} : n \in [N]\}$, a series of discrete quantizers
	\begin{equation}
		\bm{Q}_{n}^d : \mathcal{Y}_{K_{n}}^S \rightarrow \{1,\ldots,2^{SR_n}\}
	\end{equation} 
	along with the same distortion metric $d(\cdot)$ and reconstruction mapping $g$ that attains rate region $\bm{R}_K =\bm{R}$ and average distortion $D_K \ge D$, where $Y_{K_n} = Q_{K_n}(X_n)$ is built by uniformly quantizing $X_n$ into $K_n$ intervals with reconstruction levels $\{u_{k,n} : k \in [K_n]\}$. Further, $D_K$ can be arbitrarily close to $D$ for a large enough $K$, i.e.
	\begin{equation}
		\lim_{K \rightarrow \infty} D_K = D
	\end{equation}
\end{cor}
\begin{IEEEproof}
	The proof follows along the same lines as the \thmref{continuous_RD_vector}.
\end{IEEEproof}
It follows directly from \thmref{continuous_RD_vector} and \corref{continuous_RD_vector_CEO} that a tight lower bound for the continuous source distortion rate functions for the extremization problems of interest can be computed via the algorithm presented in \cite{Ku:ABA} applied to a suitably fine discretization of the continuous source.

\begin{figure*}
	\subfloat[]{\includegraphics[width=234.0pt]{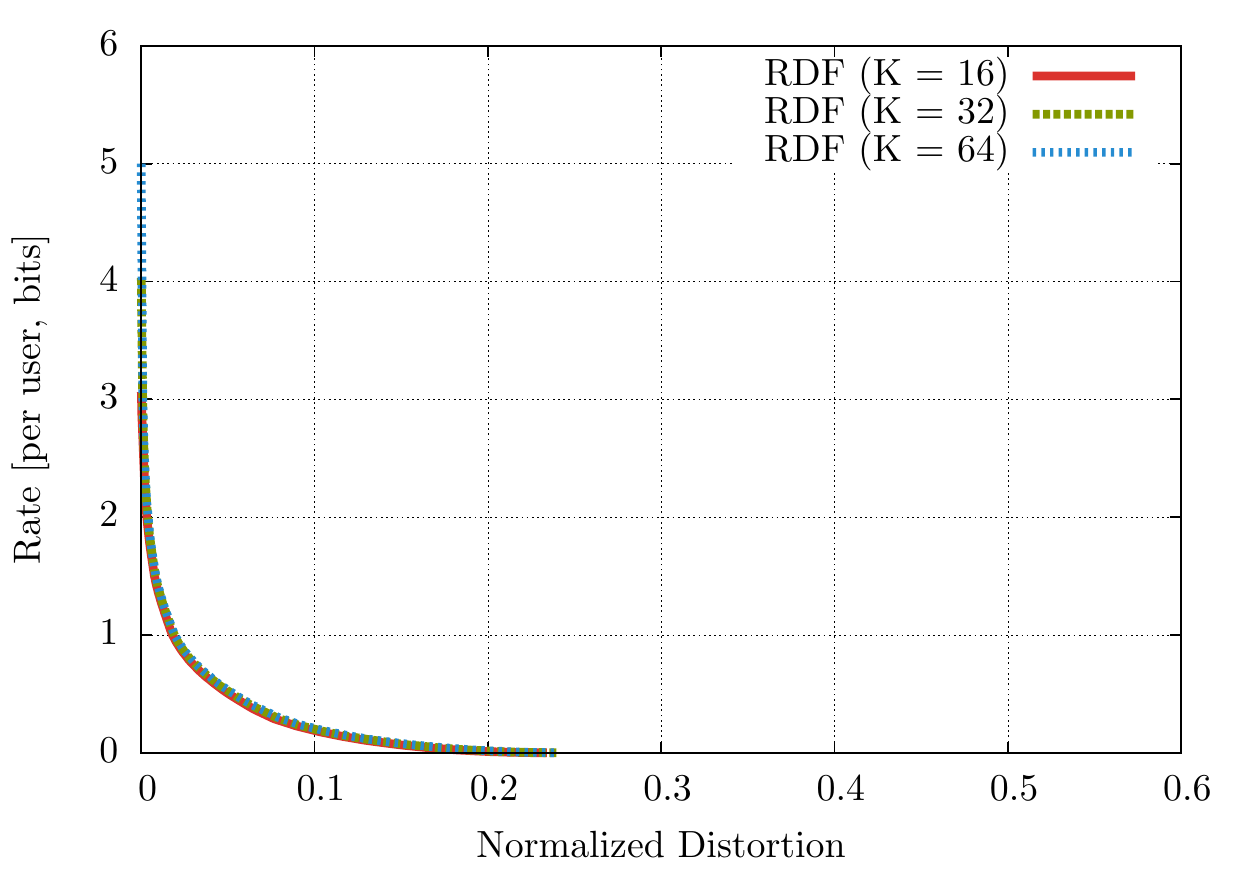}\label{fig:uni-rdf-argmax-N-2}}
	\subfloat[]{\includegraphics[width=234.0pt]{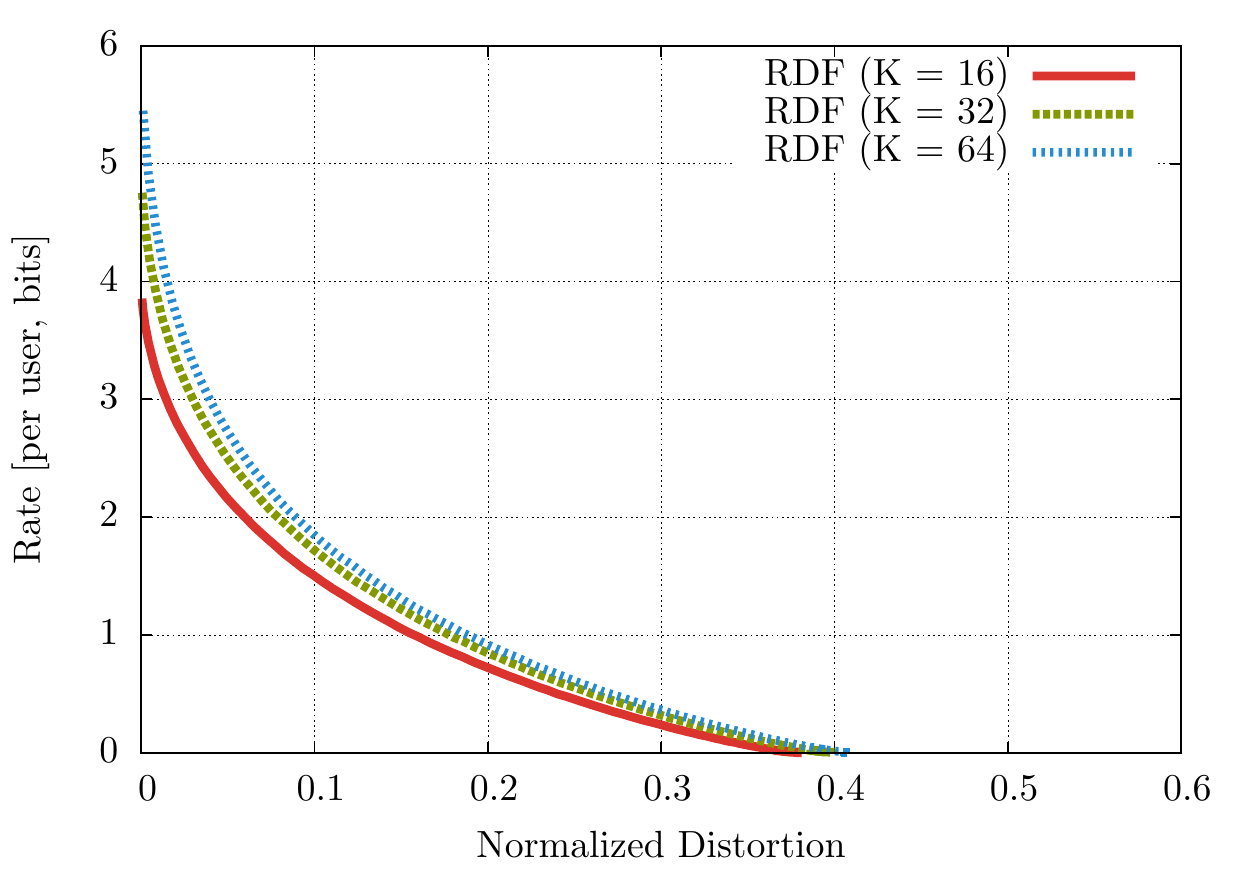}\label{fig:uni-rdf-max-N-2}}\\
	\subfloat[]{\includegraphics[width=234.0pt]{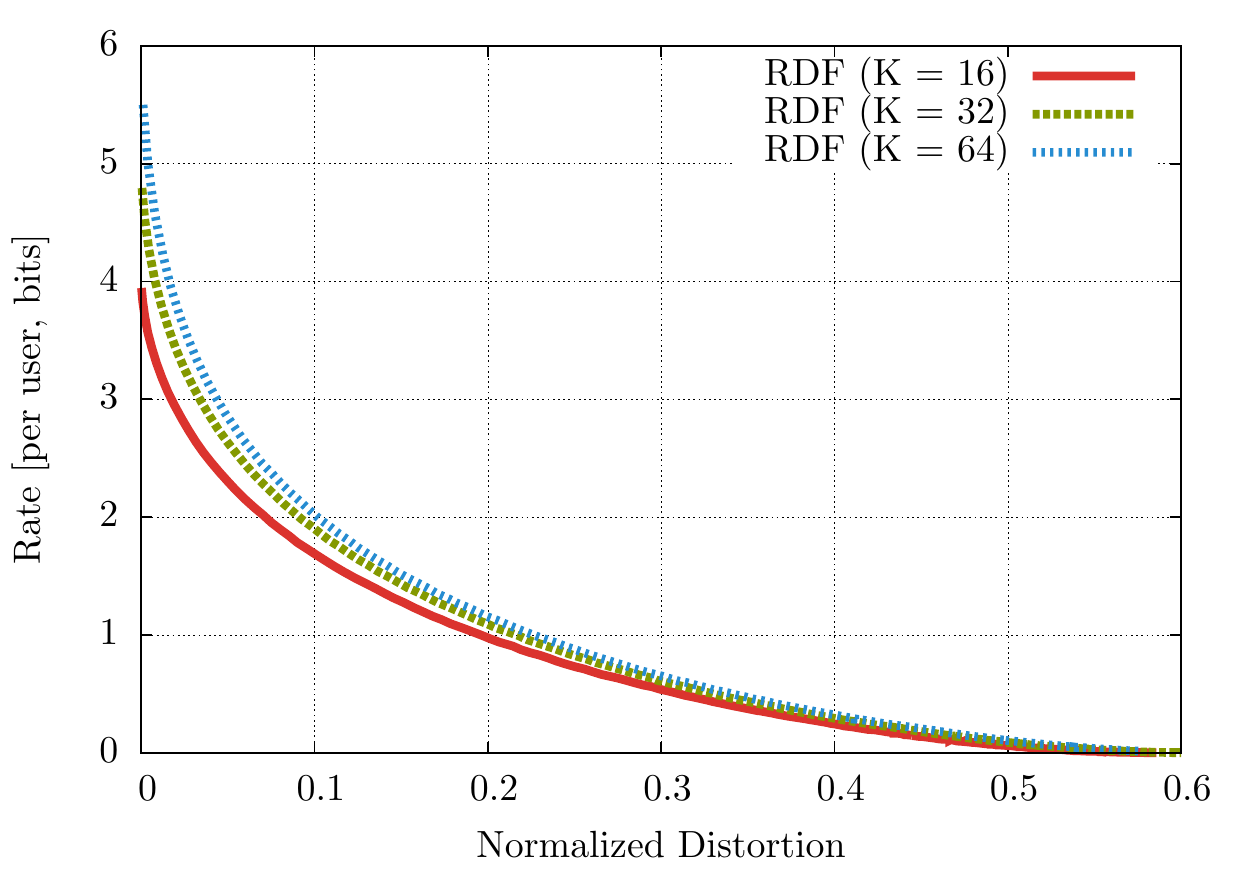}\label{fig:uni-rdf-both-N-2}}
	\subfloat[]{\includegraphics[width=234.0pt]{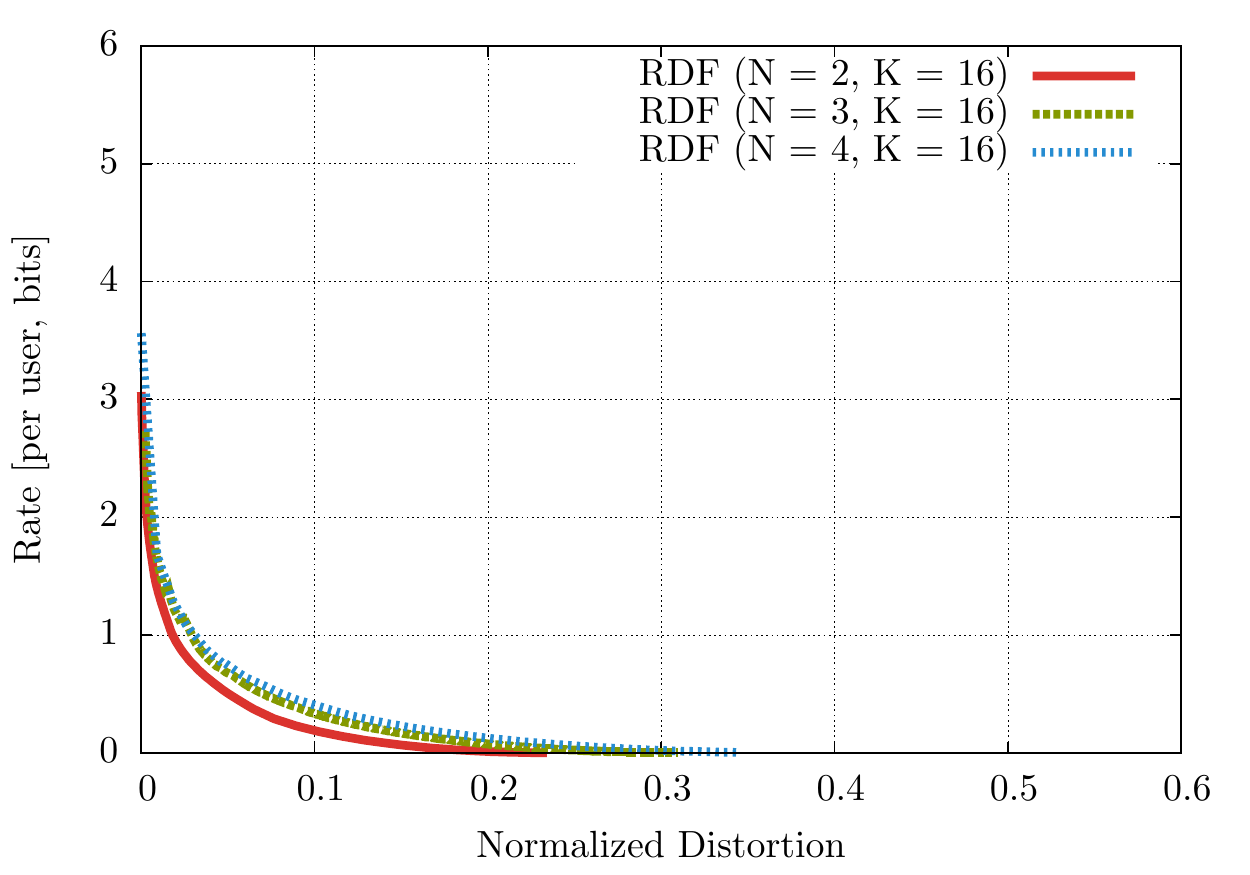}\label{fig:uni-rdf-argmax-K-16}}
	\caption[]{(a)--(c): Rate distortion function for \(N = 2\) uniformly distributed sources as a function of the discretization parameter \(K\); %
	(d): Rate distortion function for \(\argmax\) as a function of \(N\) with a fixed \(K = 16\)}\label{fig:uni-rdf-N-2}
\end{figure*}

In \fig{uni-rdf-N-2}, we show example rate distortion functions for the three extremization problems for a fixed number of users \(N = 2\) with Uniform\((0, 1)\) sources; we also show how the rate distortion function scales with the number of users \(N \in \{2,3,4\}\) for the case of the \(\argmax\) function.
Observer in the first three plots, that difference between the approximations of the continuous source rate distortion functions is rapidly diminishing with the discretization parameters \(K\).
Looking at the fourth plot, it appears that the rate distortion function scales neglibly in the number of users.
In fact, based on the performance of the \acp{SQ} discussed in \secref{sq-examples} (\fig{uniform-multiuser-sq}), we observe that the rate distortion function must \emph{decrease} as the number of users \(N\) grows large.
Note that for our plots of rate versus distortion, we normalize the distortion by the expected value of the maximum, i.e.,
\begin{equation}
	\frac{\expected{d((X_1, \ldots, X_N), \hat{z})}}{\expected{\max\{X_1, \ldots, X_N\}}}.
\end{equation}
In the rest of \secref{lossy-subband-allocation}, we will discuss quantization designs for the independent \ac{CEO} extremization problems with continuous source random variables.
The lower bounds that we compute based on \thmref{continuous_RD_vector} and \corref{continuous_RD_vector_CEO} (and shown in \fig{uni-rdf-N-2}) will be used as the fundamental limits to measure the performance of the quantization schemes we propose in \secref{arg-max-hom-sq} and \secref{two-user-het-sq}.

\subsection{Scalar Quantizers for \(\argmax\)}
\label{sec:arg-max-hom-sq}
In this section, we consider the design of \acp{SQ} as an achievable scheme and compare their performance to computed rate-distortion functions.
We first consider the case where all users are using the same quantizer and derive an expression for the resulting distortion.
Using this expression, we pose two non-linear optimization problems: first, minimize distortion for a given number of bins, and; second, minimize distortion for a given number of bins subject to a constraint on the entropy of the quantizer output.
We provide first order necessary conditions for the optimal quantizer for both non-linear optimizations.
We then argue that the same distortion performance can be achieved with a smaller sum rate by utilizing different quantizers at each user.
We show that the design of the \ac{HetSQ} can be accomplished via the same design procedure as for the \ac{HomSQ}.

Let \(X_i : i = 1, \ldots, N\) be the sources for the \(N\) users and let \(Z_A\) be the index of the user with maximum value.
Unlike previous sections, we assume \emph{continous} (instead of discrete) random variables \(X_i : i = 1, \ldots, N\).
As before, we still assume they are \iid with common PDF \(f(x)\), CDF \(F(x)\), and support set \(\mathcal{X} \subseteq \mathds{R}_+\).

\subsubsection{Homogeneous Scalar Quantizers}
Normally, a \ac{SQ} is specified as a set of \emph{decision boundaries} and \emph{reconstruction levels} \cite{Say2012}.
For the estimating the \(\argmax\), we do not need the \ac{CEO} to produce estimates for \(X_i : i = 1, \ldots, N\) or even \(X_{Z_A}\) (i.e., the value of the maximum source).
We can therefore specify the quantizer with just a set of decision boundaries \(\{\ell_k : k = 0, \ldots, K\}\) which divide the support set \(\mathcal{X}\) into \(K\) intervals
\begin{equation}
	\mathcal{L}_k = [\ell_{k-1}, \ell_k] \quad k = 1, \ldots, K
\end{equation}
where \(\ell_0 \triangleq \inf \mathcal{X}\) and \(\ell_K \triangleq \sup \mathcal{X}\).
Let \(U_i \in \{1, \ldots, K\}\) indicate the interval in which user \(i\)'s observed value lies.
The \ac{CEO} will pick user \(i\) if \(U_i > U_{i'}\) for all \(i' \neq i\) and will randomly pick a user from \(\underset{i}{\argmax} U_{i}\) otherwise; we denote the estimate so obtained as \(X_{\hat{Z}_A}\).

For notational brevity, we define the following: \(E_{j} \triangleq \expected{X \mid \ell_{j-1} \leq X \leq \ell_j}\), \(f_j = f(\ell_j)\), \(F_{j} \triangleq F(\ell_j)\), and \(p_j \triangleq \prob{\ell_{j-1} \leq X \leq \ell_{j}}\).
\begin{lem}
	Let \((X_i : i \in [N])\) be a collection of \iid random variables with cdf \(F(x)\) and pdf \(f(x)\) and
	\begin{equation}
		Z_A \triangleq \{i | X_i = \max \{X_1,\ldots,X_N\}, i \in [N]\}.
	\end{equation}
	The expected value of the max is
	\begin{equation}
		\expected{X_i | i \in Z_A} = \int_{\inf \mathcal{X}}^{\sup \mathcal{X}} x N F^{N-1}(x)f(x)\,\mathrm{d}x.
	\end{equation}
\end{lem}
\begin{IEEEproof}
	Omitted for brevity.
\end{IEEEproof}
\begin{thm}
	\label{thm:argmax-hom-sq}
	Let \((X_i : i \in [N])\) be a collection of \iid random variables with cdf \(F(x)\) and pdf \(f(x)\) and
	\begin{equation}
		Z_A \triangleq \{i | X_i = \max \{X_1,\ldots,X_N\}, i \in [N]\}.
	\end{equation}
	The expected value of the estimated \(\argmax\) when using \acp{HomSQ} with \(K\) intervals is
	\begin{equation}
		\label{eq:estimator-ev}
		\expected{X_{\hat{Z}_A}} = \sum_{j = 1}^{K} \left[E_j\left(F_j^N - F_{j-1}^N\right)\right].
	\end{equation}
\end{thm}
\begin{IEEEproof}
	\appendixproof{argmax-hom-sq-proof}
\end{IEEEproof}
Recall that for a collection of \iid random variables \(X_i : i \in [N])\), the CDF of maximum \(Z = \max_i X_i\) is given as
\begin{equation}
	F_Z(z) = F_X^N(z).
\end{equation}
We see then that an alternative and more intuitive way to view \eq{estimator-ev} is given as
\begin{equation}
	\expected{X_{\hat{Z}_A}} = \sum_{j = 1}^{K} E_j\prob{\ell_{j-1} \leq X_{Z_A} \leq \ell_j}.
\end{equation}

\begin{lem}
	\begin{subequations}
		\begin{align}
			\frac{\partial E_k}{\partial \ell_{k-1}} &= f_{k-1}\frac{E_k - \ell_{k-1}}{p_k}\\
			\frac{\partial E_k}{\partial \ell_{k}} &= f_{k}\frac{\ell_k - E_k}{p_k}
		\end{align}
	\end{subequations}
\end{lem}
\begin{IEEEproof}
	Follows from application of the quotient rule and Leibniz's rule.
\end{IEEEproof}

\begin{lem}
	\label{lem:argmax-hom-sq-derivative}
	\begin{equation}
		\frac{\partial \expected{X_{\hat{Z}_A}}}{\partial \ell_{k}} = f_k\left[\frac{(F_{k+1}^N - F_k^N)(E_{k+1} - \ell_k)}{p_{k+1}} + \frac{(F_k^N - F_{k-1}^N)(\ell_k - E_k)}{p_k} - NF_k^{N-1}(E_{k+1} - E_{k})\right]
	\end{equation}
\end{lem}
\begin{IEEEproof}
	\appendixproof{argmax-hom-sq-derivative-proof}
\end{IEEEproof}
\begin{cor}
	\label{cor:argmax-hom-sq-derivative}
	For \(N = 2\), the above simplifies to
	\begin{equation}
		\frac{\partial \expected{X_{\hat{Z}_A}}}{\partial \ell_{k}} = f_k \left[ \int_{\ell_{k-1}}^{\ell_{k+1}} \! (x - \ell_k) f(x) \, \mathrm{d}x\right].
	\end{equation}
\end{cor}
\begin{IEEEproof}
	\appendixproof{argmax-hom-sq-derivative-cor-proof}
\end{IEEEproof}

\paragraph{Minimum Distortion}
For a given number of intervals \(K\), the decision boundaries \(\{\ell_k : k = 0, \ldots, K\}\) that minimize the expected distortion are given by the solution to the following non-linear optimization:
\begin{equation}
	\label{eq:minimum-distortion}
	\begin{aligned}
		& \underset{\bm{\ell}}{\text{minimize}} & & D(\bm{\ell})\\
		& \text{subject to} & & \ell_{k-1} \leq \ell_{k} \quad k = 1, \ldots, K.
	\end{aligned}
\end{equation}
\begin{thm}
	If \(\{\ell^*_k : k = 0, \ldots, K\}\) is an optimal solution to \eq{minimum-distortion} then there exists \(\mu_K^* \geq 0\) for \(k = 1, \ldots, K\) such that
	\begin{subequations}
	\label{eq:optimal-levels}
	\begin{gather}
		f_k\left[\frac{(F_{k+1}^N - F_k^N)(\ell^*_k - E_{k+1})}{p_{k+1}} + \frac{(F_k^N - F_{k-1}^N)(E_k - \ell^*_k)}{p_k} - NF_k^{N-1}(E_{k} - E_{k+1})\right] - \mu^*_k + \mu^*_{k + 1} = 0 \label{eq:optimal-levels-1}\\
		\mu_k^*(\ell^*_{k - 1} - \ell^*_k) = 0.
	\end{gather}
\end{subequations}
\end{thm}
\begin{IEEEproof}
	The Lagrangian associated with this problem is
	\begin{equation}
		L(\bm{\ell}, \bm{\mu}) = D(\bm{\ell}) + \sum_{k = 1}^{K} \mu_k(\ell_{k-1} - \ell_k)
	\end{equation}
	Taking the derivative \wrt \(\ell_i\) gives
	\begin{equation}
		\frac{\partial L(\bm{\ell}, \bm{\mu})}{\partial \ell_k} = \frac{\partial D(\bm{\ell})}{\partial \ell_k} - \mu_{k} + \mu_{k+1}
	\end{equation}
	where
	\begin{equation}
		\frac{\partial D(\bm{\ell})}{\partial \ell_k} = -\frac{\partial \expected{X_{\hat{Z}_A}}}{\partial \ell_{k}}.
	\end{equation}
	The result follows from setting the above equal to zero and complementary slackness.
\end{IEEEproof}
\begin{cor}
	For \(N = 2\), the above simplifies to
	\begin{subequations}
		\begin{gather}
			f_k \left[ \int_{\ell^*_{k-1}}^{\ell^*_{k+1}} \! (\ell^*_k - x) f(x) \, \mathrm{d}x\right] - \mu^*_k + \mu^*_{k+1} = 0\\
			\mu_k^*(\ell_{k - 1}^* - \ell_k^*) = 0.
		\end{gather}
	\end{subequations}
\end{cor}
\begin{rem}
	In \secref{sq-examples}, we solved for the optimal decision boundaries by setting all the Lagrange multipliers to zero and solving \eq{optimal-levels-1}.
	Depending upon the distribution, \eq{optimal-levels-1} can be solved exactly or with a non-linear solver.
\end{rem}

\paragraph{Entropy-constrained minimum distortion}
The interval \(U_i\) that the \(i\)-th user's observed value lies in is a discrete random variable with probability mass function given by \(\bm{p} = (p_k : k = 1, \ldots, K)\)
and the entropy of \(U_i\) is \(H(U_i) = -\sum_{k = 1}^{K}p_k \log_2 p_k\).
The total rate needed for the \(N\) users to report their intervals is then
\begin{equation}
	R_{HomSQ}(\bm{\ell}) \triangleq \sum_{i = 1}^{N}H(U_i) = N H(U)
\end{equation}
by the \iid assumption of the sources and the homogeneity of the quantizers.
\begin{lem}
	\begin{equation}
		\frac{\partial R_{HomSQ}(\bm{\ell})}{\partial \ell_k} = N f_k \log_2\left(\frac{p_{k+1}}{p_k}\right)
	\end{equation}
\end{lem}
\begin{IEEEproof}
	\begin{equation}
		\begin{aligned}
			\frac{\partial R_{HomSQ}(\bm{\ell})}{\partial \ell_k}
			&= N\sum_{j = 1}^K \frac{\partial}{\partial \ell_k} p_j \log\left(\frac{1}{p_j}\right)\\
			&= N\left(\frac{\partial}{\partial \ell_k} p_k \log\left(\frac{1}{p_k}\right) + \frac{\partial}{\partial \ell_k} p_{k+1} \log\left(\frac{1}{p_{k+1}}\right)\right)\\
			&= N\left(-f_k + f_k\log\left(\frac{1}{p_k}\right) + f_k - f_k\log\left(\frac{1}{p_{k+1}}\right)\right)
		\end{aligned}
	\end{equation}
\end{IEEEproof}

\begin{figure}
	\centering
	\includegraphics[width=252.0pt]{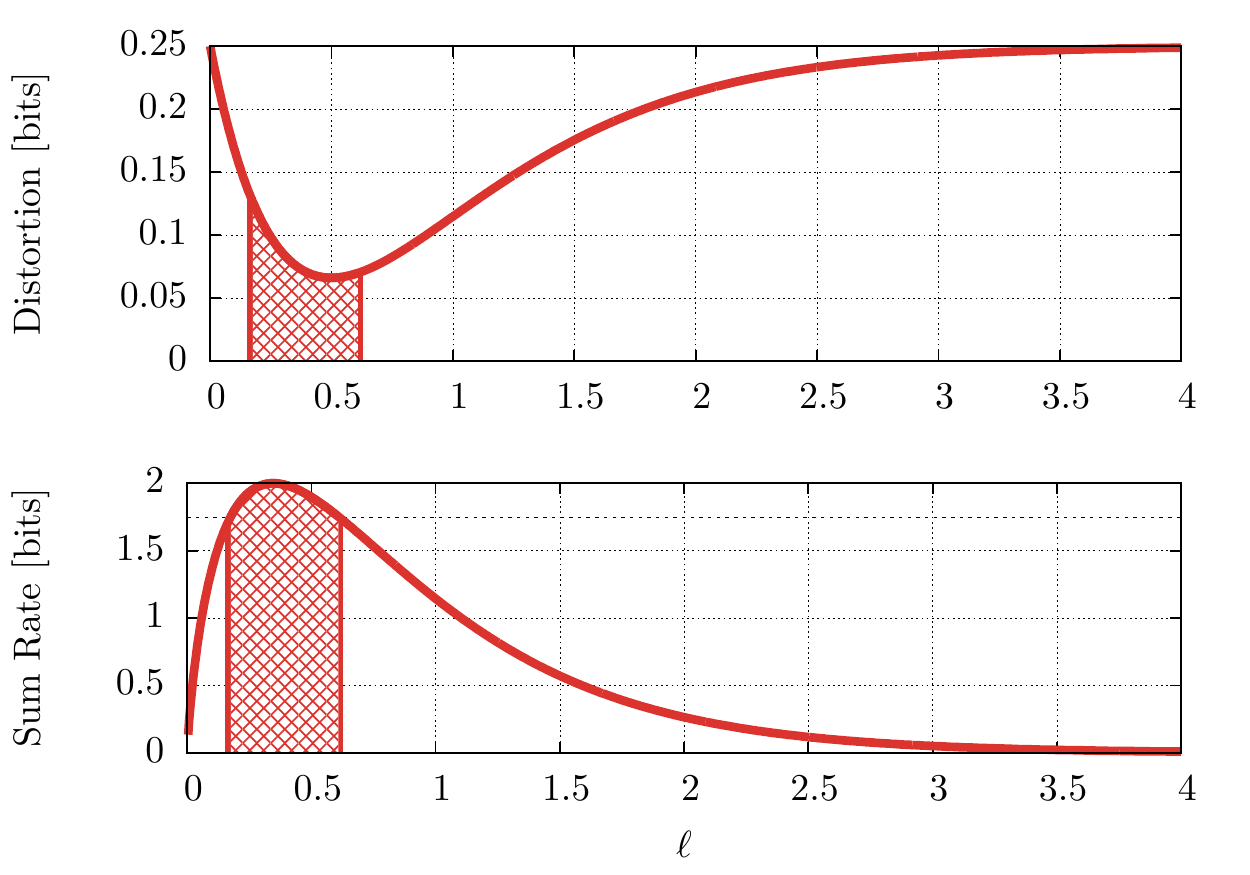}
	\caption{Plots of \(D(\ell)\) and \(R(\ell)\) as functions of \(\ell\).
	For \(R(\ell) \leq 1.75\), the set of feasible \(\ell\) is seen to be non-convex.}\label{fig:exponential-parametric}
\end{figure}

We now consider the problem of minimizing the distortion subject to an upper limit on the sum rate.
\begin{equation}
	\label{eq:entropy-constrained}
	\begin{aligned}
		& \underset{\bm{\ell}}{\text{minimize}} & & D(\bm{\ell})\\
		& \text{subject to} & & R_{HomSQ}(\bm{\ell}) \leq R_0\\
		& & & \ell_{k-1} \leq \ell_k \quad k = 1, \ldots, K
	\end{aligned}
\end{equation}
In general, this problem is \emph{not} convex.
To see this, consider \(X_i \sim\) Exp\((\lambda)\) and a single threshold \(\ell\) (two intervals: \([0, \ell), [\ell, \infty)\)).
\fig{exponential-parametric} shows a plot of \(D(\ell)\) (top) and \(R(\ell)\) (bottom) as \(\ell\) is swept from \(\inf \mathcal{X}\) to \(\sup \mathcal{X}\).
For \(R_0 = 1.75\)~bits, the range of \emph{infeasible} \(\ell\) is shown as a filled area under the rate and distortion curves and we see that the set of feasible \(\ell\) is non-convex.

\begin{thm}
	If \(\{\ell^*_k : k = 0, \ldots, K\}\) is an optimal solution to \eq{entropy-constrained}, then there exists \(\mu_K^* \geq 0\) for \(k = 1, \ldots, K\) and \(\mu_R \geq 0\) such that
	\begin{subequations}
		\label{eq:optimal-entropy-constrained}
		\begin{gather}
			\begin{gathered}
				f_k\left[\frac{(F_{k+1}^N - F_k^N)(\ell^*_k - E_{k+1})}{p_{k+1}} + \frac{(F_k^N - F_{k-1}^N)(E_k - \ell^*_k)}{p_k} - NF_k^{N-1}(E_{k} - E_{k+1}) + \mu^*_R N \log_2\left(\frac{p_{k+1}}{p_k}\right)\right]\\
				- \mu^*_k + \mu^*_{k+1} = 0\\
			\end{gathered}\\
			\mu^*_i(\ell^*_{i-1} - \ell^*_i) = 0 \text{ and } \mu^*_R(R_{HomSQ}(\bm{\ell}^*) - R_0) = 0.
		\end{gather}
	\end{subequations}
\end{thm}
\begin{IEEEproof}
	The Lagrangian associated with this problem is
	\begin{equation}
		L(\bm{\ell}, \bm{\mu}) = D(\bm{\ell}) + \mu_R(R_{HomSQ}(\bm{\ell}) - r) + \sum_{k = 1}^{K} \mu_k(\ell_{k-1} - \ell_k)
	\end{equation}
	Taking the derivative \wrt \(\ell_i\) gives
	\begin{equation}
		\frac{\partial L(\bm{\ell}, \bm{\mu})}{\partial \ell_i} = \frac{\partial D(\bm{\ell})}{\partial \ell_i} + \mu_R \frac{\partial R_{HomSQ}(\bm{\ell})}{\partial \ell_i} - \mu_{i} + \mu_{i+1}.
	\end{equation}
	The result follows from setting the above equal to zero and complementary slackness.
\end{IEEEproof}
\begin{rem}
	Solving for the optimal entropy constrained quantizer is more difficult than solving for the minimum distortion quantizer.
	Depending upon the given values of \(R_0\) and \(K\), the decision boundaries may collapse and the associated Lagrange multipliers need no longer be identically zero.
	A general solution technique for \eq{optimal-entropy-constrained} is beyond the scope of the present paper; generalizations to both Lloyd's and Max's algorithms for entropy constrained quantizer design are presented in \cite{FarMod1984}.
\end{rem}

We conclude with some observations about the rate-distortion curve for entropy-constrained quantizers.
For a given \(K\), suppose \(\bm{\ell}^*\) is a solution to \eq{minimum-distortion}.
If \(R_0 \geq R_{HomSQ}(\bm{\ell}^*)\), then the rate constraint in \eq{entropy-constrained} is not active and \(\bm{\ell}^*\) is also a solution to \eq{entropy-constrained} for the same \(K\).
On the other hand, if \(R_0 < R(\bm{\ell}^*)\) then the rate constraint in \eq{entropy-constrained} is active and \(\bm{\ell}^*\) is infeasible for \eq{entropy-constrained} \cite{FarMod1984}.
Next, consider the rate-distortion curve for a \(N\)-level entropy-constrained quantizer and the sequence of rate-distortion points given by \eq{minimum-distortion} for \(K = 1, \ldots, N\).
These rate-distortion points all lie in the rate-distortion curve for the \(N\)-level entropy-constrained quantizer.

\subsubsection{Heterogeneous Scalar Quantizers}
It is somewhat intuitive to suppose that because the sources are \iid, the quantizers at each user should be identical.
For symmetric functions (e.g., \(\max\)), Misra et al.\ consider only the design of the quantizer for a single user \cite{MisGoyVar2011}.
When the function is \emph{not} symmetric (e.g., \(\argmax\) as in our case), the assumption of \ac{HomSQ} is in fact not true.
\begin{thm}
	For an optimal \ac{HomSQ} \(\bm{\ell}^*\) that achieves a distortion \(D(\bm{\ell}^*)\), there exists a \ac{HetSQ} that achieves the same distortion with rate
	\begin{equation}
		R_{HetSQ}(\bm{\ell}) = (N-2)H(U) + \delta
	\end{equation}
	where
	\begin{equation}
		\delta = \sum_{k = 1}^{K}p_k \log\frac{1}{(p_{k-1} + p_{k})(p_{k} + p_{k+1})} \leq 2 H(U)
	\end{equation}
	and \(p_0 = 0\) and \(p_{K+1} = 0\).
\end{thm}
\begin{IEEEproof}
	We think of \ac{HomSQ} as approximating the continous distribution with a discrete one and then losslessly computing the \(\argmax\) of the quantization bin indices.
	This is exactly the problem considered in \secref{lossless-subband-allocation}.
	From \thmref{optimal_function}, we know that fewer than \(R_{HomSQ}(\bm{\ell})\) bits are needed to enable the \ac{CEO} to losslessly determine \(\argmax\) of the bin indices.
	In the proof of \thmref{optimal_function}, a code is constructed by coloring the vertices of the associated characteristic graphs for each user and entropy coding the vertex colors.
	The rate savings comes by allowing a pair of consecutive bin indices for a user to be assigned the same color, provided the pair of indices are assigned different colors for every other user.
	We can compute the colors directly, by observing that if a pair of consecutive bin indices are being assigned the same color we are merging the underlying bins into one larger bin for that user \emph{only}.
\end{IEEEproof}
\begin{rem}
	As was shown in \thmref{scale_users_argmax}, the total rate savings for losslessly determinging the \(\argmax\) of a discrete distribution is at most \(2\)~bits.
	Therefore, the rate savings of \acp{HetSQ} versus \acp{HomSQ} is also at most \(2\)~bits and the savings per user goes to zero as the number of users is increased.
\end{rem}
\begin{figure}
	\centering
	\subfloat[\ac{HomSQ}]{\includegraphics[width=0.5\textwidth]{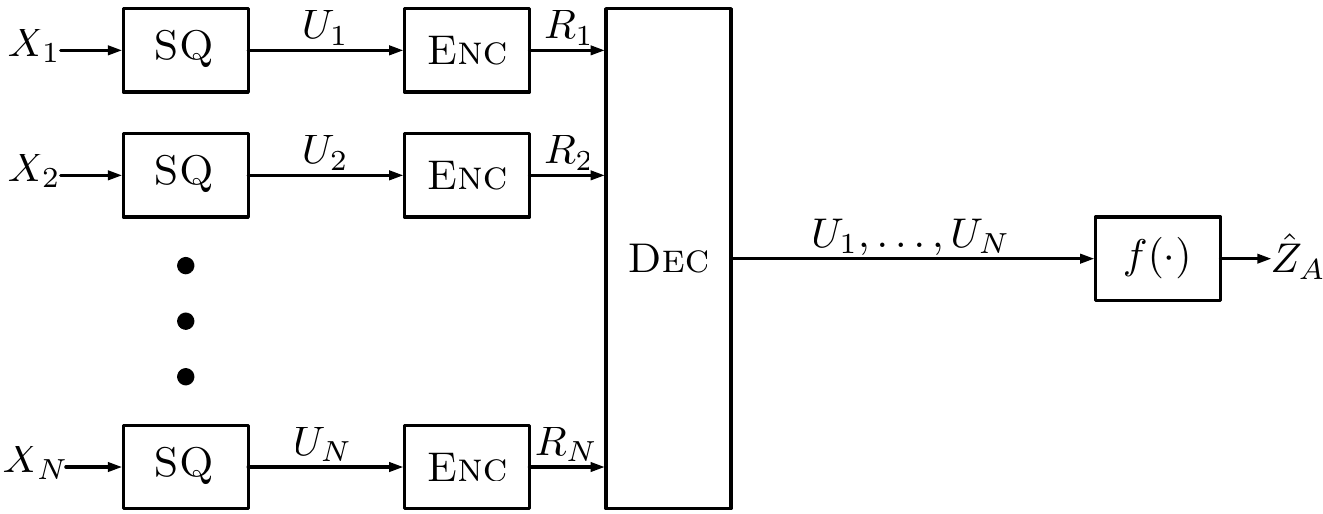}\label{fig:arg-max-hom-sq-block-diagram}}
	\subfloat[Alternative \ac{HomSQ}]{\includegraphics[width=0.5\textwidth]{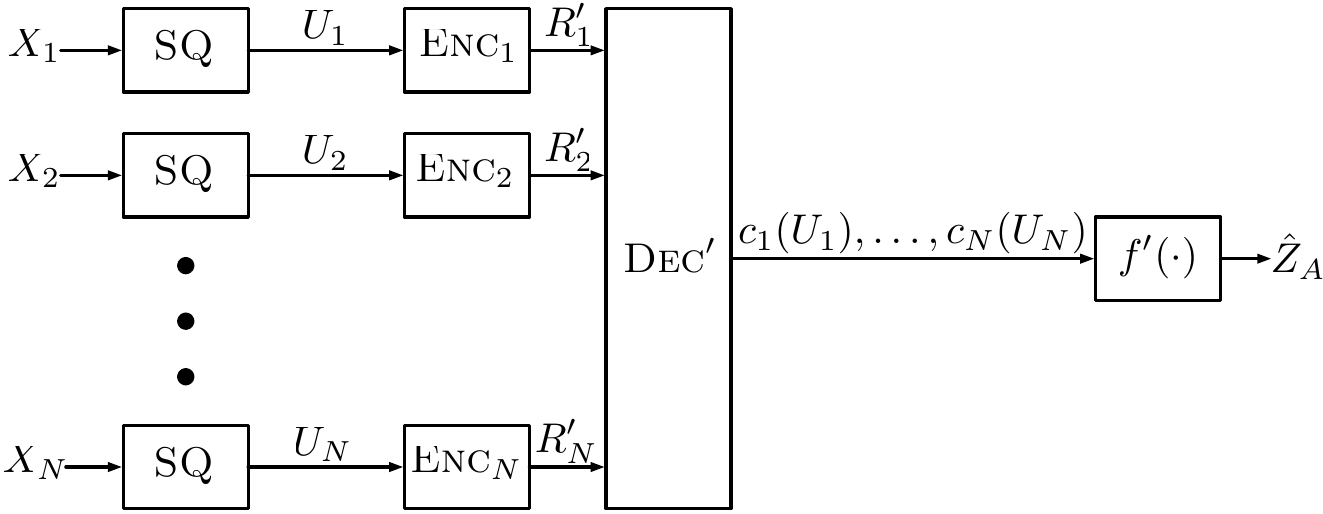}\label{fig:arg-max-alt-hom-sq-block-diagram}}\\
	\subfloat[\ac{HetSQ}]{\includegraphics[width=0.5\textwidth]{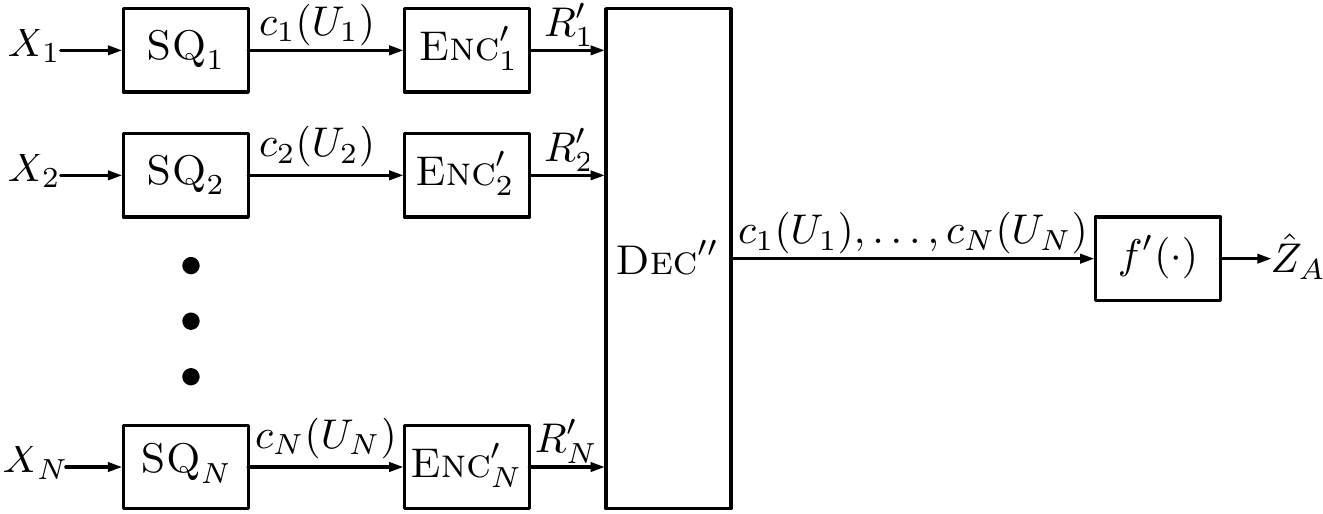}\label{fig:arg-max-het-sq-block-diagram}}
	\caption{Block diagram of possible scalar quantizers for \(\argmax\).
	For \ac{HomSQ} \protect\subref{fig:arg-max-hom-sq-block-diagram}, \(U_1, \ldots, U_N\) are \iid and the encoder is the same for each user.
	In the alternative \ac{HomSQ} scheme \protect\subref{fig:arg-max-alt-hom-sq-block-diagram}, the reduction in rate comes from coloring the vertices in a characteristic graph associated with each source.
	In general, this graph is different for each user and therefore the encoder will be different for each user.
	Finally, by having the quantization operation at each source determine the ``vertex color'', \(U'_1, \ldots, U'_N\) are independent but \emph{not} identically distributed.}
	\label{fig:arg-max-sq-block-diagrams}
\end{figure}

For \ac{HetSQ}, when \(N = 2\) and \(K = 2\) only one of the sources is sending back a bit.
We can use results from rate-distortion for the Bernoulli\((p)\) source with Hamming distortion to trace out the low-rate/high-distortion segment of the trade-off curve.
\begin{lem}
	\label{lem:hamming-estimator}
	The expected value of the estimator when a lossy source code is used to communicate the output \ac{HetSQ} for \(N = 2\) and \(K = 2\) to the CEO is given by
	\begin{equation}
		\label{eq:hamming-estimator}
		\expected{X_{\hat{Z}_A}} = (1 - \hat{p})\expected{X} + \hat{p}(D_H \expected{X | X \leq \ell} + (1 - D_H)\expected{X | \ell \leq X})
	\end{equation}
	where
	\begin{equation}
		\hat{p} = \frac{p_2 - D_H}{1 - 2 D_H}
	\end{equation}
	and the rate is given by
	\begin{equation}
		\label{eq:hamming-rdf}
		R(\ell, D_H) = \begin{dcases}
			h_2(p_2) - h_2(D_H) & D_H \leq \min\{p_2, 1 - p_2\}\\
			0 & D_H > \min\{p_2, 1 - p_2\}
		\end{dcases}
	\end{equation}
\end{lem}
\begin{IEEEproof}
	We assume that user \(1\) is sending the single indicator bit to the \ac{CEO} \wlogen and model this as a Bernoulli\((p_2)\) source with \(p_2 = \prob{X_1 \geq \ell}\) and Hamming distortion \(D_H\).
	The rate-distortion function for this subproblem is given by \eq{hamming-rdf}.
	The test channel that achieves this is a \ac{BSC}\((D_H)\) with input \(\hat{X} \sim\) Bernoulli\((\hat{p})\).
	From this we obtain an expression for the joint \ac{PMF} \(\prob{X = x, \hat{X} = \hat{x}}\) from which we can derive \eq{hamming-estimator}.
\end{IEEEproof}
\begin{rem}
	Observer that for \(D_H = 0\), we obtain the same expression as \eq{estimator-ev} for \(N = 2\) and \(K = 2\) and for \(D_H = \min\{p_2, 1 - p_2\}\), we get \(\expected{X_{\hat{Z}_A}} = \expected{X}\).
\end{rem}

\subsection{Optimal \ac{HetSQ} for \(N = 2\) Users}
\label{sec:two-user-het-sq}
Having considered \acp{SQ} as an acheivable scheme for lossy determination of the \(\argmax\) of a set of distributed sources, we investigate the use of \acp{SQ} for the scenarios where the \(\max\) and the pair \((\argmax, \max)\) need to be determined.
As was shown in the previous section, the assumption of homogeneity of the quantizers leads to suboptimal performance for \(\argmax\).
For the other two functions, we will immediately consider the design of \acp{HetSQ}.

\begin{figure}
	\centering
	\includegraphics{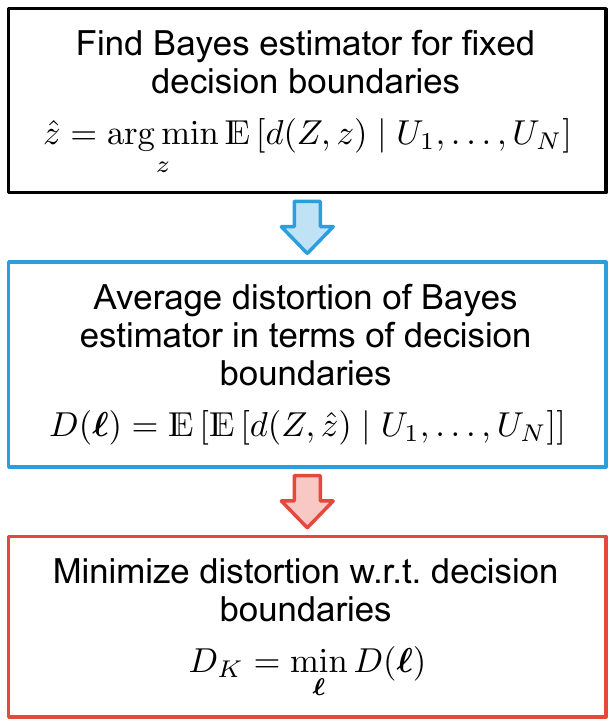}
	\caption{Method utilized to design the scalar quantizers.}\label{fig:sq-design-flow}
\end{figure}

We begin by formally stating the design process that was used implicitly in the previous section, which is depicted in \fig{sq-design-flow}.

Let \(X_i\) represent the source value of user \(i\), \(i \in \{1, \ldots, N\}\).
The quantizer for user \(i\) breaks its support \(\mathcal{X}\) into \(K\) intervals indexed by \(k_i \in \{0, \ldots, K\}\)
\begin{equation}
	\Xca = \bigcup_{k_i = 1}^K \Lca_{i, k_i}, \quad \Lca_{i, k_i}= \left[\ell_{i, k_i - 1}, \ell_{i, k_i}\right].
\end{equation}
Let \(U_i\) be the index of the quantization interval user \(i\)'s source value \(X_i\) is in (i.e., \(U_i = k_i\) if \(X_i \in \left[\ell_{i, k_i - 1}, \ell_{i, k_i}\right]\)) and let \(\Ubf = (U_i : i \in 1, \ldots, N)\).

First, the Bayes estimator \(\hat{z}(\Ubf)\) which minimizes the expected distortion given the quantization indices \(\Ubf\) is found by solving
\begin{equation}
	\label{eq:step1}
	\hat{z}(\Ubf) = \argmin_{\hat{z}} \expected{d(Z,\hat{z}) \mid U_1, \ldots, U_N}
\end{equation}
as a function of the decision boundaries \(\bm{\ell} = \left[\ell_{i, k} : k \in \{1, \ldots, K\}\right]\) and the common CDF \(F_X(x)\) of the sources.
Next, the expected distortion of the Bayes estimator is expressed in terms of the decision boundaires levels and the distribution of the source
\begin{equation}
	\label{eq:step2}
	D(\bm{\ell}) = \expected{\expected{d(Z,\hat{z}) \mid U_1, \ldots, U_N}}.
\end{equation}
Finally, this expression is numerically optimized to yield the minimum distortion \ac{HetSQ} \(\bm{\ell}\) for a given number of quantization levels \(K\)
\begin{equation}
	\label{eq:step3}
	D_K = \min_{\bm{\ell}} D(\bm{\ell}).
\end{equation}
The resulting sum-rate distortion pair for each \(K\) is then \((R_K, D_K)\) where
\begin{equation}
	R_K = -\sum_{i=1}^N \sum_{k_1=1}^{K}  \prob{U_i = k_i} \log_2\prob{U_i = k_i}
\end{equation}
which assumes that the quantization indices \(U_i\) will be block Huffman coded so as to approach a rate equal to their entropy.
Note that we do not consider the more complicated case of entropy constrained scalar quantization as the simpler minimum distortion quantizers already require calculations that are somewhat dense, and also, for the sources of interest, yield rate distortion tradeoffs close to the fundamental limits.

For the case of \(N = 2\), we provide expressions for Bayes estimator and the average distortion of the Bayes estimator as a function of the quantizer parameters \(\bm{\ell}\).
In \secref{sq-examples}, we numerically perform the optimization \eq{step3} for the case of an assumed distribution.

\begin{thm}
	\label{thm:argmax-het-sq}
	The optimal Bayes estimator for the two-user \(\argmax\) problem is
	\begin{equation}
		\hat{z}(U_1, U_2) = \begin{dcases}
			1 & \text{if } \expected{X_1 \mid U_1} \geq \expected{X_2 \mid U_2}\\
			2 & \text{if } \expected{X_1 \mid U_1} < \expected{X_2 \mid U_2}
		\end{dcases}
	\end{equation}
	and the expected distortion when using the optimal Bayes estimator is given by
	\begin{equation}
		\begin{aligned}
			D(\bm{\ell}) 
			&= \expected{d((X_{1}, X_{2}), \hat{z}(U_1, U_2))}\\
			&= \sum_{(k_1, k_2) \in \Zbf_{01}} \int_{\max(\ell_{1, k_1 - 1}, \ell_{2, k_2 - 1})}^{\ell_{2, k_2}} z f_X(z) \left[2F_X(z) - F_X(\ell_{2, k_2 - 1}) - F_X(\ell_{1, k_1 - 1})\right] \, \mathrm{d}z\\
			&+ \sum_{(k_1, k_2) \in \Zbf_{01}} \int_{\ell_{2, k_2}} ^{\ell_{1, k_1}} z f_X(z)\left[F_X(\ell_{2, k_2} ) - F_X(\ell_{2, k_2 - 1})\right] \, \mathrm{d}z\\
			&- \sum_{(k_1, k_2) \in \Zbf_{01}} \max \left(\expected{X_1 \mid U_1 = k_1}, \expected{X_2 \mid U_2 = k_2} \right)\prob{U_1 = k_1}\prob{U_2 = k_2}\\
			&+ \sum_{(k_1, k_2) \in \Zbf_{02}} \int_{\max(\ell_{1, k_1 - 1}, \ell_{2, k_2 - 1})}^{\ell_{1, k_1}} z f_X(z) \left[2F_X(z) - F_X(\ell_{1, k_1 - 1}) - F_X(\ell_{2, k_2 - 1})\right] \, \mathrm{d}z\\
			&+ \sum_{(k_1, k_2) \in \Zbf_{02}} \int_{\ell_{1, k_1}} ^{\ell_{2, k_2}} z f_X(z)\left[F_X(\ell_{1, k_1}) - F_X(\ell_{1, k_1 - 1})\right] \, \mathrm{d}z\\
			&- \sum_{(k_1, k_2) \in \Zbf_{02}} \max \left ( \expected{X_1 \mid X_1 \in \Lca_{1, k_1}}, \expected{X_2 \mid X_2 \in \Lca_{2, k_2}}\right)\prob{U_1 = k_1}\prob{U_2 = k_2}\\
		\end{aligned}
	\end{equation}
	where
	\begin{equation}
		\begin{aligned}
			\Zbf_{01} &= \{k_1, k_2 : \max(\ell_{1, k_1 - 1}, \ell_{2, k_2 - 1}) \leq \ell_{2, k_2} \leq \ell_{1, k_1}\}\\
			\Zbf_{02} &= \{k_1, k_2 : \max(\ell_{1, k_1 - 1}, \ell_{2, k_2 - 1}) \leq \ell_{1, k_1} \leq \ell_{2, k_2}\}.
		\end{aligned}
	\end{equation}
\end{thm}
\begin{IEEEproof}
	\appendixproof{argmax-het-sq-proof}
\end{IEEEproof}

We can repeat a similar procedure for case where the \ac{CEO} is interested in estimating \(\max\) of two distributed sources.
\begin{thm}
	\label{thm:max-het-sq}
	The optimal Bayes estimator for \(Z_M = \max(X_1, X_2 )\) is given by
	\begin{equation}
		\hat{z}(U_1, U_2) = \begin{dcases}
			\hat{z}_1^* & \text{if } \ell_{1, U_1 - 1} \geq \ell_{2, U_2}\\
			\hat{z}_2^* & \text{if } \ell_{2, U_2 - 1} \geq \ell_{1, U_1}\\
			\hat{z}_{01}^* & \text{if } \max(\ell_{2, U_2 - 1}, \ell_{1, U_1 - 1}) < \ell_{2, U_2} \leq \ell_{1, U_1}\\
			\hat{z}_{02}^* & \text{if } \max(\ell_{1, U_1 - 1}, \ell_{2, U_2 - 1}) < \ell_{1, U_1} \leq \ell_{2, U_2}
		\end{dcases}
	\end{equation}
	where
	\begin{equation}
		\hat{z}_1^* =
		\begin{dcases}
			\sol \left\{z : F_X(\ell_{1, U_1}) = F_X(z) + zf_X(z), \; 2f_X(z) + zf'_X(z) \geq 0\right\} & z \in \Lca_{1, U_1}\\
			\ell_{1, U_1 - 1} & \text{otherwise}
		\end{dcases}
	\end{equation}
	\begin{equation}
		\hat{z}_2^* =
		\begin{dcases}
			\sol \left\{z : F_X(\ell_{2, U_2}) = F_X(z) + zf_X(z), \; 2f_X(z) + zf'_X(z) \geq 0\right\} & z \in \Lca_{2, U_2}\\
			\ell_{2, U_2 - 1} & \text{otherwise}
		\end{dcases}
	\end{equation}
	\begin{equation}
		\label{eq:z01-thm}
		\hat{z}_{01}^* = \begin{dcases}
			\hat{z}_{11}^* & w_{11}(\hat{z}_{11}^*) \geq w_{12}(\hat{z}_{12}^*)\\
			\hat{z}_{12}^* & \text{otherwise}.
		\end{dcases}
	\end{equation}
	\begin{equation}
		\label{eq:z02-thm}
		\hat{z}_{02}^* = \begin{dcases}
			\hat{z}_{21}^* & w_{21}(\hat{z}_{21}^*) \geq w_{22}(\hat{z}_{22}^*)\\
			\hat{z}_{22}^* & \text{otherwise}.
		\end{dcases}
	\end{equation}
	\begin{equation}
		\label{eq:W1-thm}
		\begin{aligned}
			w_{11}(z) &= z\left[1 - F_{X \mid X \in \Lca_{1, U_1}}(z)F_{X \mid X \in \Lca_{2, U_2}}(z)\right]\\
			w_{12}(z) &= z\left[1 - F_{X \mid X \in \Lca_{1, U_1}}(z)\right]
		\end{aligned}
	\end{equation}
	\begin{equation}
		\label{eq:Z11-thm}
		\hat{z}_{11}^* = \begin{dcases}
			\sol \left\{z : w'_{11}(z) = 0, w''_{11}(z) \leq 0 \right\} & \max(\ell_{1, U_1 - 1}, \ell_{2, U_2 - 1}) \leq z \leq \ell_{2, U_2}\\
			\max(\ell_{1, U_1 - 1}, \ell_{2, U_2 - 1}) & \text{otherwise}
		\end{dcases}
	\end{equation}
	\begin{equation}
		\label{eq:Z12-thm}
		\hat{z}_{12} ^* = \begin{dcases}
			\sol \left\{z : F_X(\ell_{1, U_1}) = F_X(z) + zf_X(z), \; 2f_X(z) + zf'_X(z) \geq 0 \right\} & \ell_{2, U_2} \leq z \leq \ell_{1, U_1}\\
			\ell_{2, U_2} & \text{otherwise}
		\end{dcases}
	\end{equation}
	\begin{equation}		
		\label{eq:W2-thm}
		\begin{aligned}
			w_{21}(z) &= z\left[1 - F_{X \mid X \in \Lca_{1, U_1}}(z)F_{X \mid X \in \Lca_{2, U_2}}(z)\right]\\
			w_{22}(z) &= z\left[1 - F_{X \mid X \in \Lca_{2, U_2}}(z)\right]
		\end{aligned}
	\end{equation}
	\begin{equation}
		\label{eq:Z21-thm}
		\hat{z}_{21}^* = \begin{dcases}
			\sol\left\{z : w'_{21}(z) = 0, w''_{21}(z) \leq 0 \right\} & \max(\ell_{1, U_1 - 1}, \ell_{2, U_2 - 1}) \leq z \leq \ell_{1, U_1}\\
			\max(\ell_{1, U_1 - 1}, \ell_{2, U_2 - 1}) & \text{otherwise}
		\end{dcases}
	\end{equation}
	\begin{equation}
		\label{eq:Z22-thm}
		\hat{z}_{22} ^* = \begin{dcases}
			\sol\left\{z : F_X(\ell_{2, U_2}) = F_X(z) + zf_X(z), \; 2f_X(z) + zf'_X(z) \geq 0 \right\} & \ell_{1, U_1} \leq z \leq \ell_{2, U_2}\\
			\ell_{1, U_1} & \text{otherwise}
		\end{dcases}
	\end{equation}
	Furthermore, the expected distortion when using the optimal Bayes estimator is given by
	\begin{equation}
		\begin{aligned}
			D(\bm{\ell})
			&= \expected{d((X_1, X_2), \hat{z})}\\
			&= \sum_{(k_1, k_2) \in \Zbf_1} \left[\int_{\ell_{1, k_1 - 1}}^{\ell_{1, k_1}} xf_X(x) \, \mathrm{d}x - \hat{z}_1^* \left[F_X(\ell_{1, k_1}) - F_X(\hat{z}_1^*)\right]\right]\left[F_X(\ell_{2, k_2}) - F_X(\ell_{2, k_2 - 1})\right]\\
			&+ \sum_{(k_1, k_2) \in \Zbf_2} \left[\int_{\ell_{2, k_2 - 1}}^{\ell_{2, k_2}} xf_X(x) \, \mathrm{d}x - \hat{z}_2^* \left[F_X(\ell_{2, k_2}) - F_X(\hat{z}_2^*)\right]\right]\left[F_X(\ell_{1, k_1}) - F_X(\ell_{1, k_1 - 1})\right]\\
			&+ \sum_{(k_1, k_2) \in \Zbf_{01}} \int_{\max(\ell_{1, k_1 - 1}, \ell_{2, k_2 - 1})}^{\ell_{2, k_2}} zf_X(z)\left[2F_X(z) - F_X(\ell_{2, k_2 - 1}) - F_X(\ell_{1, k_1 - 1})\right] \, \mathrm{d}z\\
			&+ \sum_{(k_1, k_2) \in \Zbf_{01}} \int_{\ell_{2, k_2}}^{\ell_{1, k_1}} zf_X(z)[F_X(\ell_{2, k_2}) - F_X(\ell_{2, k_2 - 1})] \, \mathrm{d}z\\
			&- \sum_{(k_1, k_2) \in \Zbf_{01}} \max\left\{w_{11}(\hat{z}_{11}^*), w_{12}(\hat{z}_{12}^*)\right\}[F_X(\ell_{1, k_1}) - F_X(\ell_{1, k_1 - 1})][F_X(\ell_{2, k_2}) - F_X(\ell_{2, k_2 - 1})]\\
			&+ \sum_{(k_1, k_2) \in \Zbf_{02}} \int_{\max(\ell_{1, k_1 - 1}, \ell_{2, k_2 - 1})}^{\ell_{1, k_1}} zf_X(z)\left[2F_X(z) - F_X(\ell_{1, k_1 - 1}) - F_X(\ell_{2, k_2 - 1})\right] \, \mathrm{d}z\\
			&+ \sum_{(k_1, k_2) \in \Zbf_{02}} \int_{\ell_{1, k_1}}^{\ell_{2, k_2}} zf_X(z)[F_X(\ell_{1, k_1}) - F_X(\ell_{1, k_1 - 1})] \, \mathrm{d}z\\
			&- \sum_{(k_1, k_2) \in \Zbf_{02}} \max\left\{w_{21}(\hat{z}_{21}^*), w_{22}(\hat{z}_{22}^*)\right\}[F_X(\ell_{1, k_1}) - F_X(\ell_{1, k_1 - 1})][F_X(\ell_{2, k_2}) - F_X(\ell_{2, k_2 - 1})]
		\end{aligned}
	\end{equation}
	where
	\begin{equation}
		\begin{aligned}
			\Zbf_1 &= \{(k_1, k_2) : \ell_{2, k_2} \leq \ell_{1, k_1 - 1}\}\\
			\Zbf_2 &= \{(k_1, k_2) : \ell_{1, k_1} \leq \ell_{2, k_2 - 1}\}\\
			\Zbf_{01} &= \{k_1, k_2 : \max(\ell_{1, k_1 - 1}, \ell_{2, k_2 - 1}) \leq \ell_{2, k_2} \leq \ell_{1, k_1}\}\\
			\Zbf_{02} &= \{k_1, k_2 : \max(\ell_{1, k_1 - 1}, \ell_{2, k_2 - 1}) \leq \ell_{1, k_1} \leq \ell_{2, k_2}\}.
		\end{aligned}
	\end{equation}
\end{thm}
\begin{IEEEproof}
	\appendixproof{max-het-sq-proof}
\end{IEEEproof}

Finally, we consider the design of \acp{HetSQ} for the scenario where the \ac{CEO} wishes to estimate both \(Z_M = \max (X_1, X_2 )\) and \(Z_A = \arg \max (X_1, X_2 )\).
\begin{thm}
	\label{thm:pair-het-sq}
	The optimal Bayes estimator for \((Z_M, Z_A) = \left(\max(X_1, X_2), \argmax(X_1, X_2)\right)\) is given by
	\begin{equation}
		\hat{z}^*(U_1, U_2) = \begin{dcases}
			(\hat{z}_1^*, 1) & \text{if } \ell_{1, U_1 - 1} \geq \ell_{2, U_2}\\
			(\hat{z}_2^*, 2) & \text{if } \ell_{2, U_2 - 1} \geq \ell_{1, U_1}\\
			(\hat{z}_0^*, \hat{i}_0^*) & \text{if } \max(\ell_{2, U_2 - 1}, \ell_{1, U_1 - 1}) \leq \min(\ell_{1, U_1}, \ell_{2, U_2}) \leq \max(\ell_{1, U_1}, \ell_{2, U_2})
		\end{dcases}
	\end{equation}
	where
	\begin{equation}
		\hat{z}_1^* = \begin{dcases}
			\sol \left\{z : F_X(\ell_{1, U_1}) = F_X(z) + zf_X(z), 2f_{X_1}(z) + zf'_X(z) \geq 0\right\} & z \in \Lca_{1, U_1}\\
			\ell_{1, U_1 - 1} & \text{otherwise}
		\end{dcases}
	\end{equation}
	\begin{equation}
		\hat{z}_2^* = \begin{dcases}
			\sol \left\{z : F_X(\ell_{2, U_2}) = F_X(z) + zf_X(z), 2f_{X_2}(z) + zf'_{X_2}(z) \geq 0\right\} & z \in \Lca_{2, U_2}\\
			\ell_{2, U_2 - 1} & \text{otherwise}
		\end{dcases}
	\end{equation}
	\begin{equation}
		\hat{z}_0^* = \begin{dcases}
			\hat{z}_1^* & \text{if } w(\hat{z}_1^*) \geq w(\hat{z}_2^*)\\
			\hat{z}_2^* & \text{otherwise}.
		\end{dcases}
	\end{equation}
	\begin{equation}
		\hat{i}_0^* = \begin{dcases}
			1 & \text{if } w(\hat{z}_1^*) \geq w(\hat{z}_2^*)\\
			2 & \text{otherwise}
		\end{dcases}
	\end{equation}
	\begin{equation}
		w(z) = z\frac{F_X(\ell_{\hat{i}, U_{\hat{i}}}) - F_X(z)}{F_X(\ell_{\hat{i}, U_{\hat{i}}}) - F_X(\ell_{\hat{i}, U_{\hat{i}}- 1})}
	\end{equation}
	Furthermore, the expected distortion when using the optimal Bayes estimator is given by
	\begin{equation}
		\begin{aligned}
			D(\bm{\ell}) %
			&= \expected{d_{M,A}((X_1, X_2), (\hat{z},\hat{i}))}\\
			&= \sum_{(k_1, k_2) \in \Zbf_1} \left[\int_{\ell_{1, k_1 - 1}}^{\ell_{1, k_1}} xf_X(x) \, \mathrm{d}x - \hat{z}_1^*(F_X(\ell_{1, k_1}) - F_X(\hat{z}_1^*))\right](F_X(\ell_{2, k_2}) - F_X(\ell_{2, k_2 - 1}))\\
			&+ \sum_{(k_1, k_2) \in \Zbf_2} \left[\int_{\ell_{2, k_2 - 1}}^{\ell_{2, k_2}} xf_X(x) \, \mathrm{d}x - \hat{z}_2^*(F_X(\ell_{2, k_2}) - F_X(\hat{z}_2^*))\right](F_X(\ell_{1, k_1}) - F_X(\ell_{1, k_1 - 1}))\\
			&+ \sum_{(k_1, k_2) \in \Zbf_{01}} \int_{\max(\ell_{1, k_1 - 1}, \ell_{2, k_2 - 1})}^{\ell_{2, k_2}} zf_X(z)(2F_X(z) - F_X(\ell_{2, k_2 - 1}) - F_X(\ell_{1, k_1 - 1})) \, \mathrm{d}z\\
			&+ \sum_{(k_1, k_2) \in \Zbf_{01}} \int_{\ell_{2, k_2}}^{\ell_{1, k_1}} zf_X(z)(F_X(\ell_{2, k_2}) - F_X(\ell_{2, k_2 - 1})) \, \mathrm{d}z\\
			&- \sum_{(k_1, k_2) \in \Zbf_{01}} \max\left\{w(\hat{z}_{1}^*), w(\hat{z}_{2}^*)\right\}(F_X(\ell_{1, k_1}) - F_X(\ell_{1, k_1 - 1})(F_X(\ell_{2, k_2}) - F_X(\ell_{2, k_2 - 1}))\\
			&+ \sum_{(k_1, k_2) \in \Zbf_{02}} \int_{\max(\ell_{1, k_1 - 1}, \ell_{2, k_2 - 1})}^{\ell_{1, k_1}} zf_X(z)(2F_X(z) - F_X(\ell_{1, k_1 - 1}) - F_X(\ell_{2, k_2 - 1})) \, \mathrm{d}z\\
			&+ \sum_{(k_1, k_2) \in \Zbf_{02}} \int_{\ell_{1, k_1}}^{\ell_{2, k_2}} zf_X(z)(F_X(\ell_{1, k_1}) - F_X(\ell_{1, k_1 - 1})) \, \mathrm{d}z\\
			&- \sum_{(k_1, k_2) \in \Zbf_{02}} \max\left\{w(\hat{z}_{1}^*), w(\hat{z}_{2}^*)\right\}(F_X(\ell_{1, k_1}) - F_X(\ell_{1, k_1 - 1}))(F_X(\ell_{2, k_2}) - F_X(\ell_{2, k_2 - 1}))\\
		\end{aligned}
	\end{equation}
	where
	\begin{equation}
		\begin{aligned}
			\Zbf_1 &= \{(k_1, k_2) : \ell_{2, k_2} \leq \ell_{1, k_1 - 1}\}\\
			\Zbf_2 &= \{(k_1, k_2) : \ell_{1, k_1} \leq \ell_{2, k_2 - 1}\}\\
			\Zbf_{01} &= \{k_1, k_2 : \max(\ell_{1, k_1 - 1}, \ell_{2, k_2 - 1}) \leq \ell_{2, k_2} \leq \ell_{1, k_1}\}\\
			\Zbf_{02} &= \{k_1, k_2 : \max(\ell_{1, k_1 - 1}, \ell_{2, k_2 - 1}) \leq \ell_{1, k_1} \leq \ell_{2, k_2}\}.
		\end{aligned}
	\end{equation}
\end{thm}
\begin{IEEEproof}
	\appendixproof{pair-het-sq-proof}
\end{IEEEproof}

We conclude by breifly commenting on the subtle difference between between the design of the Bayes estimator for estimating the \(\max\) (\thmref{max-het-sq}) and estimating the pair \((\max, \argmax)\) (\thmref{argmax-het-sq}).
For the case of estimating the \(\max\), the Bayes estimator is given by
\begin{equation}
	\argmax_{z} z\prob{\max\{X_1, X_2\} \geq z \mid X_1 \in \Lca_{1, U_1}, X_2 \in \Lca_{2, U_2}}
\end{equation}
while for the case of estimating the pair \((\max, \argmax)\), the Bayes estimator is given by
\begin{equation}
	\argmax_{z,i} z\prob{X_i \geq z \mid X_1 \in \Lca_{1, U_1}, X_2 \in \Lca_{2, U_2}}.
\end{equation}
If \(\Lca_{1, U_1} \cap \Lca_{2, U_2} = \varnothing\), the above expressions are identical becuse the \ac{CEO} can identify the \(\argmax\) with zero error.
If, on the other hand, \(\Lca_{1, U_1} \cap \Lca_{2, U_2} \neq \varnothing\) then the two expressions are different; in the first case, the objective function is a \emph{product} of conditional CDFs and in the second case, the objective function is a conditional CDF.
For a fixed rate, the \ac{HetSQ} for \(\max\) should be able to acheive a lower distortion than the \ac{HetSQ} for the pair \((\max, \argmax)\).

\subsection{Examples}
\label{sec:sq-examples}
In this section, we consider two different continuous distributions for the sources and compare the performance of \ac{HomSQ}, homogeneous \ac{ECSQ}, and \ac{HetSQ}.
We also show results for a discrete distribution in order to gauge the performance of the \acp{SQ} relative to fundamental limit given by the rate-distortion function.

\begin{figure*}
	\centering
	\subfloat[]{\includegraphics[width=156.0pt]{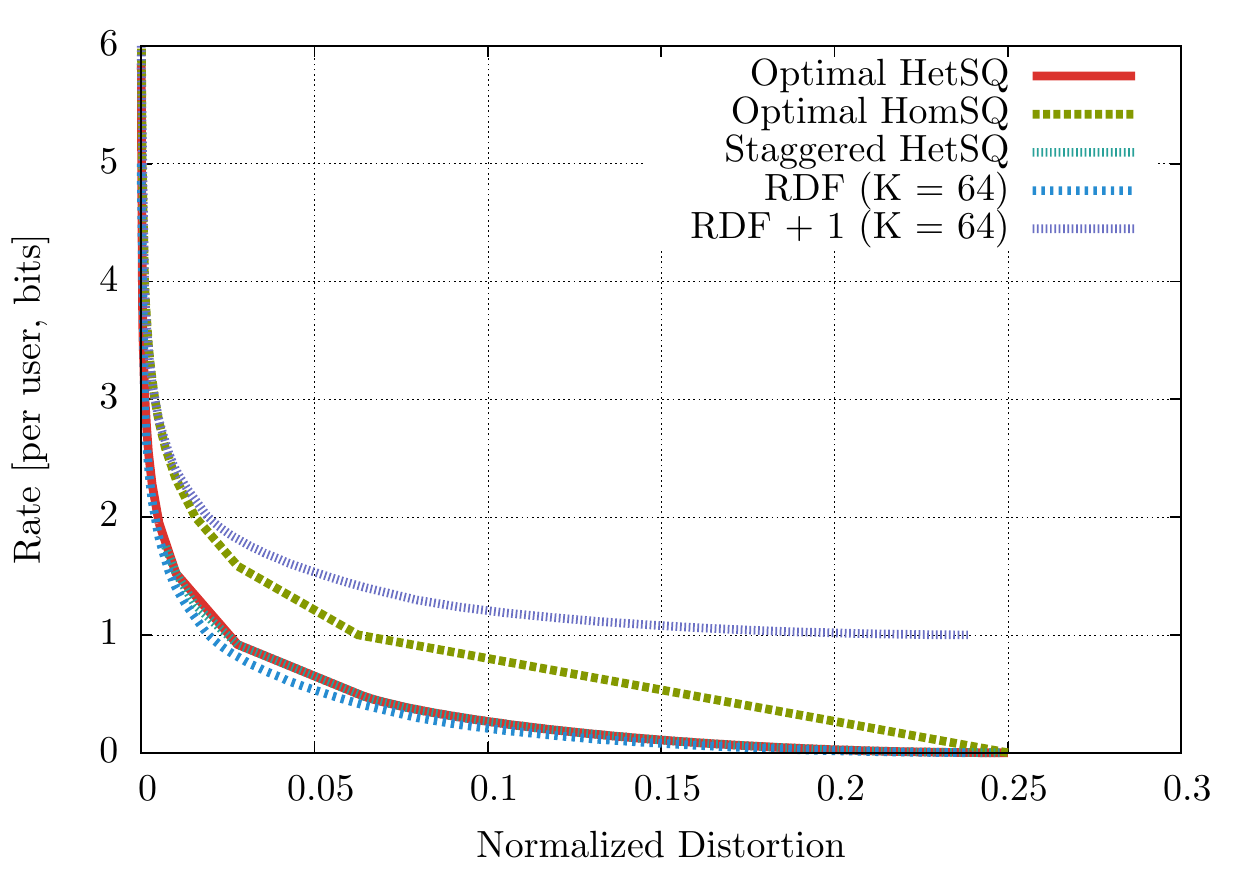}\label{fig:QuantArgmax}}
	\subfloat[]{\includegraphics[width=156.0pt]{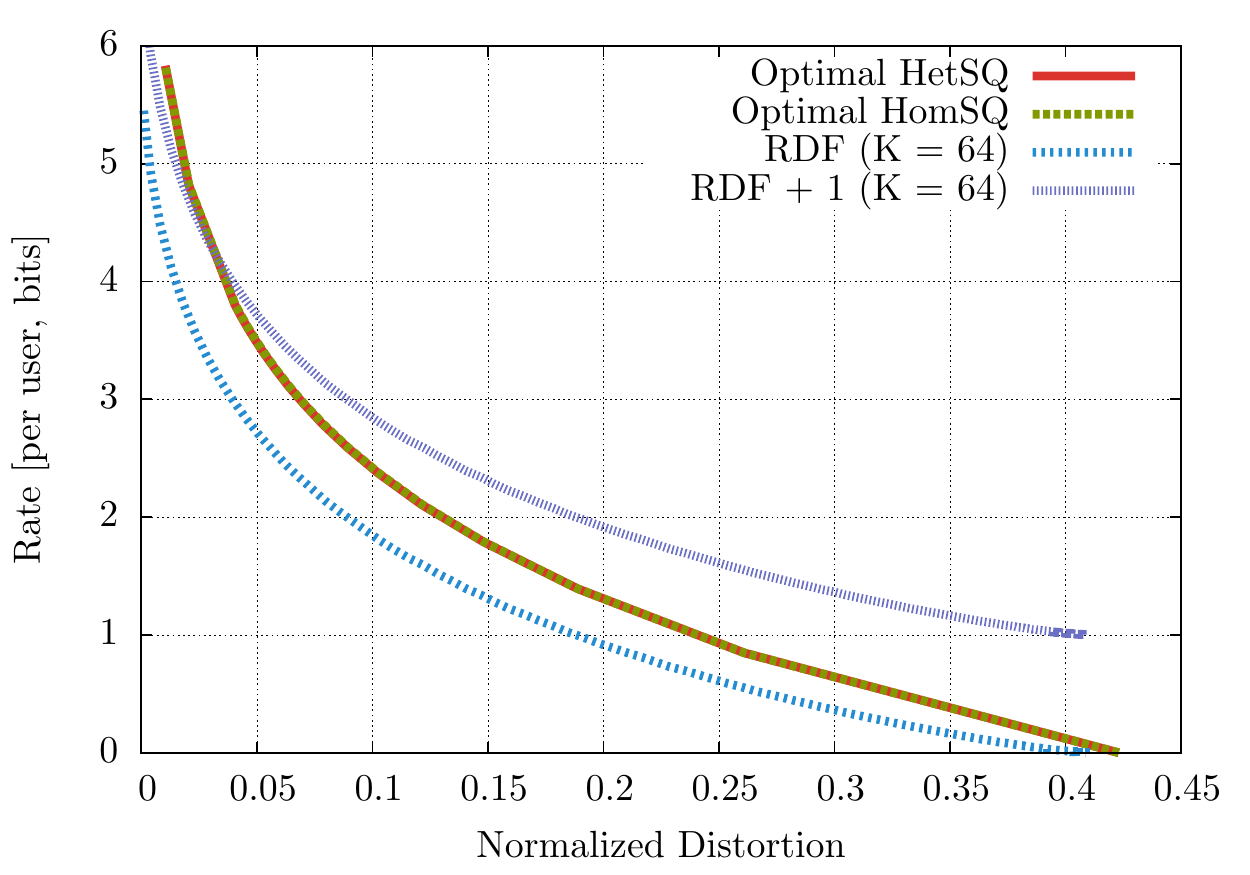}\label{fig:QuantMax}}
	\subfloat[]{\includegraphics[width=156.0pt]{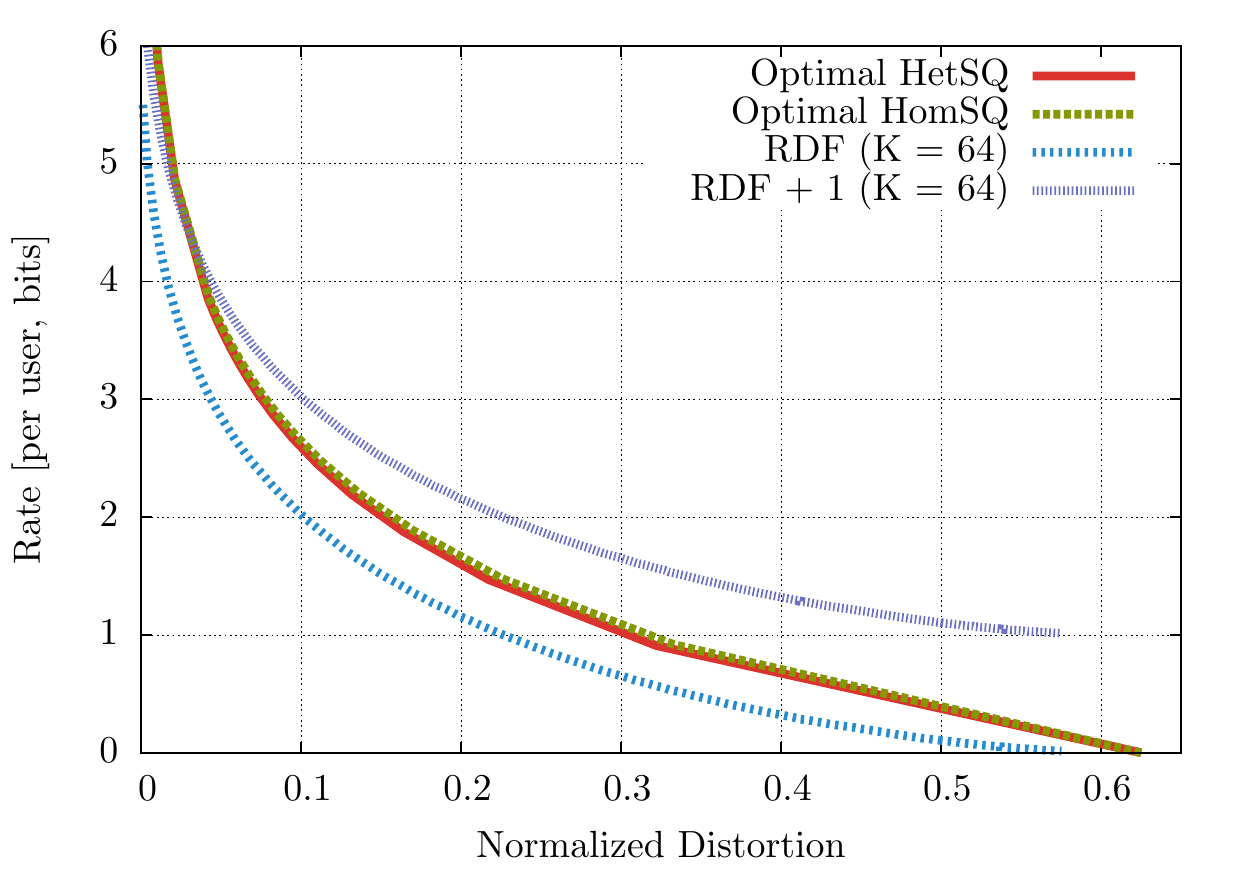}\label{fig:QuantPair}}
	\caption[]{Rate distortion tradeoff for numerically optimized \ac{HetSQ} for two users with sources distributed Uniform\((0, 1)\): \subref{fig:QuantArgmax} \(\argmax\); \subref{fig:QuantMax} \(\max\), and; \subref{fig:QuantMax} \((\argmax, \max)\). The rate-distortion performance of \ac{HomSQ} and \ac{HetSQ} is compared to the rate-distortion function. For the purposes of comparison, we have included a trendline for the rate distortion function plus a bit.}
\end{figure*}
\begin{exmp}[\(\argmax\) quantizer with Uniform\((0, 1)\)]
	When \(X_1, X_2 \sim\) Uniform\((0, 1)\), the Bayes' detector is
	\begin{equation}
		\hat{z}(U_1, U_2) = \begin{dcases}
			1 & \text{if } \frac{\ell_{1, U_1 - 1} + \ell_{1, U_1}}{2} \geq \frac{\ell_{2, U_2 - 1} + \ell_{2, U_2}}{2}\\
			2 & \text{if } \frac{\ell_{1, U_1 - 1} + \ell_{1, U_1}}{2} < \frac{\ell_{2, U_2 - 1} + \ell_{2, U_2}}{2}
		\end{dcases}
	\end{equation}
	For \((U_1 = k_1 , U_2 = k_2) \in \Zbf_{01}\), the expected distortion \(\expected{d((X_1, X_2), \hat{z}(U_1, U_2)) \mid U_1, U_2}\prob{U_1, U_2}\) is
	\begin{equation}
		\begin{split}
			\left[\frac{2}{3}z^3 - \frac{1}{2}(\ell_{1,k_1 - 1} + \ell_{2,k_2 - 1})z^2\right] \Big|_{\max(\ell_{1, k_1 - 1}, \ell_{2, k_2 - 1})}^{\ell_{2,k_2}} + \frac{1}{2}(\ell_{2,k_2} - \ell_{2,k_2 -1})z^2\Big|_{\ell_{2,k_2}}^{\ell_{1,k_1}}\\
			- (\ell_{1,k_1} - \ell_{1, k_1 - 1})(\ell_{2,k_2} - \ell_{2, k_2 - 1})\max\left\{\frac{\ell_{1,k_1} + \ell_{1,k_1 - 1}}{2}, \frac{\ell_{2,k_2} + \ell_{2,k_2 - 1}}{2}\right\}
		\end{split}
	\end{equation}
	For \((U_1 = k_1, U_2 = k_2) \in \Zbf_{02}\), the expected distortion \(\expected{d((X_1, X_2), \hat{z}(U_1, U_2)) \mid U_1, U_2}\prob{U_1, U_2}\) is
	\begin{equation}
		\begin{split}
			\left[\frac{2}{3}z^3 - \frac{1}{2}(\ell_{1,k_1 - 1} + \ell_{2,k_2 - 1})z^2\right] \Big|_{\max(\ell_{1, k_1 -1}, \ell_{2, k_2 - 1})}^{\ell_{1,k_1}} + \frac{1}{2}(\ell_{2,k_2} - \ell_{2,k_2 - 1})z^2\Big|_{\ell_{1,k_1}} ^{\ell_{2,k_2}}\\
			- (\ell_{1,k_1} - \ell_{1, k_1 - 1})(\ell_{2,k_2} - \ell_{2, k_2 - 1})\max\left\{\frac{(\ell_{1,k_1} + \ell_{1,k_1 - 1})}{2}, \frac{(\ell_{2,k_2} + \ell_{2,k_2 - 1})}{2}\right\}
		\end{split}
	\end{equation}
	Summing the above two expressions over \(\Zbf_{01}\) and \(\Zbf_{02}\) gives expression for the distortion when the \ac{CEO} utilizes a Bayes estimator a function of quantization decision boundaries \(\bm{\ell}\) that are heterogeneous across users.
	We then numerically optimize this expression to obtain the rate-distortion paris \((R_K, D_K)\) for the given number \(K\) of quantization bins.
	
	For the case of \ac{HomSQ}, we have
	\begin{equation}
		\ell^*_k = \frac{k}{K},\; \mu^*_k = 0 \quad k \in 1, \ldots, K
	\end{equation}
	as a solution to \eq{optimal-levels}.
	As expected, the optimal quantizer for a uniform distribution is uniform.
	Substituting into the expressions for distortion and rate we obtain
	\begin{equation}
		D_K = \frac{1}{6K^2},\; R_K = N\log_2 K.
	\end{equation}

	\fig{QuantArgmax} shows the per user rate and the normalized distortion of both \ac{HomSQ}, staggered \ac{HetSQ}, and the optimal \ac{HetSQ} along with a numerically computed approximation of the rate-distortion function for function \(\argmax\).
	Interestingly, while the approach of staggering \ac{HomSQ} decision boundaries across users to effect potentially suboptimal \ac{HetSQ} design, we observe here that the optimal \ac{HetSQ} we have derived yields nearly identical performances, at least for the two user case under investigation.	
	Here, a large improvement is achieved by passing between the \ac{HomSQ} and \ac{HetSQ} designs.
	Additionally, all of the \(16\), \(32\), and \(64\) level fundamental limits are right on top of one another and have already effectively converged to the continuous limit.
	Finally, we observe that the designed practical scalar scheme is right up against the fundamental overhead performance tradeoff.
\end{exmp}
\begin{exmp}[Two user \(2\)-level distributed \ac{HomSQ} for \(\max\) with Uniform\((0, 1)\)]
	For \(2\)-level scalar quantizer, we set up the qunatizer parameters as follows
	\begin{align*}
		\Lca_{1,1} &= [\ell_{1, 0}, \ell_{1, 1}], \qquad \Lca_{1, 2} = [\ell_{1, 1}, \ell_{1, 2}]\\
		\Lca_{2,1} &= [\ell_{2, 0}, \ell_{2, 1}], \qquad \Lca_{2, 2} = [\ell_{2, 1}, \ell_{2, 2}]
	\end{align*}
	where \(\ell_{1,0} = \ell_{2,0} = 0\) and \(\ell_{1,2} = \ell_{2,2} = 1\).
	We want to find \(l_{1,1}\) and \(l_{2,1}\) minimizing the expected distortion.
	For convenience, we analyze a homogeneous scalar quantizer which has same parameters between users.
	In this case, we set \(\ell_{1,1} = \ell_{2,1} = \ell\).
	First, we solve for the Bayes estimator \(\hat{z}(U_1, U_2)\) as a function of \(\ell\).
	Based on the \thmref{max-het-sq}, we obtain the following expression for the Bayes estimator
	\begin{equation}
		\hat{z}(U_1, U_2) = \begin{dcases}
			\frac{\ell}{\sqrt{3}} & \text{if } U_1 = U_2 = 1\\
			\max\left\{\frac{1}{2}, \ell\right\} & \text{if } U_1 \neq U_2\\
			\frac{2\ell + \sqrt{4\ell^2 - 6\ell + 3}}{3} & \text{if } U_1 = U_2 = 2
		\end{dcases}
	\end{equation}
	Next, we substitute the above expression in the expression for conditional expected distortion to obtain, then we solve an optimization problem.
	\begin{multline}
		\expected{d((X_1, X_2), \hat{z}(U_1, U_2)) \mid U_1, U_2}\prob{U_1, U_2} = \\
		\begin{dcases}
			\frac{2(\sqrt{3}-1)}{3\sqrt{3}}\ell^3 & \text{if } U_1 = U_2 = 1\\
			\frac{(1 - 2\ell^2)\ell}{2}  \mathds{1}_{\ell \leq \frac{1}{2}} + (1-\ell)^2 \ell \mathds{1}_{\ell > \frac{1}{2}}  & \text{if } U_1 \neq U_2\\
			\frac{-7\ell^3 + 36 \ell^2 - 45\ell + 18 - 2(4 \ell^2 - 6\ell +3)^{\frac{3}{2}}}{27} & \text{if } U_1 = U_2 = 2.
		\end{dcases}
	\end{multline}

	Finally, we numerically optimize the expected distortion as a function \(\ell\).
	For \(0 < l \leq \frac{1}{2}\), the average distortion is
	\begin{equation}
		D(\ell) = \frac{2(\sqrt{3}-1)}{3\sqrt{3}}\ell^3 + \frac{(1 - 2\ell^2)\ell}{2} + \frac{-7\ell^3 + 36 \ell^2 - 45\ell + 18 - 2(4 \ell^2 - 6\ell +3)^{\frac{3}{2}}}{27}
	\end{equation}
	which has a minimum value of \(0.2204\) at \(\ell = 0.5\).
	In \(\frac{1}{2} \leq l < 1\), the average distortion is
	\begin{equation}
		D(\ell) = \frac{2(\sqrt{3}-1)}{3\sqrt{3}}\ell^3 + (1-\ell)^2 \ell + \frac{-7\ell^3 + 36 \ell^2 - 45\ell + 18 - 2(4 \ell^2 - 6\ell +3)^{\frac{3}{2}}}{27}
	\end{equation}
	which has a minimum value of \(0.1742\) at \(\ell = 0.7257\).
	Therefore, we should to choose the \ac{HomSQ} parameter \(\ell = 0.7257\) to attain a minimum distortion of \(0.1742\).
\end{exmp}
\begin{exmp}[Two user \(K\)-level Distributed Scalar Quantizer for \(\max\) with Uniform\((0, 1)\)]
	When the two users' source is Uniform\((0, 1)\), an expected minimum distortion is
	\begin{equation}
		\begin{aligned}
			D(\bm{\ell}) %
			&= \sum_{(k_1 , k_2) \in \Zbf_1}
			\begin{dcases}
				\frac{(\ell_{1, k_1}^2 - 2\ell_{1, k_1 - 1}^2)(\ell_{2, k_2} - \ell_{2, k_2 - 1})}{4} & \ell_{1, k_1} \geq 2 \ell_{1, k_1 - 1}\\
				\frac{(\ell_{1, k_1} - \ell_{1, k_1 - 1})^2 (\ell_{2, k_2} - \ell_{2, k_2 - 1})}{2} & \ell_{1, k_1} < 2 \ell_{1, k_1 - 1}
			\end{dcases}\\
			&+ \sum_{(k_1 , k_2) \in \Zbf_2}
			\begin{dcases}
				\frac{(\ell_{1, k_1} - \ell_{1, k_1 - 1})(\ell_{2, k_2}^2 - 2\ell_{2, k_2 - 1}^2)}{4} & \ell_{2, k_2} \geq 2 \ell_{2, k_2 - 1}\\
				\frac{(\ell_{1, k_1} - \ell_{1, k_1 - 1})(\ell_{2, k_2} - \ell_{2, k_2 - 1})^2}{2} & \ell_{2, k_2} < 2 \ell_{2 ,k_2 - 1}
			\end{dcases}\\
			&+ \sum_{(k_1, k_2) \in \Zbf_{01}} \left[\left(\frac{2}{3}\hat{z}_{01}^3 - \frac{(\ell_{1, k_1 - 1} + \ell_{2, k_2 - 1})}{2}\hat{z}_{01}^2\right) \Big|_{\max\{\ell_{1, k_1 - 1}, \ell_{2, k_2 - 1}\}}^{\ell_{2, k_2}} + \frac{(\ell_{2, k_2} - \ell_{2, k_2 - 1})}{2}\hat{z}_{01}^2 \Big|_{\ell_{2, k_2}}^{\ell_{1, k_1}} \right]\\
			&- \sum_{(k_1, k_2) \in \Zbf_{01}} \left[(\ell_{1, k_1} - \ell_{1, k_1 - 1})(\ell_{2, k_2 } - \ell_{2, k_2 - 1}) \max\left\{w_{11}(\hat{z}_{11}^*), w_{12}(\hat{z}_{12}^*)\right\}\right]\\
			&+ \sum_{(k_1, k_2) \in \Zbf_{02}} \left[\left(\frac{2}{3}\hat{z}_{02}^3 - \frac{(\ell_{1, k_1 - 1} + \ell_{2, k_2 - 1})}{2}\hat{z}_{02}^2\right) \Big|_{\max\{\ell_{1, k_1 - 1}, \ell_{2, k_2 - 1}\}}^{\ell_{1, k_1}} + \frac{(\ell_{1, k_1} - \ell_{1, k_1 - 1})}{2}\hat{z}_{02}^2 \Big|_{\ell_{1, k_1}}^{\ell_{2, k_2}}\right]\\
			&- \sum_{(k_1, k_2) \in \Zbf_{02}} \left[(\ell_{1, k_1} - \ell_{1, k_1 - 1})(\ell_{2, k_2 } - \ell_{2, k_2 - 1}) \max\left\{w_{21}(\hat{z}_{21}^*), w_{22}(\hat{z}_{22}^*)\right\}\right]
		\end{aligned}
	\end{equation}
	where \(\hat{z}^*_{01}\), \(w_{11}(\cdot)\), \(\hat{z}^*_{11}\), \(w_{12}(\cdot)\), \(\hat{z}^*_{12}\), \(\hat{z}^*_{02}\), \(w_{21}(\cdot)\), \(\hat{z}^*_{21}\), \(w_{22}(\cdot)\), and \(\hat{z}^*_{22}\) are given by (\ref{eq:z01-thm}--\ref{eq:Z22-thm}).
	
	\fig{QuantMax} shows the per user rate and normalized distortion for \ac{HomSQ} and \ac{HetSQ} along with a numerically computed approximation of the rate-distortion function for estimating the \(\max\) of two distributed users with sources distributed Uniform\((0, 1)\).
	Numerically optimizing the expected distortion for \ac{HomSQ} and \ac{HetSQ} yields rate-distortion pairs \((R_K, D_K)\) that are nearly identical.
	The achievable \ac{SQ} schemes are not particularly far from the fundamental limit, leaving only a small gain possible from a better designed scalar of vector quantizer.
\end{exmp}
\begin{exmp}[\(\max\) and \(\argmax\) Quantizer with Uniform\((0, 1)\)]
	When two users' source is Uniform\((0, 1)\), the expected minimum distortion in region \((U_1 = k_1 , U_2 = k_2) \in \Zbf_1\) is
	\begin{equation}
		\begin{aligned}
			D(\bm{\ell}) %
			&= \sum_{(k_1 , k_2) \in \Zbf_1}
			\begin{dcases}
				\frac{(\ell_{1, k_1}^2 - 2\ell_{1, k_1 - 1}^2)(\ell_{2, k_2} - \ell_{2, k_2 - 1})}{4} & \ell_{1, k_1} \geq 2\ell_{1, k_1 - 1}\\
				\frac{(\ell_{1, k_1} - \ell_{1, k_1 - 1})^2 (\ell_{2, k_2} - \ell_{2, k_2 - 1})}{2} & \ell_{1, k_1} < 2\ell_{1, k_1 - 1}
			\end{dcases}\\
			&+ \sum_{(k_1 , k_2) \in \Zbf_2}
			\begin{dcases}
				\frac{(\ell_{1, k_1} - \ell_{1, k_1 - 1})(\ell_{2, k_2}^2 - 2\ell_{2, k_2 - 1}^2)}{4} & \frac{\ell_{2, k_2}}{2} \geq \ell_{2, k_2 - 1}\\
				\frac{(\ell_{1, k_1} - \ell_{1, k_1 - 1})(\ell_{2, k_2} - \ell_{2, k_2 - 1})^2}{2} & \frac{\ell_{2, k_2}}{2} < \ell_{2, k_2 - 1}
			\end{dcases}\\
			&+ \sum_{(k_1 , k_2) \in \Zbf_{01}} \left[\frac{2}{3}z^3 - \frac{(\ell_{1, k_1 - 1} + \ell_{2, k_2 - 1})}{2} z^2\right] \Big|_{\max\{\ell_{1, k_1 - 1}, \ell_{2, k_2 - 1}\}}^{\ell_{2, k_2}} + \frac{(\ell_{2, k_2} - \ell_{2, k_2 - 1})}{2}z^2 \Big|_{\ell_{2, k_2}}^{\ell_{1, k_1}}\\
			&- \sum_{(k_1 , k_2) \in \Zbf_{01}} (\ell_{1, k_1} - \ell_{1, k_1 - 1})(\ell_{2, k_2} - \ell_{2, k_2 - 1})\max\left\{z_{11}^*\frac{\ell_{1, k_1} - z_{11}^*}{\ell_{1, k_1} - \ell_{1, k_1 - 1}}, z_{12}^*\frac{\ell_{2, k_2} - z_{12}^*}{\ell_{2, k_2} - \ell_{2, k_2 - 1}}\right\}\\
			&+ \sum_{(k_1 , k_2) \in \Zbf_{02}} \left[\frac{2}{3}z^3 - \frac{(\ell_{1, k_1 - 1} + \ell_{2, k_2 - 1})}{2} z^2\right] \Big|_{\max\{\ell_{1, k_1 - 1}, \ell_{2, k_2 - 1}\}}^{\ell_{1, k_1}} + \frac{(\ell_{2, k_2} - \ell_{2, k_2 - 1})}{2}z^2 \Big|_{\ell_{1, k_1}}^{\ell_{2, k_2}}\\
			&- \sum_{(k_1 , k_2) \in \Zbf_{02}} (\ell_{1, k_1} - \ell_{1, k_1 - 1})(\ell_{2, k_2} - \ell_{2, k_2 - 1})\max\left\{z_{21}^*\frac{\ell_{1, k_1} - z_{21}^*}{\ell_{1, k_1} - \ell_{1, k_1 - 1}}, z_{22}^*\frac{\ell_{2, k_2} - z_{22}^*}{\ell_{2, k_2} - \ell_{2, k_2 - 1}}\right\}
		\end{aligned}
	\end{equation}
	where
	\begin{align}
		z_{11}^* &=
		\begin{dcases}
			\frac{\ell_{1, k_1}}{2} & \text{if } \ell_{1, k_1} \ge 2 \ell_{1, k_1 - 1}\\
			\ell_{1, k_1 - 1} & \text{otherwise}
		\end{dcases}\\
		z_{12}^* &=
		\begin{dcases}
			\frac{\ell_{2, k_2}}{2} & \text{if } \ell_{2, k_2} \ge 2 \ell_{2, k_2 - 1}\\
			\ell_{2, k_2 - 1} & \text{otherwise}
		\end{dcases}\\
		z_{21} ^* &=
		\begin{dcases}
			\frac{\ell_{1, k_1}}{2} & \text{if } \ell_{1, k_1} \ge 2 \ell_{1, k_1 - 1}\\
			\ell_{1, k_1 - 1} & \text{otherwise}
		\end{dcases}\\
		z_{22} ^* &=
		\begin{dcases}
			\frac{\ell_{2, k_2}}{2} & \text{if } \ell_{2, k_2} \ge 2 \ell_{2, k_2 - 1}\\
			\ell_{2, k_2 - 1} & \text{otherwise}
		\end{dcases}
	\end{align}	

	\fig{QuantPair} shows the per user rate and normalized distortion for \ac{HomSQ} and \ac{HetSQ} along with a numerically computed approximation of the rate-distortion function for estimating both the \(\argmax\) and \(\max\) of two distributed users with sources distributed Uniform\((0, 1)\).
	Unlike the previous example (estimating just the \(\max\)), we do observe that \ac{HetSQ} has a better performance than \ac{HomSQ} although the improvement is not as marked as for estimating \(\argmax\).
\end{exmp}
\begin{exmp}[\(\argmax\) \ac{SQ} for \(N > 2\) Uniform\((0, 1)\)]
	We now consider the design of a \ac{HomSQ} for estimating the \(\argmax\) from \(N > 2\) sources \iid Uniform\((0, 1)\).
	From \eq{estimator-ev} we obtain the following expression for the expected distortion
	\begin{equation}
		D(\bm{\ell}) = \frac{N}{N+1} - \sum_{j = 1}^{K} \frac{(\ell_{j - 1} + \ell_j)(\ell_j^N - \ell_{j - 1}^N)}{2}
	\end{equation}
	from which we solve for optimal quantizer parameter \(\bm{\ell}^*\)
	\begin{equation}
		\ell_{j - 1}^{*N - 1} = \frac{\ell_{j + 1}^{*N} - \ell_{j - 1}^{*N}}{N(\ell^*_{j + 1} - \ell^*_{j - 1})}.
	\end{equation}

	\fig{uniform-multiuser} shows the per user rate and normalized distortion for \ac{HomSQ} and the staggered \ac{HetSQ} derived from the optimal \ac{HomSQ} for estimating the \(\argmax\) of a collection of distributed users with sources \iid Uniform\((0, 1)\).
	The left subplot is for \(N = 2\) users, the middle subplot for \(N = 4\) users, and \(N = 8\) users.
	We observe immediately that the performance gains of the staggered \ac{HetSQ} over \ac{HomSQ} diminish as the number of users increases.
	Additionally, while the zero rate distortion is increasing in the number of users, we observe that the required rate per user to acheive a specified normalized distortion is non-montonic in the number of users.
	For example, fixing \(D = 0.01\) we observe per user rate for \ac{HomSQ} is \(2.32\)~bits for \(N = 2\), \(2.32\)~bits for \(N = 4\), and \(1.86\)~bits for \(N = 8\).
	The per user rate for \ac{HetSQ} is \(1.52\)~bits for \(N = 2\), \(1.95\)~bits for \(N = 4\), and \(1.71\)~bits for \(N = 8\).
	\begin{figure*}
		\centering
		\subfloat[]{\includegraphics[width=156.0pt]{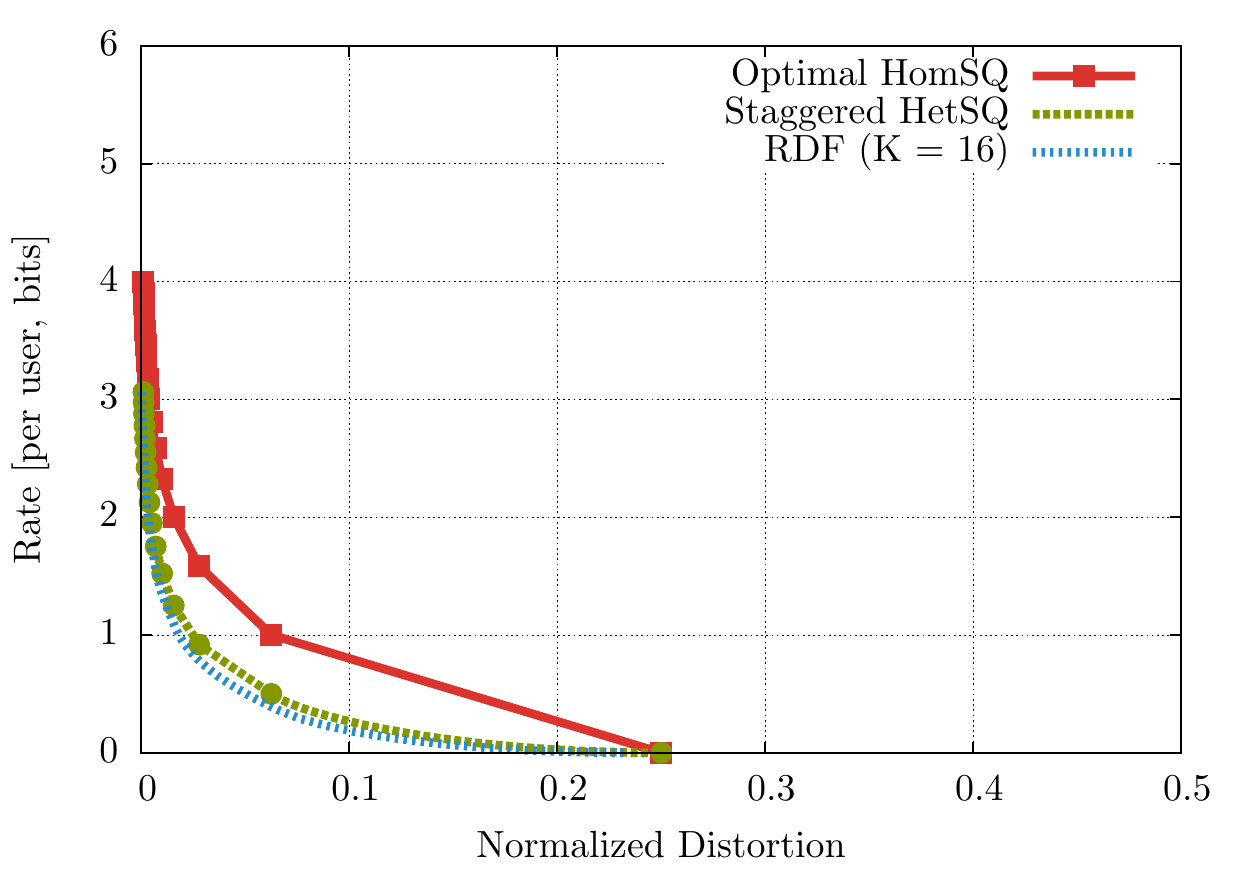}\label{fig:uni-argmax-N-2}}
		\subfloat[]{\includegraphics[width=156.0pt]{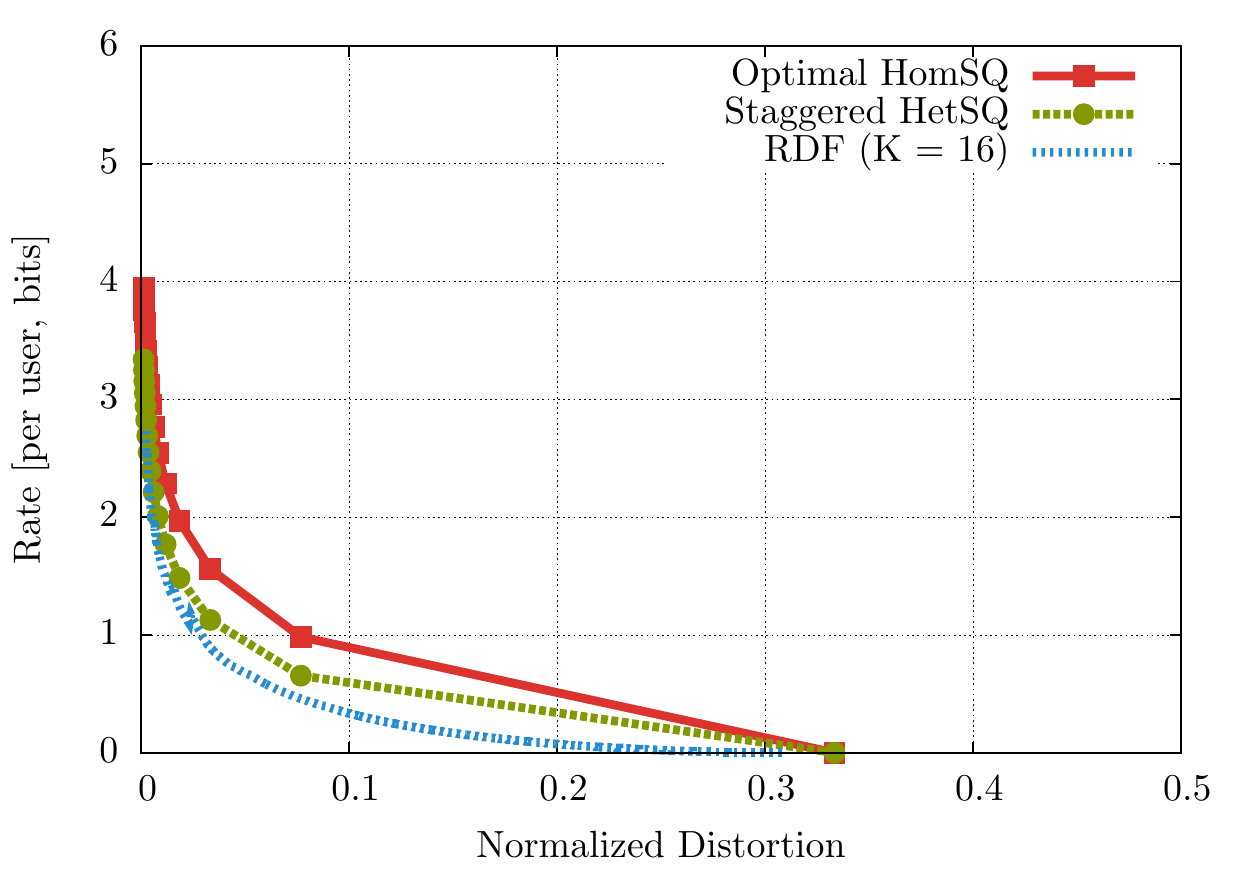}\label{fig:uni-argmax-N-3}}
		\subfloat[]{\includegraphics[width=156.0pt]{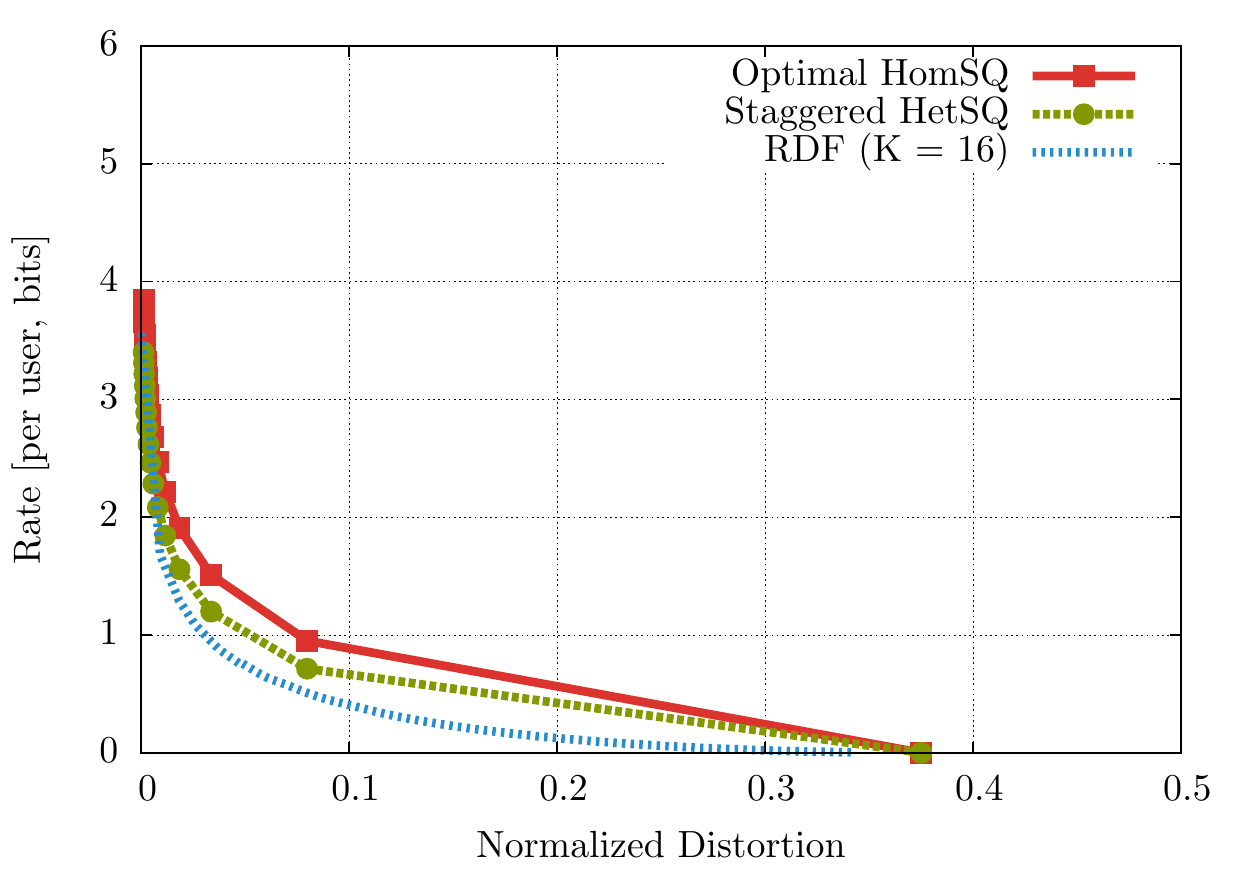}\label{fig:uni-argmax-N-4}}
		\caption[]{Comparison of the rate-distortion tradeoff for \ac{HomSQ} vs.\ \ac{HetSQ} for: \subref{fig:uni-argmax-N-2} \(N = 2\) users; \subref{fig:uni-argmax-N-3} \(N = 3\) users, and; \subref{fig:uni-argmax-N-4} \(N = 4\) users.}
		\label{fig:uniform-multiuser}
	\end{figure*}

	Computing the rate distortion bounds becomes computationally expensive for a larger number of users \(N\); however, we can investigate the scaling behavior of the presented achievable schemes for a wider range of \(N\).
	We see in \fig{uniform-multiuser-sq} that there is very little difference between the curves for \(N = 2\) and \(N = 4\), which matches with the behavior observed in \fig{uni-rdf-N-2}.
	For larger values of \(N\), we see that the per-user rate required to obtain a given distortion rapidly decreases with \(N\).
	This proves in turn that the rate distortion function must also posses this property.
	\begin{figure*}
		\centering
		\includegraphics[width=234.0pt]{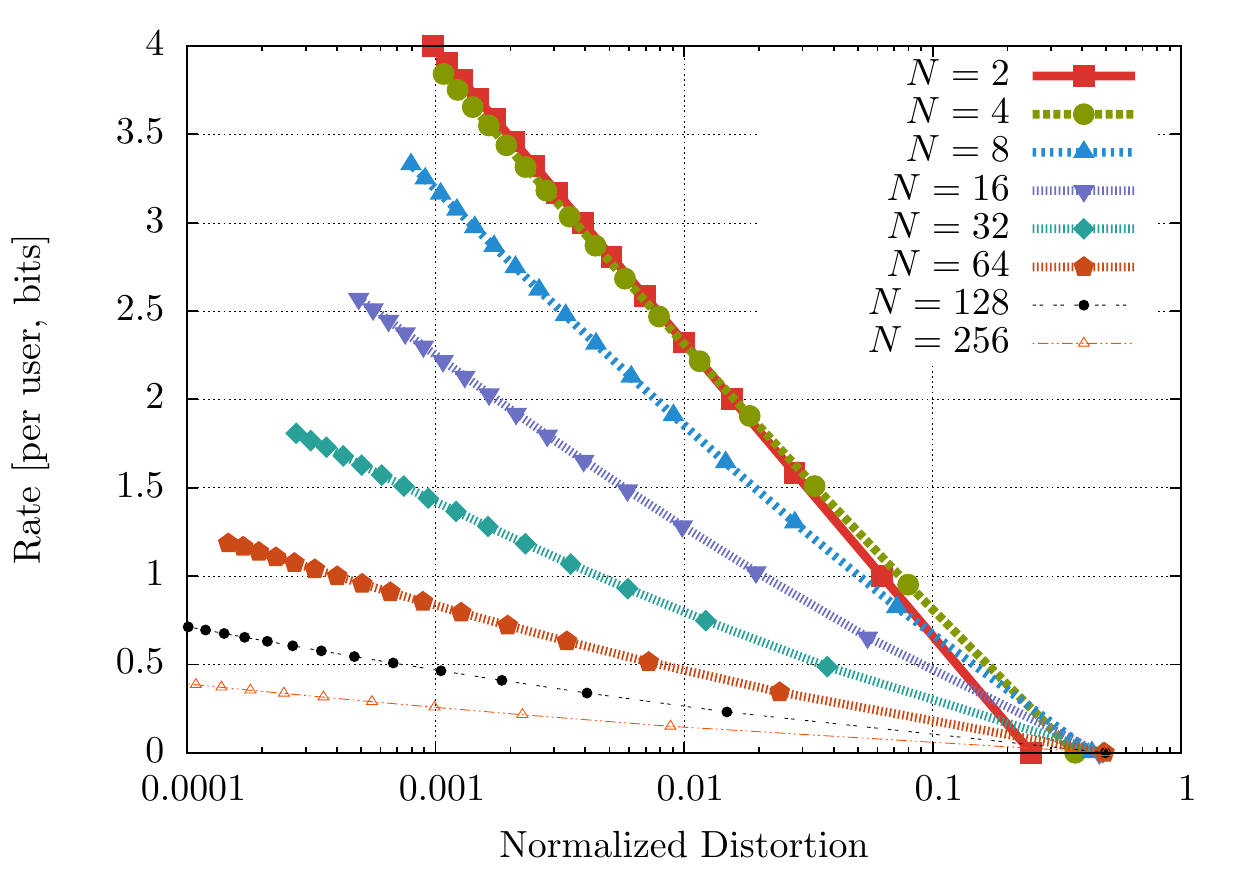}
		\caption[]{Comparison of the rate-distortion tradeoff for optimal \ac{HomSQ} as the number of users is increased.}
		\label{fig:uniform-multiuser-sq}
	\end{figure*}
\end{exmp}

Our investigation of the rate distortion tradeoff for the \ac{CEO} to compute the extremization functions in a lossy manner was motivated by the minimal rate savings shown in \secref{lossless-subband-allocation}.
Shown in \tblref{rate-savings} are the rate savings of \ac{SQ} for a small increase in tolerable distortion when the sources are distributed Uniform\((0, 1)\).
\begin{table}
	\centering
	\begin{tabular}{ccccccccc}
		\toprule
		& \(N = 2\) & \(N = 4\) & \(N = 8\) & \(N = 16\) & \(N = 32\) & \(N = 64\) & \(N = 128\) & \(N = 256\) \\
		\cmidrule(r){2-9}
		optimal \ac{HomSQ} & 41.72\% & 39.51\% & 41.62\% & 43.18\% & 43.00\% & 41.92\% & 40.59\% & 41.80\% \\
		staggered \ac{HetSQ} & 50.05\% & 42.75\% & 43.12\% & 43.94\% & 43.40\% & 42.13\% & 40.70\% & 41.86\% \\
		\bottomrule
	\end{tabular}
	\caption{Rate savings for uniform sources when the allowable normalized distortion is increased from \(0.001\) to \(0.01\)}
	\label{tbl:rate-savings}
\end{table}
We see an average savings of about 43\% accoss \ac{SQ} type and number of users.
We conclude that by incurring small increase in estimation error, a significant rate savings can be realized and that these savings do not appear to diminish as the number of users is increased.

\section{INTERACTIVE EXTREMIZATION}
\label{sec:interactive-subband-allocation}
Comparing with the straightforward scheme in which each user uses a \ac{SW} code to forward its observations to the CEO to enable it to learn the $\argmax$, we showed in \secref{lossless-subband-allocation} that it is possible to save some rate by applying graph coloring. However we showed that the maximum possible such savings is small: one can save at most $2$ bits for independent and identically distributed sources and the per user saving as the number of users goes to infinity will be $0$. This motivated us to investigate other coding strategies capable of delivering a larger reduction in rate. While the previous section considered strategies that enabled this rate reduction by relaxing the requirement that the CEO compute the extremizations perfectly to computing them in a lossy manner, here we will revert to the requirement that the extremizations are computed losslessly and focus instead on rate savings obtainable through interactive communication.

\emph{Interactive communication} is defined to be a method of communication that allows message passing forward and backward multiple times between two or more terminals \cite{InteractiveTheory}. It has been shown that interactive communication can provide substantial rate savings over non-interactive communication in some distributed function computation problems \cite{InteractiveFunctionComputation}. Here, we will apply interactive communication to the extremization problems, and show that a large reduction in rate is possible relative to the non-interactive lossless limits presented in \secref{lossless-subband-allocation}. While we will not discuss any fundamental limits as they are not yet available in the literature for the interactive CEO problems under investigation, we will demonstrate that through interaction we can obtain substantial rate savings. 

Inspired by the selective multiuser diversity (SMUD)\cite{SMUD} scheme as well as the multi-predefined thresholds \cite{MultiThresholds} scheme which is an extension of SMUD, we propose here the \ac{MTIS} between the CEO and the users that efficiently encodes the feedback necessary for the lossless computation of the extremization problems. We show that the \ac{MTIS} achieves a large reduction in the rate when interaction is utilized when compared with the rate results of \thmref{N-user-achievability} in \secref{lossless-subband-allocation} in which each user sends its own message to the CEO by graph coloring.

 Here we will model the observations of the users as identically distributed discrete random variables with support set $\mathcal{X}=\{\alpha_1,\ldots,\alpha_L\}\ \text{s.t.}\ 0 < \alpha_1 < \alpha_2 < \ldots < \alpha_L$, and cumulative distribution function $F_x(x)$. The users each initially occupy a fraction of a bandwidth to communicate to the CEO. The CEO knows the user index and the part of the bandwidth that it corresponds to at the beginning. The interactive communication will occur over multiple rounds indexed by $t$. During each round, only a subset of the users called the online users will participate in the communication, and the CEO will know which users are offline by the information it exchanges with the online users.  For instance, in the $\argmax$ case, a user remains online only while it is still possible to be the $\argmax$ based on the information it has received up until this round, and is offline otherwise.  The part of communication bandwidth associated with offline users is freed up for use by other communications and is thus not wasted. During round $t$, given the CDF $F_t(x)$, the support set $\mathcal{X}_t=\left\{\alpha_1^t,\ldots,\alpha_{L(t)}^t\right\}$ and the $N_t$ conditioned on the information that the CEO obtained about the online users thus far, it will determine and send a common message $V_t$ to declare a threshold to each of the online users, and each online user $i$ responds with a message $U^{i}_{t}$ to let the CEO know whether or not it is above this threshold for all $i \in [N_t]$. The user will stay online for the next round if it feeds back a 1.  Alternatively, if a user feeds back a 0, but the next threshold $\lambda_{t+1}$ is lower than $\lambda_{t}$ (which indicating that all users replied $0$ at round $t$), it will also stay online, otherwise this user becomes offline.
 After receiving all of the feedback bits, the CEO can obtain the information $F_{t+1}(x), \mathcal{X}_{t+1}$ and $N_{t+1}$ for next round's communication. If there is only one user above the threshold $\lambda_T$ at the round $T$, this user is the $\argmax$ and the communication process stops. Similarly, if $\left|\mathcal{X}_T\right| =1$, then all of the online users in the next round attain the $\max$, and the communication process stops since the CEO can pick any one of these users to be the $\argmax$. If more than one online user replies a $1$, then conditioned on all the information received thus far, the new channel distribution parameters for the next round are
\begin{equation} \label{eq:analy1}
\begin{aligned}
& N_{t+1} = \sum_{i=1}^{N_t} \mathds{1}_{x_i \ge \lambda_t}\\
& \alpha_{1}^t=\lambda_t\\
& \alpha_{L(t+1)}^t=\alpha_{L(t)}^t\\
& F_{t+1}(x)=\frac{F_t(x)-F_t(\lambda_t)}{F_t(\alpha_{L(t)}^t)-F_t(\lambda_t)}
\end{aligned}
\end{equation}
While if all users reply $0$, then conditioned on all the information received thus far at the CEO, the new channel distribution parameters for the next round are
\begin{equation} \label{eq:analy2}
\begin{aligned}
& N_{t+1}=N_{t}\\
& \alpha_1^{t+1}=\alpha_1^t\\
& \alpha_{L(t+1)}^{t+1}=\lambda_t\\
& F_{t+1}(x)=\frac{F_t(x)-F_t(\alpha_1^t)}{F_t(\lambda_t)-F_t(\alpha_1^t)}
\end{aligned}
\end{equation}
The threshold for next round can be generated based on the new information.  Hence the algorithm of \ac{MTIS} operates as follows.

\begin{algorithm}[H]
\DontPrintSemicolon
\KwResult{Let the CEO decide the $\argmax$}
initialization:\ number of online users $N_1 = N$, the support set and the CDF of the discrete source random variables $\mathcal{X}_1=\mathcal{X}, F_1(x) = F_x(x)$\;
\While{$N_t>1$ \& $|\mathcal{X}_t|>1$}{
step 1) CEO sends threshold $\lambda_t$ to all users\;
step 2) online users generate the parameters $\mathcal{X}_t$ and $F_t(x)$ according to \eq{analy1} \eq{analy2}, and decide to stay online or not\;
step 3) online users send $U_t^i=\mathds{1}_{x_i \ge \lambda}$ for all $i \in [N_t]$\;
step 4) CEO generates the parameters $N_{t+1}$,$\mathcal{X}_{t+1}$ and $F_{t+1}(x)$ according to \eq{analy1} \eq{analy2}\;
}
\caption{Muti-Thresholds Interactive Scheme\label{algorithm:MTIS}}
\end{algorithm}
\subsection{Analysis}
Our aim in this subsection is to determine the optimal choice of the thresholds in the interactive scheme in the sense of minimizing the average total amount of rates must incur.
\begin{figure*}[ht] 
	\centering
	\subfloat[]{\includegraphics[width=234.0pt]{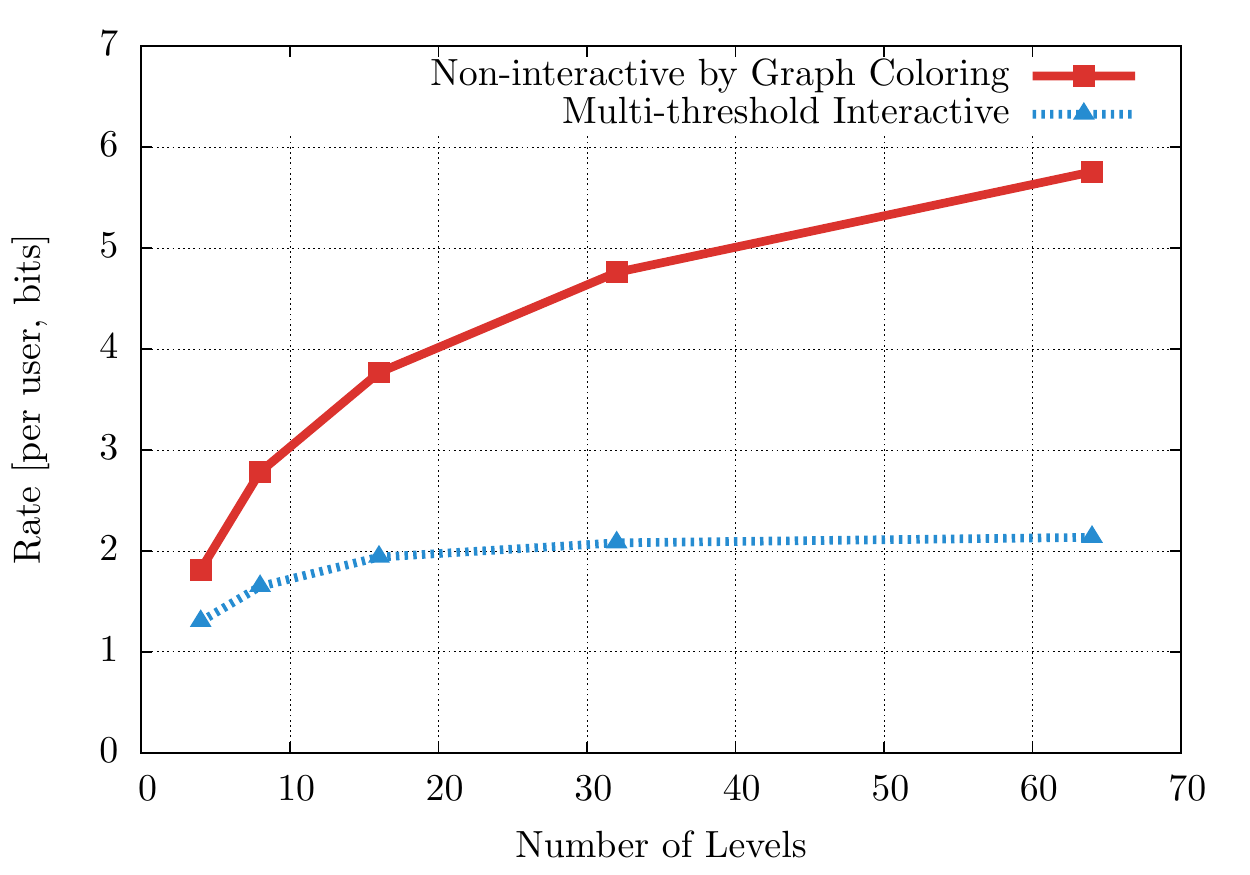}\label{fig:user8}}
	\subfloat[]{\includegraphics[width=234.0pt]{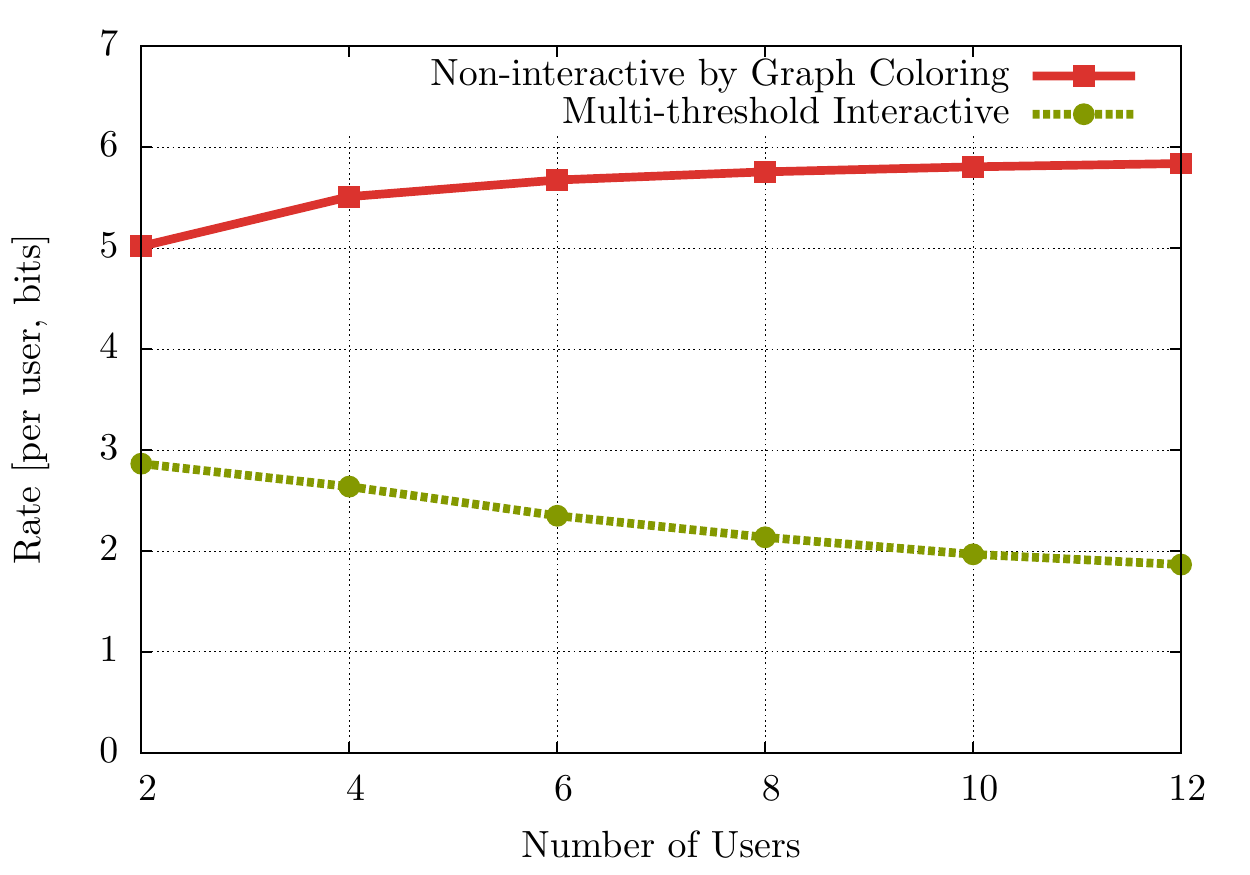}\label{fig:level64_2}}
	\caption[]{\ac{MTIS} Vs. non interactive for the $\argmax$ problem with \subref{fig:user8} $N_1=8$, and; \subref{fig:level64_2} $\left|\mathcal{X}_1\right| = 64$}
	\label{fig:fix-level}
\end{figure*}

Define $R$ to be the total expected number of overhead bits exchanged when using the series of threshold levels $\lambda_1,\lambda_2,\cdots$, and define $R^*$ to be
\begin{equation}
\label{eq:optimal-R}
R^*= \min_{\lambda_1, \lambda_2, \ldots} R(\lambda_1,\lambda_2,\ldots)
\end{equation}
It is clear that $R^*$ will be a function of the initial number of users $N_1$ (all of whom are initially online) and $\mathcal{X}_1$. We will need the following theorem to solve the optimization problem.
\begin{thm} \label{thm:DP}
Problem (\ref{eq:optimal-R}) is a dynamic programming problem.
\end{thm}
\begin{IEEEproof}
We first show there will be a finite stop T for \eq{optimal-R}. The threshold $\lambda_t$ is picking from the support set of the sources $\mathcal{X}=\left\{\alpha_1^t,\ldots,\alpha_{L(t)}^t\right\}$. After each round of communication, the support set will be updated to either $\left\{\alpha_1^t, \ldots, \lambda_t\right\}$ or $\left\{\lambda_t, \ldots, \alpha_{L(t)}^t\right\}$, hence the size of the support set is monotone decreasing. Therefore finite rounds are needed to decrease the support set to be $1$ and the communication stops.

Also, we observe that if  policy $\lambda^*_1,\cdots,\lambda^*_T$ is the optimal choice of thresholds for initial condition $N_1$, $\left\{\alpha_1^1,\ldots,\alpha_{L(1)}^1\right\}$ and $F_1(x)$ then the truncated policy $\lambda^*_t,\cdots,\lambda^*_T$ will be the optimal choice of thresholds for initial condition $N_t$, $\left\{\alpha_1^t,\ldots,\alpha_{L(t)}^t\right\}$ and $F_t(x)$, and thus the problem has the form of a dynamic programming problem.
\end{IEEEproof}

In order to solve this problem, we begin with a one round analysis in which we assume to pick $\lambda_t$ as the threshold for round t and that the thresholds after round t have been optimized already. Define $R_t(\lambda_t)$ as the expected aggregate rate from round t to the end, then
\begin{equation}
\label{eq:dynamic-rate}
R_t(\lambda_t) = H(\lambda_t|\lambda_1,\mathcal{X}_1,N_1,\cdots,\lambda_{t-1},\mathcal{X}_{t-1},N_{t-1})+N_t+ \mathbb{E}[R_{t+1}]
\end{equation}
where the first term represents the minimum number of bits needed to let the users know the threshold in round t, the second term represents the total number of bits of feedback from  the $N_t$ users, and the last term represents the expected rate cost for future rounds which can be further expressed as
\begin{equation} \label{eq:E(R)}
\mathbb{E}[R_{t+1}]=\sum_{i=0}^{N_t} p_i\mathbb{E}[R_{t+1}|i]=(F_t(\lambda_t))^{N_t}R^*(N_t,\alpha_1^t,\lambda_t)+\sum_{i=1}^{N_t}(1-F_t(\lambda_t))^iF_t(\lambda_t)^{N_t-i}\frac{N_t!}{i!(N_t-i)!}R^*(i,\lambda_t,\alpha_{L(t)}^t)
\end{equation}
where $p_i$ represents the probability of i users reply 1 at round t. The optimal choice of threshold at round $t$ then must satisfy
\begin{equation}
\label{eq:dynamic-policy}
\lambda^*_t= \argmin_{\lambda_t} R_t(\lambda_t)
\end{equation}
\eq{dynamic-rate} \eq{E(R)} and \eq{dynamic-policy} together form a policy iteration algorithm\cite{DynamicProgramming} for this dynamic programming problem.

\subsection{Thresholds vs.\ Number of Users}
Let us now consider several possible methods of encoding the threshold, and hence several possible values for the quantity $H(\lambda_t|\lambda_1,\mathcal{X}_1,N_1,\cdots,\lambda_{t-1},\mathcal{X}_{t-1},N_{t-1})$ in \eq{dynamic-rate}. Based on \ac{SW} codes, the minimum information the CEO needs to broadcast should be the conditional entropy of the threshold given all previous knowledges that the online users have.

For the purposes of comparison, and ease of the associated algorithm encoder design, let us also consider two additional coding strategies which are easy to implement.  We will see that these two strategies also require less communication than the non-interaction scheme. The first strategy is to encode the threshold with no conditioning
\begin{equation}
U_t = H(\lambda_t)=\log_2 |\mathcal{X}_t|
\end{equation}
Motivated by the idea that the users may calculate the optimal choice of threshold themselves rather than receiving it, we provide the second strategy that the BS broadcasts the number of currently online users. Observe that the optimal policy $\lambda^*$ at each round is determined by the information the CEO has, including $N_t$, $f_t(x)$ and $\mathcal{X}_t=\left\{\alpha_1^t,\ldots,\alpha_{L(t)}^t\right\}$. We show that it is enough to let the users calculate the threshold by broadcasting $N_t$ by induction.
\begin{thm}
\label{thm:user-number}
The number of online users $N_t$ is a sufficient statistic of the optimal threshold $\lambda_t^*$.
\end{thm}
\begin{IEEEproof}
\eq{dynamic-rate} \eq{E(R)} \eq{dynamic-policy} show that the  CEO determines the $\lambda_t^*$ by the information of $\left\{\left(F_i(x), \mathcal{X}_i, N_i\right) : i\in [t]\right\}$, hence it suffices to show that the users can learn $F_t(x)$ and $\mathcal{X}_t$ by knowing $N_t$ at round $t$. We prove it by induction.  
At round $1$, each user has the CDF $F_1(x)$, the support set $\mathcal{X}_1=\left\{\alpha_1^1,\ldots,\alpha_{L(1)}^1\right\}$ and its own value $x_i$, hence the optimal threshold $\lambda_1^*$ can be calculated after receiving the initial number of the online users $N_1$. Suppose that at round $t-1$ the users successfully compute the threshold $\lambda_{t-1}^*$ by the information $N_{t-1}$, $F_{t-1}(x)$ and $\mathcal{X}_{t-1}=\left\{\alpha_1^{t-1},\ldots,\alpha_{L(t-1)}^{t-1}\right\}$. Now at round $t$ for any user $i \in [N_{t-1}]$, if it receives $N_t=N_{t-1}$ and its value is below the threshold $\lambda_{t-1}$ which means it replied a $0$ at previous round, then it knows that every user must be below the previous threshold and $\mathcal{X}_t = \{\alpha_1^{t-1},\lambda_{t-1}^*\}$; similarly if it receives $N_t=N_{t-1}$ and its value is above the threshold $\lambda_{t-1}$, then it knows that every user must be above the previous threshold and $\mathcal{X}_t=\{\lambda_{t-1}^*,\alpha_{L(t-1)}^{t-1}\}$. Therefore the $\mathcal{X}_t$ can be renewed at each user by the following rules
\begin{equation}
\mathcal{X}_t=\left\{
\begin{aligned}
& \left\{\lambda_{t-1}^*,\alpha_{L(t-1)}^{t-1}\right\}&& if\ N_t<N_{t-1}\\
& \left\{\alpha_1^{t-1}, \lambda_{t-1}^*\right\}&& if\ N_t=N_{t-1}\ and\ \lambda_{t-1}^*>x_i\\
& \left\{\lambda_{t-1}^*,\alpha_{L(t-1)}^{t-1}\right\}&& if\ N_t=N_{t-1}\ and\ \lambda_{t-1}^*\le x_i
\end{aligned}.
\right.
\end{equation}
Note that the user will turn offline if $N_t<N_{t-1}$ and $\lambda_{t-1}^*>x_i$ and stay online otherwise. The updated CDF $F_t(x)$ can be get by \eq{analy1} \eq{analy2} once $\mathcal{X}_t$ has been renewed. Therefore, the threshold $\lambda_t^*$ can be determined after each user receiving the $N_{t}$.
\end{IEEEproof}
\begin{figure}
	\centering
	\includegraphics[width=252.0pt]{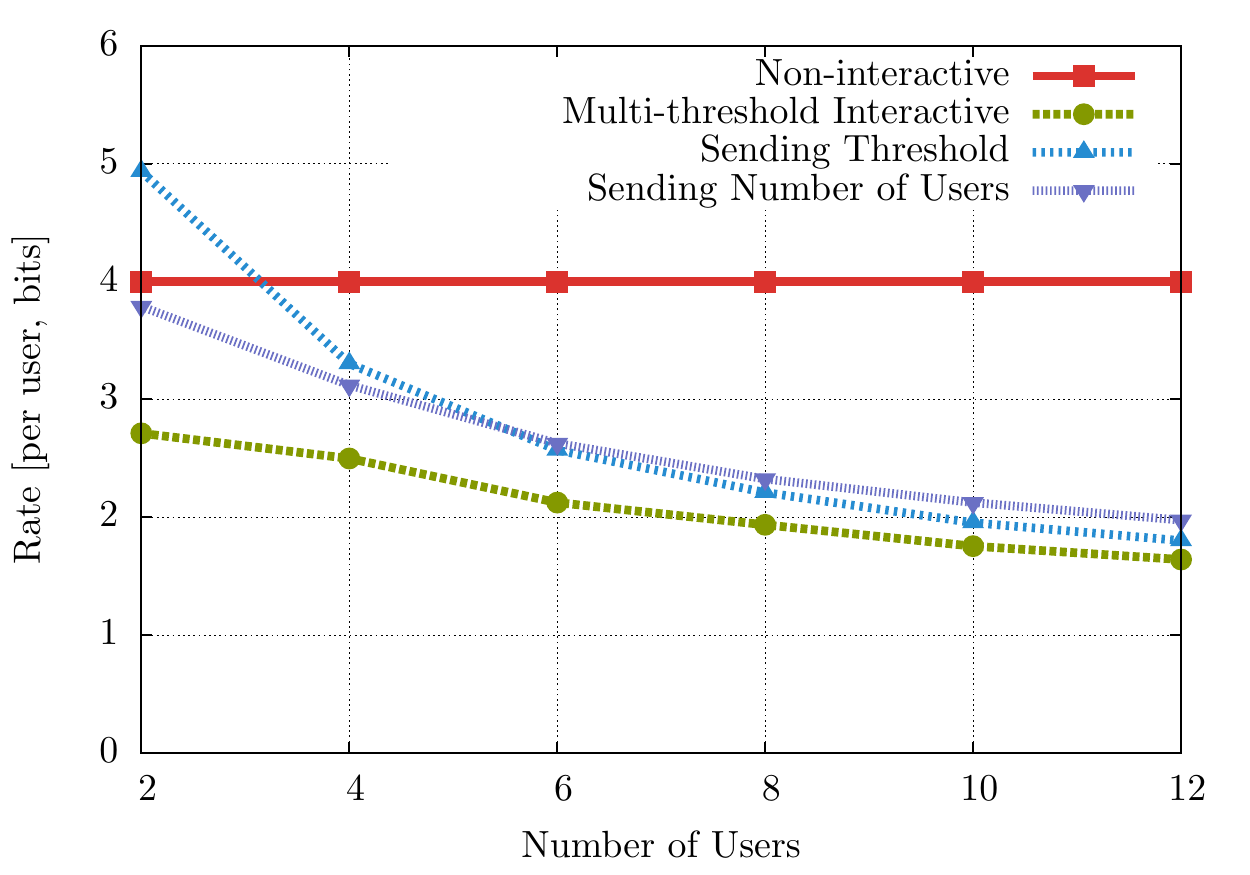}
	\caption{A comparison of sending thresholds and sending number of users with $\left|\mathcal{X}_1\right|=16$.}
	\label{fig:thresholdVSnumber}
\end{figure}
\subsection{Results---Interaction in the $\argmax$ case}
Having identified the policy iteration form of the problem of minimizing the expected aggregate rate exchanged for the \ac{MTIS} scheme for determining the user with the $\argmax$, we now solve the policy iteration for the various methods of communicating the thresholds. We will measure the amount of communication required in each case and compare with the amount of information which must be transmitted without interaction.
As we mentioned before, \eq{dynamic-rate} \eq{E(R)} \eq{dynamic-policy} can be solved by iteration with the boundary condition
\begin{equation} \label{eq:boundary_case2}
R^*(N_t,\mathcal{X}_t)= 0 
\end{equation}
if $N_t=1$ or $\left|\mathcal{X}_t\right|=1$.  Fig. \ref{fig:user8}, \ref{fig:level64_2}, \ref{fig:users8total}, \ref{fig:levels16total}, \ref{fig:thresholdVSnumber}, \ref{fig:max_user8} and \ref{fig:max_level16} present the number of bits communicated under the various schemes when the sources are uniformly distributed. \fig{user8} compares the bits communicated by \ac{MTIS}, with \ac{SW} coded thresholds achieving the conditional entropies \eq{dynamic-rate}, and the non-interactive scheme with $N_1=8$, while \fig{level64_2} performs the same comparison with $\mathcal{X}_1=64$.  From both figures we can see significant rate savings through interaction when calculating the $\argmax$.

As mentioned in previous subsection, we suggested two simple encoding strategies for the base station to broadcast which include Huffman encoding the $\lambda_t^*$ with no conditioning on previous thresholds and Huffman encoding the $N_t$. \fig{thresholdVSnumber} shows the number of bits that must be exchanged when these methods are used. The strategy of sending the threshold outperforms the strategy of sending the number of users in the situation that the initial number $N_1$ is large; while when $N_1$ is small, the latter shows better performance. The minimum between these two schemes requires an amount of communication close to the best scheme, which \ac{SW} encodes the thresholds.
 
 \subsection{Results--$\max$ and $(\arg\max,\max)$ Case}
We can also apply the achievable interaction scheme in the problem that the exact maximum value need to be decided as well as the problem that both the $\max$ and $\argmax$ need to be decided, following the same analysis as \eq{analy1} to \eq{dynamic-policy} with the only difference being the boundary conditions. Instead of \eq{boundary_case2}, we will have the following condition for determining the $\max$ and the pair
\begin{equation}
R^*(N_t,\mathcal{X}_t) = 0 \iff \left|\mathcal{X}_t\right|=1.
\end{equation}

\begin{figure*}
	\centering
	\subfloat[]{\includegraphics[width=234.0pt]{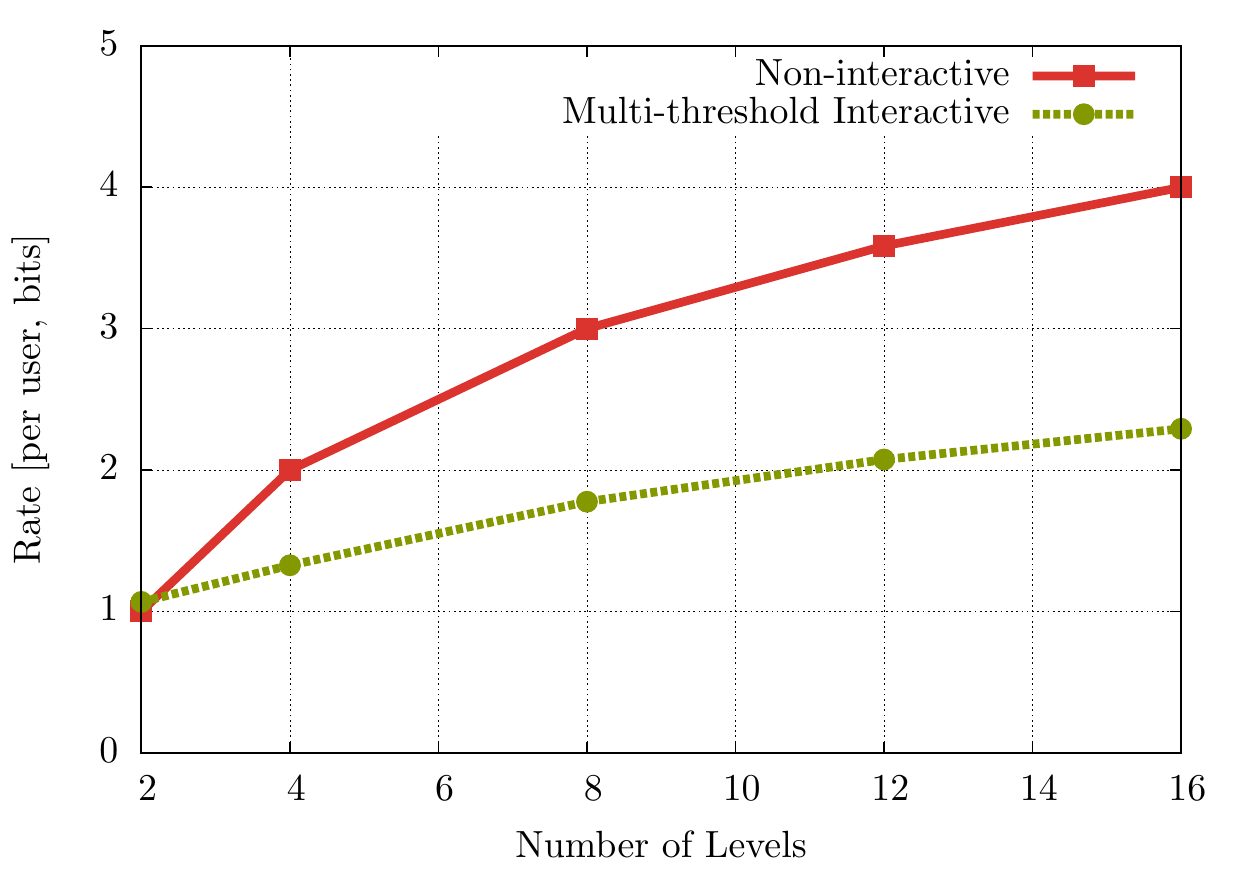}\label{fig:max_user8}}
	\subfloat[]{\includegraphics[width=234.0pt]{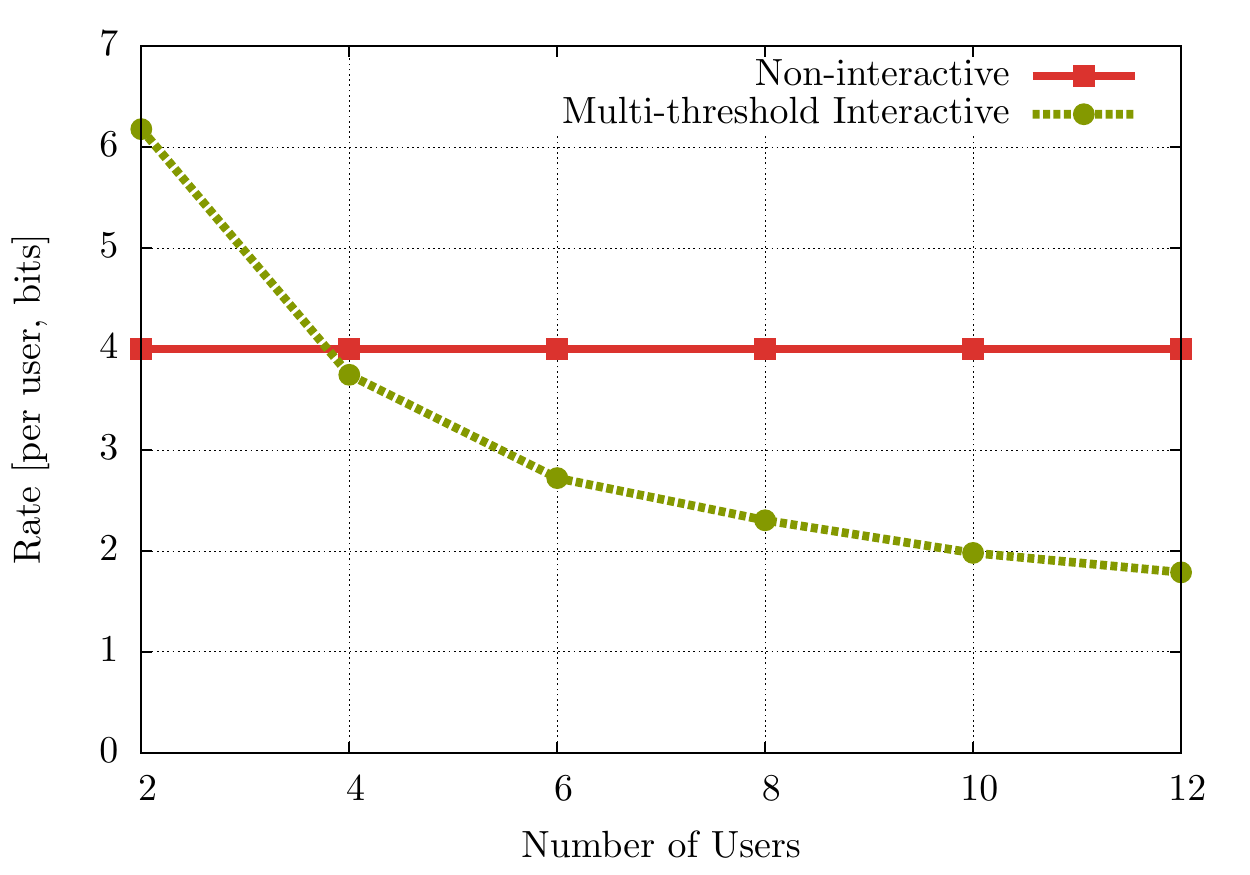}\label{fig:max_level16}}
	\caption[]{\ac{MTIS} Vs. non-interactive for the $\max$ problem with \subref{fig:max_user8} $N_1=8$, and; \subref{fig:max_level16} $\left|\mathcal{X}_1\right| = 16$}
\end{figure*}
For the problem that the CEO wants to learn the $\max$ or the pair $(\max,\argmax)$, \fig{max_user8} compares the bits communicated by \ac{MTIS}, with \ac{SW} coded thresholds achieving the conditional entropies  \eq{dynamic-rate}, and the non-interactive scheme with $N_1=8$, while \fig{max_level16} performs the same comparison with $\mathcal{X}_1=16$.
Note that case 1 and case 3 share the same boundary conditions and hence have the same rates because once the CEO knows the maximum value, it can pick any one of the online users that achieves the maximum. Also note that by \thmref{converse_max}, the one-way fundamental limit of determining the $\max$ is $NH(X)$ because we have selected  $\min \mathcal{X} > 0$. 

\subsection{Scaling Laws}
We have shown for the lossless non-interactive communication, one can have at most $2$ bits saving for the $\argmax$ case, and the per user saving goes to $0$ as the number of users goes to infinity. Now we will see our proposed interactive scheme will exhibit a better scaling law.
\begin{thm}
For the case that two users each observe uniformly distributed independent discrete sources, the aggregate expected rate required to losslessly determine the $\argmax$ by interactive communication satisfies
\begin{equation}
R^* < 6 - 6\left(\frac{1}{2}\right)^{\lceil\log_2 L\rceil} <6
\end{equation}
hence the per-user rate goes to $0$ as $N$ goes to infinity.
\end{thm}
\begin{IEEEproof}
We will derive an upper bound on the amount of information exchanged by \ac{MTIS} by choosing non-optimal thresholds and transmitting $N_t$ instead of the threshold.  The users, instead of computing $\lambda_t^*$ by dynamic programming, will always pick the median of $\mathcal{X}_t$ as the threshold and send a $1$ bit message indicating whether its observation is in $\{\alpha_1^t,\lambda_{t-1}\}$ or $\{\lambda_t,\ldots,\alpha_{L(t)}^{t}\}$. The CEO then also replies a 1 bit message indicating whether or not the two users are in the same region. The communication process stops if the two users are not in the same region, otherwise the problem degenerates to a 2-user $\argmax$ problem with support set shrinking to a half of the original size. Define $R(L)$ as the expected aggregate rate by this interactive scheme with support set $\{\alpha_1,\ldots,\alpha_L\}$ in the 2-user arg-max problem. 
\begin{equation}\label{eq:2user_argmax}
\begin{aligned}
R(L) &\overset{(a)}{=} 2 + 1 + (p_1^t p_2^t)\left(R\left(\lceil L/2\rceil\right)+R\left(\lfloor L/2 \rfloor\right)\right) \\
        &\le 3 + 2p_1^t p_2^t R\left(\lceil L/2 \rceil\right) \\
        &\le 3 + 0.5 \left(R\left(\lceil L/2 \rceil\right)\right) 
\end{aligned}
\end{equation}
where $p_1^t = \mathbb{P}\left(x \in \{\mathcal{X}_1^t,\lambda_{t-1}\}\right)$, $p_2^t = \mathbb{P}\left(x \in \{\lambda_{t-1},\mathcal{X}_{L(t)}^t\}\right)$. Where the $2$ in \eq{2user_argmax} stands for the $2$ bits communicated by the two users in this round, the $1$ stands for the replied bit from the CEO, and the last term stands for the case that both users either reply $1$ or $0$. 
As \eq{boundary_case2} suggests, we have $R(1) = 0$, hence for any $\mathcal{X} = \{\alpha_1,\ldots,\alpha_L\}$ we have
\begin{equation}
\begin{aligned}
R(L) - 6 &\le \frac{1}{2} \left(\left(R\left(\lceil L/2 \rceil\right)\right) - 6\right) \\
	& \le \left(\frac{1}{2}\right)^m (R(1) - 6) \\
	& = -6 \left(\frac{1}{2}\right)^m
\end{aligned}
\end{equation}
where $2^{m-1}\le L \le 2^m$, and therefore
\begin{equation} \label{eq:rate-scheme2}
R^* \le R(L) \le 6 - 6 \left(\frac{1}{2}\right)^{m} < 6.
\end{equation}
\end{IEEEproof}
\begin{thm} \label{thm:interactive_scale_N}
Let $\Delta_A = R_{A}^*- R^*$ be the rate saving of the proposed interactive scheme w.r.t. the lossless non-interactive limit $R_A^*$ in the $\argmax$ problem, the per-user saving $\Delta_A/N$ satisfies
\begin{equation}
\lim_{N \rightarrow \infty} \frac{\Delta_A}{N} \ge H(X) - 1
\end{equation}
\end{thm}
\begin{IEEEproof}
We propose an interactive scheme which will derive an upper bound on the amount of information exchanged by \ac{MTIS} by choosing $\lambda = \max \mathcal{X}$. 
Define $R_{U}(\mathcal{X},N)$ as the expected aggregate rate of this scheme, we know $R_U \ge R^*$, and
\begin{equation} \label{eq:send_max}
\begin{aligned}
R_{U}(\{\alpha_1,\ldots,\alpha_L\},N) &= (1-p_L)^NR_U\left(\left\{\alpha_1,\ldots,\alpha_{L-1}\right\},N\right) + (1-(1-p_L)^N)\cdot 0 + H(X) + N\\
 &\le (1-p_L)^N R_U\left(\left\{\alpha_1,\ldots,\alpha_L\right\},N\right) + H(X) + N \\
 &\le (1-p_L)^N NH(X) + H(X) +N\\
\end{aligned}
\end{equation} 
where the first two terms in \eq{send_max} stand for the expected rate cost for future rounds, $p_L=\mathbb{P}(X=\alpha_L)$, $H(X)$ stands for the bits required to send the threshold $\lambda = \max \mathcal{X}$ and $N$ stands for the bits replied by the $N$ users. Hence by \eq{musers_rate}, \eq{send_max} and the fact that $\lim_{N \rightarrow \infty} (1-p_L)^N = 0$, we have
\begin{equation}
\begin{aligned}
\lim_{N \rightarrow \infty} \frac{\Delta_A}{N} &\ge \lim_{N \rightarrow \infty} \frac{1}{N} \left((N-2)H(X) - \sum_{i=1}^{L-1}p_{i,i+1}\log_2 p_{i,i+1} - p_1\log_2 p_1 - p_L\log_2 p_L - (1-p_L)^N\cdot NH(X) - H(X) -N \right) \\
& = H(X) - 1. \\
\end{aligned}
\end{equation}
\end{IEEEproof}
\subsection{Compare with Other Interactive Schemes}
As an interesting point of comparison, we compare the \ac{MTIS} with another two interactive schemes. Both of the two schemes are given in\cite{InteractiveFunctionComputation} as examples that show interaction can enable rate savings relative to non-interactive schemes  in distributed function computation problems.  In both schemes, it is assumed that when the user sends an message, the CEO knows without cost which user this message is from. Additionally, in the first scheme, referred to as \ac{RIS}, the users transmit sequentially with one user transmitting at a time for reception by the next user. The second scheme, called \ac{NBIS}, has an additional constraint that all communication must occur between the CEO and users and the CEO can only communicate to one user at a time. Here we illustrate the schemes for $3$ users. Pseudocode for the two schemes is provided in Algorithms \ref{algorithm:RIS} and \ref{algorithm:NBIS} respectively.
\begin{figure*}
	\centering
	\subfloat[]{\includegraphics[width=234.0pt]{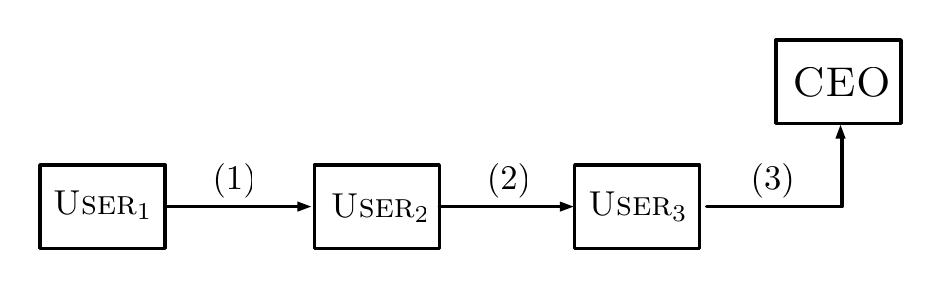}\label{fig:RIS}}
	\subfloat[]{\includegraphics[width=234.0pt]{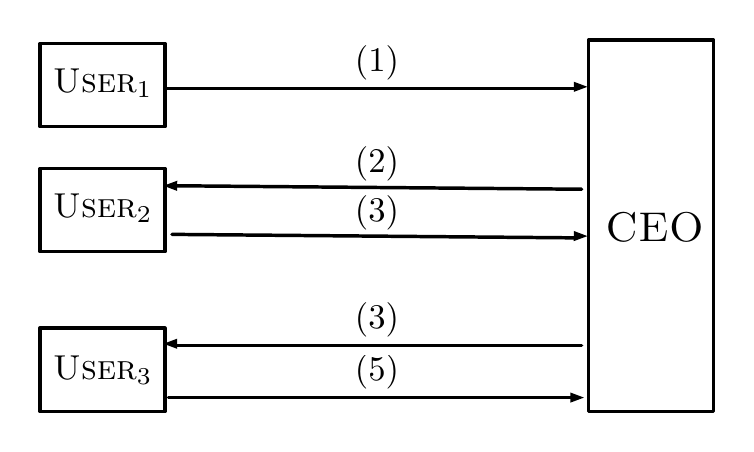}\label{fig:NBIS}}
	\caption[]{The $3$ users case of the CEO extremization problem with \subref{fig:RIS} \ac{RIS}, and; \subref{fig:NBIS} \ac{NBIS}}
\end{figure*}

\vspace{3mm}
\begin{algorithm}[H]
\KwResult{Let the CEO decide the $\argmax$}
initialization:\ number of users $N$, the support set and the CDF of the source random variables $\mathcal{X} = \{\alpha_1,\ldots,\alpha_L\}, F_x(x)$\;
step 1) user $1$ sends its value to user $2$\;
step 2) user $2$ computes $\max \{x_1,x_2\}$ and sends it with its index (the $\argmax$) to user $3$\;
$\cdots$ \;
step L-1) user $N-1$ computes $\max \{x_1,\ldots,x_{N-1}\}$ and sends it with its index to user $N$\;
step L) user $N$ computes $\max \{x_1,\ldots,x_{N}\}$ and sends its index to the CEO\;
\caption{Relay Interaction Scheme\label{algorithm:RIS}}
\end{algorithm}
\vspace{3mm}
\begin{algorithm}[H] 
\KwResult{Let the CEO decide the $\argmax$}
initialization:\ number of users $N$, the support set and the CDF of the source random variables $\mathcal{X}=\{\alpha_1,\ldots,\alpha_L\}, F_x(x)$\;
step 1) user $1$ sends its value to the CEO\;
step 2) CEO forwards user 1's value to user 2\; 
step 3) user $2$ computes $\max \{x_1,x_2\}$ and sends it to the CEO\;
step 4) CEO learns both the $\argmax$ and the $\max$ of the first $2$ users and forwards $\max \{x_1,x_2\}$ to user 3\;
$\cdots$ \;
step 2N-3) user $N-1$ computes $\max \{x_1,\ldots,x_{N-1}\}$ and sends it to the CEO\;
step 2N-2) CEO learns both the $\argmax$ and the $\max$ of the first $N-1$ users and forwards $\max \{x_1,\ldots, x_{N-1}\}$ to user N\;
step 2N-1) user $N$ computes $\max \{x_1,\ldots,x_{N}\}$ and sends it to the CEO\;
\caption{Non-Broadcasting Interaction Scheme\label{algorithm:NBIS}}
\end{algorithm}
\vspace{3mm}
\begin{figure*}
	\centering
	\subfloat[]{\includegraphics[width=234.0pt]{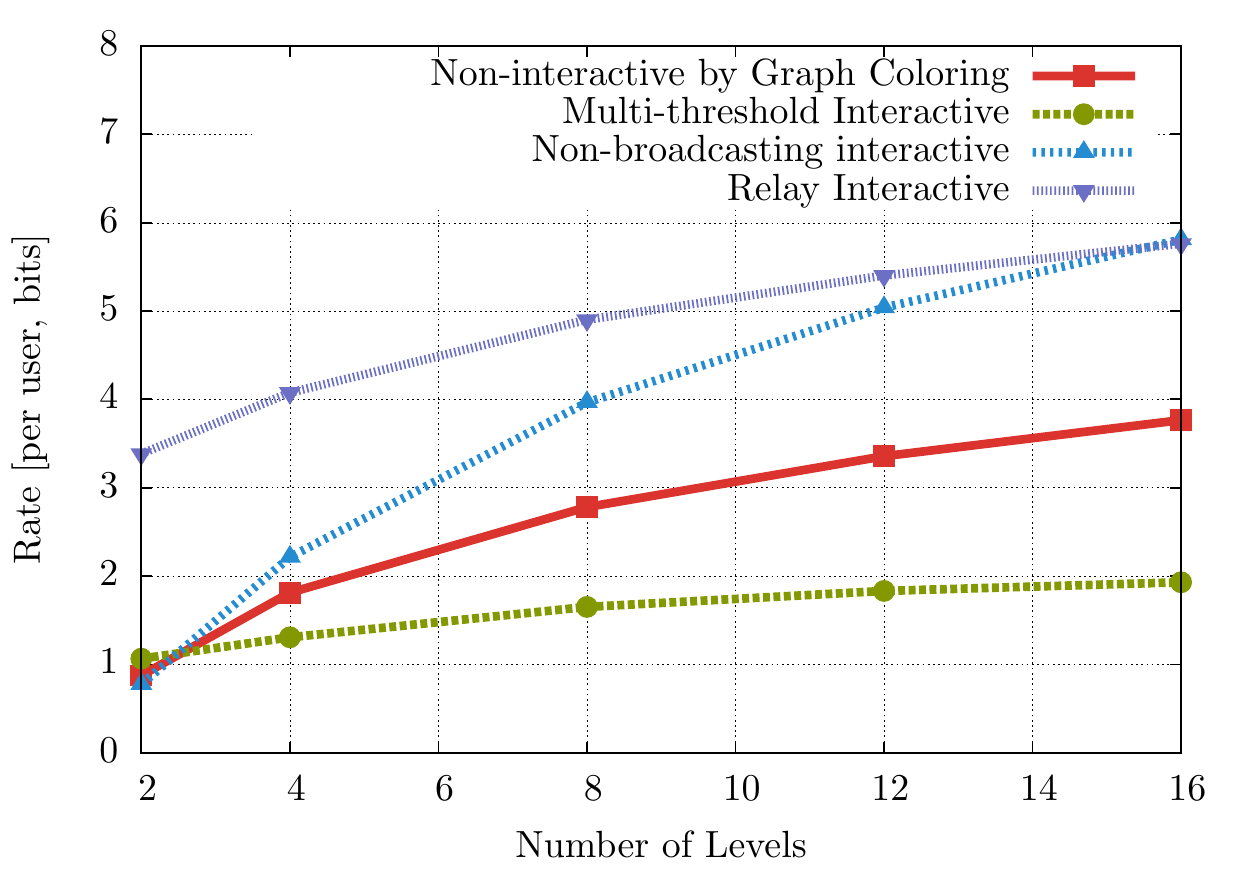}\label{fig:users8total}}
	\subfloat[]{\includegraphics[width=234.0pt]{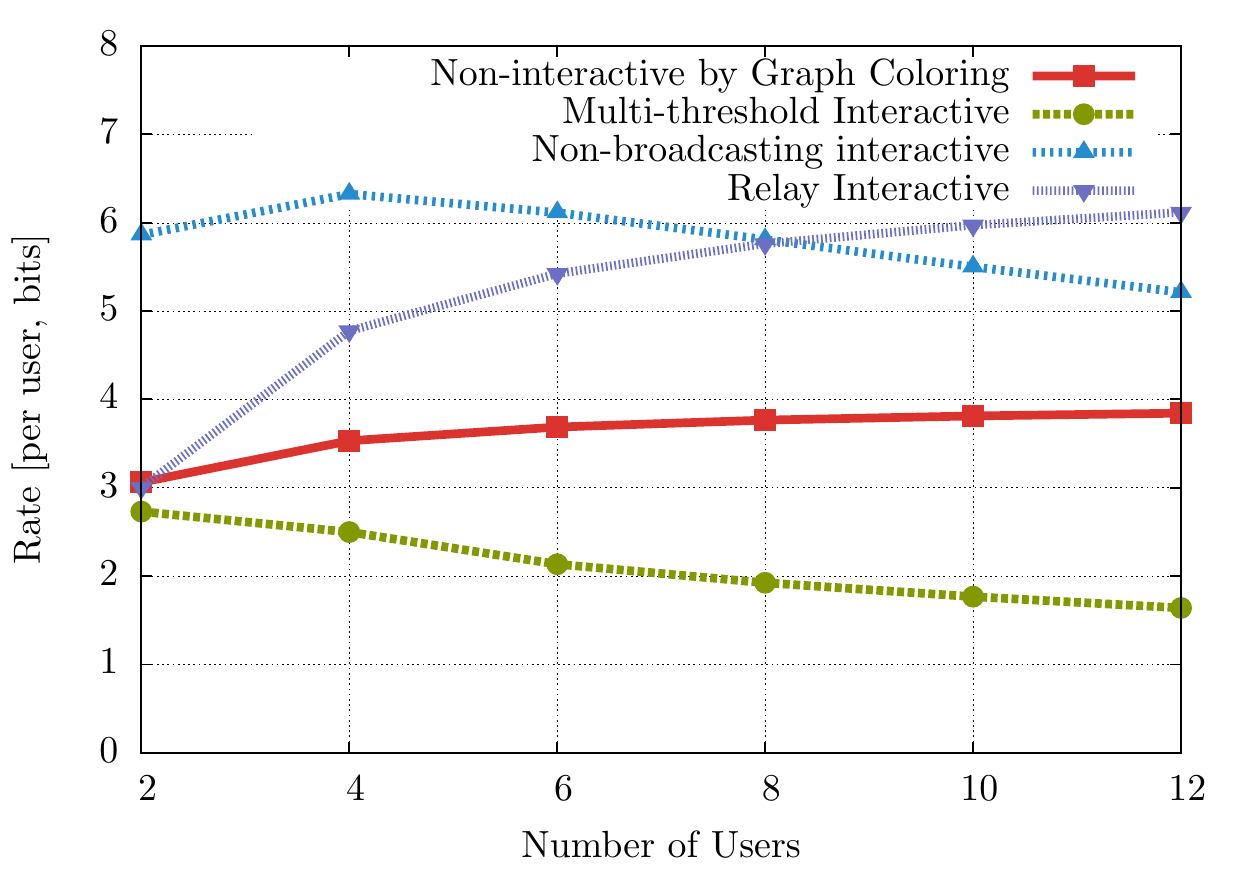}\label{fig:levels16total}}
	\caption[]{A comparison of the three interactive schemes with \subref{fig:users8total} $N_1=8$, and; \subref{fig:levels16total} $\left|\mathcal{X}_1\right| = 16$}
\end{figure*}
In \fig{users8total} and \fig{levels16total}, we see that for uniformly distributed sources, the \ac{NBIS} has better performance than \ac{MTIS} when there is only two users. For most of the cases, the \ac{MTIS} utilizes fewer overhead bits than the other two schemes. 

In summary, we observe from \fig{user8} -- \fig{levels16total} that the \ac{MTIS} provides a substantial saving in sum rate relative to the the non-interactive scheme as well as the \ac{RIS} and the \ac{NBIS} while still obtaining the answer losslessly. In fact, we observe from \thmref{interactive_scale_N} that the per-user rate goes to $1$ as the number of users goes to infinity, which is a very large reduction relative to the minimum necessary communication if non-interaction is required.

\section{Conclusion}
\label{sec:conclusion}

In this paper, we considered resource allocation problems in which a resource controller needs to compute an extremization function (one or both of the functions $\max,\arg\max$) over a series of $N$ remote users.  Designs were developed that minimized the amount of information exchange necessary for this remote function computation.  We first showed that, in most of the cases where the extremization must be computed losslessly, at most two bits can be saved relative to the direct scheme in which the users simply forward their metrics to the controller which computes the function.  In contrast to this lossless case, we observed that substantial rate savings can be achieved if the controller tolerates even a small amount of distortion in computing the function.  In particular, we developed simple quantizers for remote extremization whose rate distortion performance closely matches the optimal rate distortion curve.  Alternatively, if no distortion can be tolerated, we demonstrate that substantial rate savings can still be achieved if the controller and the users are allowed to interactively communicate.  An attractive feature of both the interactive and lossy paradigms for remote extremization is that the rate saving obtained improve with the number of users.  An important direction for future work is to further reduce the rate necessary via lossy interactive computation, by building a hybrid combining the developed lossy and interactive schemes.

\bibliographystyle{IEEEtran}
\bibliography{IEEEabrv,sources}

\begin{thebibliography}{10}
\providecommand{\url}[1]{#1}
\csname url@samestyle\endcsname
\providecommand{\newblock}{\relax}
\providecommand{\bibinfo}[2]{#2}
\providecommand{\BIBentrySTDinterwordspacing}{\spaceskip=0pt\relax}
\providecommand{\BIBentryALTinterwordstretchfactor}{4}
\providecommand{\BIBentryALTinterwordspacing}{\spaceskip=\fontdimen2\font plus
\BIBentryALTinterwordstretchfactor\fontdimen3\font minus
  \fontdimen4\font\relax}
\providecommand{\BIBforeignlanguage}[2]{{%
\expandafter\ifx\csname l@#1\endcsname\relax
\typeout{** WARNING: IEEEtran.bst: No hyphenation pattern has been}%
\typeout{** loaded for the language `#1'. Using the pattern for}%
\typeout{** the default language instead.}%
\else
\language=\csname l@#1\endcsname
\fi
#2}}
\providecommand{\BIBdecl}{\relax}
\BIBdecl

\bibitem{BoyWalWeb2014}
B.~D. Boyle, J.~M. Walsh, and S.~Weber, ``Distributed scalar quantizers for
  subband allocation,'' in \emph{Conf. Information Sciences and Systems
  (CISS)}, March 2014.

\bibitem{RenWal2014}
J.~Ren and J.~M. Walsh, ``Interactive communication for resource allocation,''
  in \emph{Conf. Information Sciences and Systems (CISS)}, March 2014.

\bibitem{VariableRateCC}
S.~Verdu and S.~Shamai, ``Variable-rate channel capacity,'' \emph{{IEEE} Trans.
  Inf. Theory}, vol.~56, no.~6, pp. 2651--2667, Jun. 2010.

\bibitem{LubyLT}
M.~Luby, ``Lt codes,'' in \emph{Proceedings of the 43rd Symposium on
  Foundations of Computer Science}, 2002, pp. 271--280.

\bibitem{RatelessAWGN}
U.~Erez, M.~D. Trott, and G.~W. Wornell, ``Rateless coding for {G}aussian
  channels,'' \emph{{IEEE} Trans. Inf. Theory}, vol.~58, no.~2, pp. 530--547,
  Feb. 2012.

\bibitem{TS36211-8}
\emph{E-UTRA; Physical Channels and Modulation (Release 8)}, 3GPP Technical
  Specification TS36\,211-890, Dec. 2009.

\bibitem{KatWro2000}
D.~Katabi and J.~Wroclawski, ``A framework for scalable global ip-anycast
  ({GIA}),'' \emph{SIGCOMM Comput. Commun. Rev.}, vol.~30, no.~4, pp. 3--15,
  2000.

\bibitem{ProSal1994}
J.~G. Proakis and M.~Salehi, \emph{Communication Systems Engineering}.\hskip
  1em plus 0.5em minus 0.4em\relax Prentice Hall, 1994.

\bibitem{Var1992}
H.~R. Varian, \emph{Microeconomic Analysis}.\hskip 1em plus 0.5em minus
  0.4em\relax W.M. Norton, 1992.

\bibitem{Raw1971}
J.~Rawls, \emph{A Theory of Justics}.\hskip 1em plus 0.5em minus 0.4em\relax
  Harvard University Press, 1971.

\bibitem{Kri2010}
V.~Krishna, \emph{Auction Theory}, 2nd~ed.\hskip 1em plus 0.5em minus
  0.4em\relax Academic Press, 2010.

\bibitem{ButMorSan2014}
I.~Butun, S.~D. Morgera, and R.~Sankar, ``A survey of intrusion detection
  systems in wireless sensor networks,'' \emph{{IEEE} Commun. Surveys Tuts.},
  vol.~16, no.~1, pp. 266--282, 2014.

\bibitem{SW}
D.~Slepian and J.~K. Wolf, ``Noiseless coding of correlated information
  sources,'' \emph{{IEEE} Trans. Inf. Theory}, vol.~19, no.~4, pp. 471--480,
  Jul. 1973.

\bibitem{InteractiveFunctionComputation}
N.~Ma and P.~Ishwar, ``Some results on distributed source coding for
  interactive function computation,'' \emph{{IEEE} Trans. Inf. Theory},
  vol.~57, no.~9, pp. 6180--6195, Sep. 2011.

\bibitem{Witsenhausen}
H.~S. Witsenhausen, ``The zero-error side information problem and chromatic
  numbers,'' \emph{{IEEE} Trans. Inf. Theory}, vol.~22, no.~5, pp. 592--593,
  Sep. 1976.

\bibitem{Orlitsky1}
A.~Orlitsky and J.~R. Roche, ``Coding for computing,'' \emph{{IEEE} Trans. Inf.
  Theory}, vol.~47, no.~3, pp. 903--917, Mar. 2001.

\bibitem{FunctionComputing}
\BIBentryALTinterwordspacing
M.~Sefidgaran and A.~Tchamkerten, ``Distributed function computation over a
  rooted directed tree,'' {submitted to \emph{IEEE Trans. Inf. Theory}}.
  [Online]. Available: \url{{http://arxiv.org/pdf/1312.3631v1.pdf}}
\BIBentrySTDinterwordspacing

\bibitem{FunctionCompression}
V.~Doshi, D.~Shah, M.~Medard, and S.~Jaggi, ``Functional compression through
  graph coloring,'' \emph{{IEEE} Trans. Inf. Theory}, vol.~56, no.~8, pp.
  3901--3917, Aug. 2010.

\bibitem{CEO}
T.~Berger, Z.~Zhang, and H.~Viswanathan, ``The {CEO} problem [multiterminal
  source coding],'' \emph{{IEEE} Trans. Inf. Theory}, vol.~42, no.~3, pp.
  887--902, May 1996.

\bibitem{Sha:Cod}
C.~E. Shannon, ``Coding theorems for a discrete source with a fidelity
  criterion,'' \emph{IRE Nat. Conv. Rec.}, vol.~7, no.~4, pp. 142--163, 1959.

\bibitem{Ber:Rat}
T.~Berger, \emph{Rate Distortion Theory: A Mathematical Basis for Data
  Compression}.\hskip 1em plus 0.5em minus 0.4em\relax Englewood Cliffs, N.J.:
  Prentice-Hall, 1971.

\bibitem{Gal:Inf}
R.~G. Gallager, \emph{Informattion Theory and Reliable Communication}.\hskip
  1em plus 0.5em minus 0.4em\relax New York: Wiley, 1968.

\bibitem{Ber:Mul}
T.~Berger, ``Multiterminal source coding,'' \emph{The Information Theory
  Approach to Communications}, vol.~22, 1977.

\bibitem{Tun:Mul}
S.~Tung, ``Multiterminal source coding,'' Ph.D. dissertation, Cornell
  University, 1978.

\bibitem{Ari:Alg}
S.~Arimoto, ``An algorithm for computing the capacity of arbitrary discrete
  memoryless channels,'' \emph{{IEEE} Trans. Inf. Theory}, vol.~18, no.~1, pp.
  14--20, 1972.

\bibitem{Bla:Com}
R.~E. Blahut, ``Computation of channel capacity and rate-distortion
  functions,'' \emph{{IEEE} Trans. Inf. Theory}, vol.~18, no.~4, pp. 460--473,
  1972.

\bibitem{Csi:Com}
I.~Csisz{\'a}r, ``On the computation of rate-distortion functions (corresp.),''
  \emph{{IEEE} Trans. Inf. Theory}, vol.~20, no.~1, pp. 122--124, 1974.

\bibitem{Bou:Upp}
P.~Boukris, ``An upper bound on the speed of convergence of the {B}lahut
  algorithm for computing rate-distortion functions (corresp.),'' \emph{{IEEE}
  Trans. Inf. Theory}, vol.~19, no.~5, pp. 708--709, 1973.

\bibitem{Ooh:Rat}
Y.~Oohama, ``Rate-distortion theory for {G}aussian multiterminal source coding
  system with several side informations at the decoder,'' \emph{{IEEE} Trans.
  Inf. Theory}, vol.~51, no.~7, pp. 2577--2593, Jul. 2005.

\bibitem{Pra:Rat}
V.~Prabhakaran, D.~Tse, and K.~Ramchandran, ``Rate region of the quadratic
  {G}aussian {CEO} problem,'' in \emph{Proc. Int. Symp. Inform. Theory (ISIT)},
  Jun. 2004, p. 117.

\bibitem{Wag:Imp}
A.~B. Wagner and V.~Anantharam, ``An improved outer bound for multiterminal
  source coding,'' \emph{{IEEE} Trans. Inf. Theory}, vol.~54, no.~5, pp.
  1919--1937, May 2008.

\bibitem{Wyn:Rat}
A.~D. Wyner and J.~Ziv, ``The rate-distortion function for source coding with
  side information at the decoder,'' \emph{{IEEE} Trans. Inf. Theory}, vol.
  IT-22, no.~1, pp. 1--10, Jan. 1976.

\bibitem{Ku:ABA}
G.~Ku, J.~Ren, and J.~M. Walsh, ``Computing the rate distortion region for the
  {CEO} problem with independent sources,'' \emph{{IEEE} Trans. Signal
  Process.}, 2014, submitted.

\bibitem{MisGoyVar2011}
V.~Misra, V.~K. Goyal, and L.~R. Varshney, ``Distributed scalar quantization
  for computing: High-resolution analysis and extensions,'' \emph{{IEEE} Trans.
  Inf. Theory}, vol.~57, no.~8, pp. 5298--5325, 2011.

\bibitem{ZamBer1999}
R.~Zamir and T.~Berger, ``Multiterminal source coding with high resolution,''
  \emph{{IEEE} Trans. Inf. Theory}, vol.~45, no.~1, pp. 106--117, 1999.

\bibitem{WagTavVis2008}
A.~B. Wagner, S.~Tavildar, and P.~Viswanath, ``Rate region of the quadratic
  {G}aussian two-encoder source-coding problem,'' \emph{{IEEE} Trans. Inf.
  Theory}, vol.~54, no.~5, pp. 1938--1961, 2008.

\bibitem{Ser2005}
S.~D. Servetto, ``Achievable rates for multiterminal source coding with scalar
  quantizers,'' in \emph{Conf. Rec. of the 39th Asilomar Conf. Signals, Systems
  and Computers}, 2005, pp. 1762--1766.

\bibitem{InteractiveTheory}
A.~H. Kaspi, ``Two-way source coding with a fidelity criterion,'' \emph{{IEEE}
  Trans. Inf. Theory}, vol.~31, no.~6, pp. 735--740, Nov. 1985.

\bibitem{InteractionWynerZiv}
N.~Ma and P.~Ishwar, ``Interaction strictly improves the wyner-ziv
  rate-distortion function,'' in \emph{International Symposium on Information
  Theory}, 2010, pp. 61--65.

\bibitem{SMUD}
D.~Gesbert and M.~S. Alouini, ``How much feedback is multi-user diversity
  really worth?'' in \emph{Proc. IEEE Int. Conf. Commun.}, 2004, pp. 234--238.

\bibitem{MultiThresholds}
V.~Hassel, M.~S. Alouini, D.~Gesbert, and G.~E. Oien, ``Exploiting multiuser
  diversity using multiple feedback thresholds,'' in \emph{Proc. IEEE Veh.
  Technol. Conf.}, 2005, pp. 1302--1306.

\bibitem{FunctionCompressionSideInfo}
V.~Doshi, D.~Shah, M.~Medard, and S.~Jaggi, ``Graph coloring and conditional
  graph entropy,'' in \emph{40th Asilomar Conf. on Signals, Systems, and
  Computers}, 2006, pp. 2137--2141.

\bibitem{InteractionDP}
J.~Ren and J.~M. Walsh, ``Interactive communication for resource allocation,''
  Drexel University, Dept. of ECE, Tech. Rep., Dec. 2013.

\bibitem{Say2012}
K.~Sayood, \emph{Introduction to Data Compression}, 4th~ed.\hskip 1em plus
  0.5em minus 0.4em\relax Elsevier, 2012.

\bibitem{FarMod1984}
N.~Farvardin and J.~W. Modestino, ``Optimum quantizer performance for a class
  of non-{G}aussian memoryless sources,'' \emph{{IEEE} Trans. Inf. Theory},
  vol.~30, no.~3, pp. 485--497, 1984.

\bibitem{DynamicProgramming}
D.~P. Bertsekas, \emph{Dynamic Programming and Optimal Control}.\hskip 1em plus
  0.5em minus 0.4em\relax Athena Scientific, 2005.

\end{thebibliography}
\section*{Disclaimer}
The views and conclusions contained herein are those of the authors and should not be interpreted as necessarily representing the official policies or endorsements, either expressed or implied, of the Air Force Research Laboratory or the U.S.\ Government.

\appendix
\subsection{Proofs}

\subsubsection{Proof of \thmref{optimal_function}}
\label{sec:proof_optimal_function}
\begin{IEEEproof}
First we prove 1. In user $1$'s characteristic graph $G_1(V_1,E_1)$ where $V_1=\mathcal{X}=\{\alpha_1,\ldots,\alpha_L\}$, by \lemref{diffColors}, we must have $\{\alpha_i,\alpha_j\} \in G_1$ if  $|i-j| \ge 2$. Now we consider the pair of vertices $\{\alpha_i,\alpha_{i+1}\}$ for any $i \in \{1,\ldots,L-1\}$.  When $\mod (n,2) \ne \mod (i,2)$, there exists a sequence 
$
\boldsymbol{x}_{\setminus\{1\}} = (\alpha_i,\ldots,\alpha_i)$ satisfying
$
f^*_n(\alpha_i,\boldsymbol{x}_{\setminus\{1\}}) = f^*_{n-1} (\boldsymbol{x}_{\setminus\{1\}})+1
$
by \eq{NDisAmb2} , and
$
f^*_n(\alpha_{i+1},\boldsymbol{x}_{\setminus\{1\}}) = 1
$ by \eq{NDisAmb1}. Note that $f^*_n (\boldsymbol{x}) > 0$ for all $n$. This implies $f^*_n(\alpha_i,\boldsymbol{x}_{\setminus\{1\}})>f^*_n(\alpha_{i+1},\boldsymbol{x}_{\setminus\{1\}})$, hence $\{\alpha_i,\alpha_{i+1}\} \in G_1$. Next we will prove
$
\{\alpha_i,\alpha_{i+1}\} \not \in G_1(f^*_n) 
$ if $\mod (n,2) = \mod (i,2)$. Since 
\(
\mathcal{X}^{n-1} = \bigr\{\boldsymbol{x}_{\setminus\{1\}}\bigr | \max \{\boldsymbol{x}_{\setminus\{1\}}\} < \alpha_{i+1}\bigr\} \bigcup \bigr\{\boldsymbol{x}_{\setminus\{1\}}\bigr | \max \{\boldsymbol{x}_{\setminus\{1\}}\} = \alpha_{i+1}\bigr\} \bigcup \bigr\{\boldsymbol{x}_{\setminus\{1\}}\bigr | \max \{\boldsymbol{x}_{\setminus\{1\}}\} > \alpha_{i+1}\bigr\}
\), 
it suffices to show that for any given $\boldsymbol{x}_{\setminus\{1\}}$ in these three sets, the function will not differ when $\mod (n,2) = \mod (i,2)$. 
For 
$\boldsymbol{x}_{\setminus\{1\}} \in \bigr\{\boldsymbol{x}_{\setminus\{1\}}\bigr | \max \{\boldsymbol{x}_{\setminus\{1\}}\} < \alpha_{i+1}\bigr\} $,
we observe that $f^*_n(\alpha_{i+1},\boldsymbol{x}_{\setminus\{1\}})=f^*_n(\alpha_i,\boldsymbol{x}_{\setminus\{1\}})=1$ 
by \eq{NDisAmb1} and \eq{NDisAmb2}.
For $\boldsymbol{x}_{\setminus\{1\}} \in \bigr\{\boldsymbol{x}_{\setminus\{1\}}\bigr | \max \{\boldsymbol{x}_{\setminus\{1\}}\} = \alpha_{i+1}\bigr\}$,
we observe that $ f^*_n(\alpha_{i+1},\boldsymbol{x}_{\setminus\{1\}})=f^*_n(\alpha_i,\boldsymbol{x}_{\setminus\{1\}}) = f^*_{n-1} (\boldsymbol{x}_{\setminus\{1\}})+1$ by \eq{NDisAmb2} and \eq{NDisAmb3}. Finally for $\boldsymbol{x}_{\setminus\{1\}} \in \bigr\{\boldsymbol{x}_{\setminus\{1\}}\bigr | \max \{\boldsymbol{x}_{\setminus\{1\}}\} > \alpha_{i+1}\bigr\}$, we also observe that $f^*_n(\alpha_i,\boldsymbol{x}_{\setminus\{1\}}) = f^*_n(\alpha_{i+1},\boldsymbol{x}_{\setminus\{1\}})= f^*_{n-1} (\boldsymbol{x}_{\setminus\{1\}})+1$ by \eq{NDisAmb3}.
Therefore in user $1$'s characteristic graph $G_1$, we have
\begin{equation}
\{\alpha_i,\alpha_j\}\not \in G_1 \ \iff\ \mod(n,2)=\mod(i,2)\ \&\ j=i+1.
\end{equation}
In user $2$'s characteristic graph $G_2(V_2,E_2)$ where $V_2=\mathcal{X}=\{\alpha_1,\ldots,\alpha_L\}$, similarly by \lemref{diffColors}, we must have $\{\alpha_i,\alpha_j\} \in G_2$ if  $|i-j| \ge 2$. Now we consider the pair of vertices $\{\alpha_i,\alpha_{i+1}\}$ for any $i \in \{1,\ldots,L-1\}$.  When $\mod (n,2) = \mod (i,2)$, there exists a sequence 
$
(\alpha_i,\ldots,\alpha_i)
$ 
satisfying
$
f^*_n(\alpha_i,\ldots,\alpha_i) = 1
$
by \eq{NDisAmb2} , and another sequence
$
(\alpha_i,\alpha_{i+1},\alpha_i,\ldots,\alpha_i)
$
satisfying
$
f^*_n(\alpha_i,\alpha_{i+1},\alpha_i,\ldots,\alpha_i) = 2
$ by \eq{NDisAmb3}. This implies $\{\alpha_i,\alpha_{i+1}\} \in G_2$ if $\mod (n,2) = \mod (i,2)$. Next we will prove
$
\{\alpha_i,\alpha_{i+1}\} \not \in G_2(f^*_n) 
$ if $\mod (n,2) \ne \mod (i,2)$. Since 
\(
\mathcal{X}^{n-2} = \bigr\{\boldsymbol{x}_{\setminus\{1,2\}}\bigr | \max \{\boldsymbol{x}_{\setminus\{1,2\}}\} < \alpha_{i+1}\bigr\} \bigcup \bigr\{\boldsymbol{x}_{\setminus\{1,2\}}\bigr | \max \{\boldsymbol{x}_{\setminus\{1,2\}}\} = \alpha_{i+1}\bigr\} \bigcup \bigr\{\boldsymbol{x}_{\setminus\{1,2\}}\bigr | \max \{\boldsymbol{x}_{\setminus\{1,2\}}\} > \alpha_{i+1}\bigr\}
\), 
it suffices to show that for any given $x_1 \in \mathcal{X}$ and $\boldsymbol{x}_{\setminus\{1,2\}}$ in these three sets, the function will not differ when $\mod (n,2) \ne \mod (i,2)$.
For
$x_1 < \max \{\boldsymbol{x}_{\setminus\{1,2\}}\}$ and $\boldsymbol{x}_{\setminus\{1,2\}} \in \bigr\{\boldsymbol{x}_{\setminus\{1,2\}}\bigr | \max \{\boldsymbol{x}_{\setminus\{1,2\}}\} < \alpha_{i+1}\bigr\}$, we observe that 
\begin{equation}
f_n^*(x_1,\alpha_i,\boldsymbol{x}_{\setminus\{1,2\}}) \overset{(a.1)}{=}  f_{n-1}^*(\alpha_i,\boldsymbol{x}_{\setminus\{1,2\}})+1 \overset{(a.2)}{=} 2
\end{equation}
and
\begin{equation}
f_n^*(x_1,\alpha_{i+1},\boldsymbol{x}_{\setminus\{1,2\}}) \overset{(a.3)}{=} f_{n-1}^*(\alpha_{i+1},\boldsymbol{x}_{\setminus\{1,2\}})+1 \overset{(a.4)}{=} 2
\end{equation} 
where (a.1) and (a.3) hold by \eq{NDisAmb3}, (a.2) hold by \eq{NDisAmb1} if $\max \{\boldsymbol{x}_{\setminus\{1,2\}}\}<\alpha_i$, and by \eq{NDisAmb2} and the fact that $\mod (n,2) \ne \mod (i,2)$ implies $\mod (n-1,2) = \mod (i,2)$ if $\max \{\boldsymbol{x}_{\setminus\{1,2\}}\}=\alpha_i$, and (a.4) hold by \eq{NDisAmb1}.
For
$x_1 < \max \{\boldsymbol{x}_{\setminus\{1,2\}}\}$ and $\boldsymbol{x}_{\setminus\{1,2\}} \in \bigr\{\boldsymbol{x}_{\setminus\{1,2\}}\bigr | \max \{\boldsymbol{x}_{\setminus\{1,2\}}\} = \alpha_{i+1}\bigr\}$, we observe that 
\begin{equation}
f_n^*(x_1,\alpha_i,\boldsymbol{x}_{\setminus\{1,2\}}) \overset{(b.1)}{=}  f_{n-1}^*(\alpha_i,\boldsymbol{x}_{\setminus\{1,2\}})+1 \overset{(b.2)}{=}  f_{n-2}^*(\boldsymbol{x}_{\setminus\{1,2\}})+2
\end{equation}
and
\begin{equation}
f_n^*(x_1,\alpha_{i+1},\boldsymbol{x}_{\setminus\{1,2\}}) \overset{(b.3)}{=} f_{n-1}^*(\alpha_{i+1},\boldsymbol{x}_{\setminus\{1,2\}})+1 \overset{(b.4)}{=} f_{n-2}^*(\boldsymbol{x}_{\setminus\{1,2\}})+2
\end{equation} 
where (b.1) and (b.3) hold by by \eq{NDisAmb3}, (b.2) hold by \eq{NDisAmb3}, and (b.4) hold by \eq{NDisAmb2} and the fact that $\mod (n,2) \ne \mod (i,2)$ implies $\mod (n-1,2) \ne \mod (i+1,2)$.
For
$x_1 < \max \{\boldsymbol{x}_{\setminus\{1,2\}}\}$ and $\boldsymbol{x}_{\setminus\{1,2\}} \in \bigr\{\boldsymbol{x}_{\setminus\{1,2\}}\bigr | \max \{\boldsymbol{x}_{\setminus\{1,2\}}\} > \alpha_{i+1}\bigr\}$, we observe that 
\begin{equation}
f_n^*(x_1,\alpha_i,\boldsymbol{x}_{\setminus\{1,2\}}) \overset{(c.1)}{=}  f_{n-1}^*(\alpha_i,\boldsymbol{x}_{\setminus\{1,2\}})+1 \overset{(c.2)}{=}  f_{n-2}^*(\boldsymbol{x}_{\setminus\{1,2\}})+2
\end{equation}
and
\begin{equation}
f_n^*(x_1,\alpha_{i+1},\boldsymbol{x}_{\setminus\{1,2\}}) \overset{(c.3)}{=} f_{n-1}^*(\alpha_{i+1},\boldsymbol{x}_{\setminus\{1,2\}})+1 \overset{(c.4)}{=} f_{n-2}^*(\boldsymbol{x}_{\setminus\{1,2\}})+2
\end{equation} 
where (c.1) (c.2) (c.3) (c.4) all hold by \eq{NDisAmb3}.
For
$x_1 = \max \{\boldsymbol{x}_{\setminus\{1,2\}}\}$ and $\boldsymbol{x}_{\setminus\{1,2\}} \in \bigr\{\boldsymbol{x}_{\setminus\{1,2\}}\bigr | \max \{\boldsymbol{x}_{\setminus\{1,2\}}\} < \alpha_{i+1}\bigr\}$, we observe that 
\begin{equation}
f_n^*(x_1,\alpha_i,\boldsymbol{x}_{\setminus\{1,2\}}) \overset{(d.1)}{=}  f_{n-1}^*(\alpha_i,\boldsymbol{x}_{\setminus\{1,2\}})+1 \overset{(d.2)}{=}  2
\end{equation}
and
\begin{equation}
f_n^*(x_1,\alpha_{i+1},\boldsymbol{x}_{\setminus\{1,2\}}) \overset{(d.3)}{=} f_{n-1}^*(\alpha_{i+1},\boldsymbol{x}_{\setminus\{1,2\}})+1 \overset{(d.4)}{=} 2
\end{equation}
where (d.1) holds by \eq{NDisAmb2} if $x_1=\alpha_i$ and by \eq{NDisAmb3} if $x_1 < \alpha_i$, (d.2) holds by \eq{NDisAmb2} if $\max \{\boldsymbol{x}_{\setminus\{1,2\}}\} = \alpha_i$ and by \eq{NDisAmb1} if $\max \{\boldsymbol{x}_{\setminus\{1,2\}}\} < \alpha_i$, (d.3) holds by \eq{NDisAmb3}, and (d.4) holds by \eq{NDisAmb1}.  
For
$x_1 = \max \{\boldsymbol{x}_{\setminus\{1,2\}}\}$ and $\boldsymbol{x}_{\setminus\{1,2\}} \in \bigr\{\boldsymbol{x}_{\setminus\{1,2\}}\bigr | \max \{\boldsymbol{x}_{\setminus\{1,2\}}\} = \alpha_{i+1}\bigr\}$, we observe that 
\begin{equation}
f_n^*(x_1,\alpha_i,\boldsymbol{x}_{\setminus\{1,2\}}) \overset{(e.1)}{=} 1
\end{equation}
and
\begin{equation}
f_n^*(x_1,\alpha_{i+1},\boldsymbol{x}_{\setminus\{1,2\}}) \overset{(e.2)}{=} 1
\end{equation} 
where (e.1) (e.2) both hold by by \eq{NDisAmb2} and the fact that $\mod (n,2) \ne \mod (i,2)$ implies $\mod (n,2) = \mod (i+1,2)$.
For
$x_1 = \max \{\boldsymbol{x}_{\setminus\{1,2\}}\}$ and $\boldsymbol{x}_{\setminus\{1,2\}} \in \bigr\{\boldsymbol{x}_{\setminus\{1,2\}}\bigr | \max \{\boldsymbol{x}_{\setminus\{1,2\}}\} > \alpha_{i+1}\bigr\}$, which means $x_1 = \max \{\boldsymbol{x}\}$, we observe that if $x_1= \alpha_j$ where $\mod (j,2) = \mod(n,2)$, then
\begin{equation}
f_n^*(x_1,\alpha_i,\boldsymbol{x}_{\setminus\{1,2\}}) \overset{(f.1)}{=} 1
\end{equation}
and
\begin{equation}
f_n^*(x_1,\alpha_{i+1},\boldsymbol{x}_{\setminus\{1,2\}}) \overset{(f.2)}{=} 1
\end{equation} 
where (f.1) (f.2) both hold by \eq{NDisAmb2}.
If $x_1= \alpha_j$ where $\mod (j,2) \ne \mod(n,2)$, then
\begin{equation}
f_n^*(x_1,\alpha_i,\boldsymbol{x}_{\setminus\{1,2\}}) \overset{(f.3)}{=}  f_{n-1}^*(\alpha_i,\boldsymbol{x}_{\setminus\{1,2\}})+1 \overset{(f.4)}{=}  f_{n-2}^*(\boldsymbol{x}_{\setminus\{1,2\}})+2
\end{equation}
and
\begin{equation}
f_n^*(x_1,\alpha_{i+1},\boldsymbol{x}_{\setminus\{1,2\}}) \overset{(f.5)}{=} f_{n-1}^*(\alpha_{i+1},\boldsymbol{x}_{\setminus\{1,2\}})+1 \overset{(f.6)}{=} f_{n-2}^*(\boldsymbol{x}_{\setminus\{1,2\}})+2
\end{equation} 
where (f.3) (f.4) both hold by \eq{NDisAmb2}, and (f.5) (f.6) both hold by \eq{NDisAmb3}.
For
$x_1 > \max \{\boldsymbol{x}_{\setminus\{1,2\}}\}$ and $\boldsymbol{x}_{\setminus\{1,2\}} \in \bigr\{\boldsymbol{x}_{\setminus\{1,2\}}\bigr | \max \{\boldsymbol{x}_{\setminus\{1,2\}}\} < \alpha_{i+1}\bigr\}$, we observe that if $x_1 \le \alpha_i$, then
\begin{equation}
f_n^*(x_1,\alpha_i,\boldsymbol{x}_{\setminus\{1,2\}}) \overset{(g.1)}{=}  2
\end{equation}
and
\begin{equation}
f_n^*(x_1,\alpha_{i+1},\boldsymbol{x}_{\setminus\{1,2\}}) \overset{(g.2)}{=} 2
\end{equation} 
where (g.1) holds by \eq{NDisAmb2} if $x_1 = \alpha_i$, and by \eq{NDisAmb3} if $x_1 < \alpha_i$, and (g.2) holds by \eq{NDisAmb3}.
If $x_1 > \alpha_i$, then
\begin{equation}
f_n^*(x_1,\alpha_i,\boldsymbol{x}_{\setminus\{1,2\}}) \overset{(g.3)}{=}  1
\end{equation}
and
\begin{equation}
f_n^*(x_1,\alpha_{i+1},\boldsymbol{x}_{\setminus\{1,2\}}) \overset{(g.4)}{=} 1
\end{equation} 
where (g.3) holds by \eq{NDisAmb1}, and (g.4) holds by \eq{NDisAmb2} and the fact that $\mod (n,2) \ne \mod (i,2)$ implies $\mod (n,2) = \mod (i+1,2)$ if $x_1 =\alpha_{i+1}$, and by \eq{NDisAmb1} if $x_1>\alpha_{i+1}$.
For
$x_1 > \max \{\boldsymbol{x}_{\setminus\{1,2\}}\}$ and $\boldsymbol{x}_{\setminus\{1,2\}} \in \bigr\{\boldsymbol{x}_{\setminus\{1,2\}}\bigr | \max \{\boldsymbol{x}_{\setminus\{1,2\}}\} \ge \alpha_{i+1}\bigr\}$, we observe that 
\begin{equation}
f_n^*(x_1,\alpha_i,\boldsymbol{x}_{\setminus\{1,2\}}) \overset{(h.1)}{=} 1
\end{equation}
and
\begin{equation}
f_n^*(x_1,\alpha_{i+1},\boldsymbol{x}_{\setminus\{1,2\}}) \overset{(h.2)}{=} 1
\end{equation} 
where (h.1) (h.2) both hold by \eq{NDisAmb1}.
Therefore in user $2$'s characteristic graph $G_1$, we have
\begin{equation}
\{\alpha_i,\alpha_j\}\not \in G_2 \ \iff\ \mod(n,2)\ne \mod(i,2)\ \&\ j=i+1.
\end{equation}
By \lemref{2} and user $1$ and $2$'s characteristic graphs, we know user $i$'s characteristic graph will be complete for all $i \in [N]\setminus \{1,2\}$.
Therefore if $N$ is odd and $L$ is odd, the set of all maximal independent sets for each of the users will be
\begin{equation} \label{eq:max_ind_L_odd_N_odd}
\begin{aligned}
& \Gamma(1) = \{\{\alpha_1,\alpha_2\},\{\alpha_3,\alpha_4\},\{\alpha_5,\alpha_6\},\ldots,\{\alpha_{L-2},\alpha_{L-1}\},\{\alpha_L\}\}\\
& \Gamma(2) = \{\{\alpha_1\},\{\alpha_2,\alpha_3\},\{\alpha_4,\alpha_5\},\ldots,\{L-1,L\}\} \\
& \Gamma(i) = \{\{\alpha_1\},\{\alpha_2\},\ldots,\{\alpha_L\} \}\ \forall i \in [N], i \not\in \{1,2\}. 
\end{aligned}
\end{equation}
If $N$ is odd and $L$ is even, the set of all maximal independent sets for each of the users will be
\begin{equation} \label{eq:max_ind_L_even_N_odd}
\begin{aligned}
& \Gamma(1) = \{\{\alpha_1,\alpha_2\},\{\alpha_3,\alpha_4\},\{\alpha_5,\alpha_6\},\ldots,\{\alpha_{L-1},\alpha_L\}\}\\
& \Gamma(2) = \{\{\alpha_1\},\{\alpha_2,\alpha_3\},\{\alpha_4,\alpha_5\},\ldots,\{L-2,L-1\},\{L\}\} \\
& \Gamma(i) = \{\{\alpha_1\},\{\alpha_2\},\ldots,\{\alpha_L\} \}\ \forall i \in [N], i \not\in \{1,2\}.
\end{aligned}
\end{equation}
If $N$ is even and $L$ is odd, the set of all maximal independent sets for each of the users will be
\begin{equation} \label{eq:max_ind_L_odd_N_even}
\begin{aligned}
& \Gamma(1) = \{\{\alpha_1\},\{\alpha_2,\alpha_3\},\{\alpha_4,\alpha_5\},\ldots,\{\alpha_{L-1},\alpha_L\}\}\\
& \Gamma(2) = \{\{\alpha_1,\alpha_2\},\{\alpha_3,\alpha_4\},\ldots,\{L-2,L-1\},\{L\}\} \\
& \Gamma(i) = \{\{\alpha_1\},\{\alpha_2\},\ldots,\{\alpha_L\} \}\ \forall i \in [N], i \not\in \{1,2\}.
\end{aligned}
\end{equation}
If $N$ is even and $L$ is even, the set of all maximal independent sets for each of the users will be
\begin{equation} \label{eq:max_ind_L_even_N_even}
\begin{aligned}
& \Gamma(1) = \{\{\alpha_1\},\{\alpha_2,\alpha_3\},\{\alpha_4,\alpha_5\},\ldots,\{\alpha_{L-2},\alpha_{L-1}\},\{L\}\}\\
& \Gamma(2) = \{\{\alpha_1,\alpha_2\},\{\alpha_3,\alpha_4\},\ldots,\{L-1,L\}\} \\
& \Gamma(i) = \{\{\alpha_1\},\{\alpha_2\},\ldots,\{\alpha_L\} \}\ \forall i \in [N], i \not\in \{1,2\}.
\end{aligned}
\end{equation}
Note that in each $\Gamma(n), n\in [N]$, no vertex belongs to two maximal independent sets. Also note that $\{\alpha_i,\alpha_{i+1}\}$ appears exactly once in $\{\Gamma(n)|n \in [N]\}$ for all $\alpha_i \in \mathcal{X}$. To achieve the minimum sum-rate, the optimal coloring method would be assigning a color for each of the independent sets (see \fig{OptcharGraph}). For the case that both $L$ and $N$ are odd, we have
\begin{equation}
\begin{aligned}
R_A(f^*_n) & = \sum_{i=1}^{n} \min_{c_i \in G_i(f^*_n)}  H(c_i(X_i)) \\
& = (n-2)H(X) + \min_{c_{1} \in \mathcal{C}(G_1(f^*_n))} H(c_{1}(X_1)+\min_{c_{2} \in \mathcal{C}(G_2(f^*_n))} H(c_{2}(X_2))\\
& = -(n-2) \left(\sum_{i=1}^{L} p_i \log_2 p_i\right) - \left(\sum_{i=1}^{\frac{L-1}{2}}p_{2i-1,2i}\log_2 p_{2i-1,2i}\right) - p_{L}\log_2 p_{L} -p_{1}\log_2 p_1 - \left(\sum_{i=1}^{\frac{L-1}{2}}p_{2i,2i+1} \log_2 p_{2i,2i+1}\right) \\
& = - (n-2) \left(\sum_{i=1}^{L} p_i\log_2 p_i\right) - \left(\sum_{i=1}^{L-1}p_{i,i+1}\log_2 p_{i,i+1}\right) - p_1\log_2 p_1 - p_L\log_2 p_L.\\
\end{aligned}
\end{equation}
For the other cases, we will get the exact same expression of the sum-rate although there exists a minor variation on the argument. 

Now we will show that for any function $f_n \in \mathcal{F}_{A,n}, n \in [N]$ the sum-rate under $f_n$ will be no lower than \eq{musers_rate} and $f^*_{n} \in \mathcal{F}_{A,n}^*$.
By \lemref{diffColors} we know that no three vertices can be assigned the same color and hence only the neighbor pair can share the color. By applying \lemref{2} N-1 times, we know that if a neighbor pair $\{\alpha_i,\alpha_{i+1}\}$ are given the same color in user 1's characteristic graph then they have to have distinct colors in all other users' graph. Therefore for all candidate $\argmax$ functions, we can have at most $L-1$ different consecutive pairs $\{\alpha_1,\alpha_2\},\cdots,\{\alpha_{L-1},\alpha_L\}$ that share the color, all other vertices have to have their own colors, and by assigning distinct colors to each of the $L-1$ node paris and all other single nodes and encoding the colors by \ac{SW} coding, \eq{musers_rate} is achieved. 
\end{IEEEproof}

\subsubsection{Proof of \thmref{scale_users_argmax}}
\label{sec:proof_scale_users_argmax}
\begin{IEEEproof}
By \thmref{optimal_function} we have
\begin{equation}
\begin{aligned}
\Delta_A &= NH(X) - (N-2)H(X) + \sum_{i=1}^{L-1} \left(p_{i,i+1}\log_2 p_{i,i+1}\right) + p_1\log_2 p_1 + p_L\log_2 p_L\\
	&=-2 \sum_{i=1}^{L} \left(p_i \log_2 p_i\right) + \sum_{i=1}^{L-1} \left(p_i \log_2 p_{i,i+1}\right) + \sum_{i=1}^{L-1} \left(p_{i+1} \log_2 p_{i,i+1}\right)+ p_1 \log_2 p_1 + p_L \log_2 p_L \\
	&= -\sum_{i=1}^{L-1} \left(p_i \log_2 p_i\right)-\sum_{i=2}^{L} \left(p_i \log_2 p_i\right)+\sum_{i=1}^{L-1} \left(p_i \log_2 p_{i,i+1}\right)+\sum_{i=1}^{L-1} \left(p_{i+1} \log_2 p_{i,i+1}\right) \\
	&= -\sum_{i=1}^{L-1} \left(p_i \log_2 \frac{p_i}{p_{i,i+1}}\right)-\sum_{i=1}^{L-1} \left(p_{i+1} \log_2 \frac{p_{i+1}}{p_{i,i+1}}\right) \\
	& = \sum_{i=1}^{L-1} (p_i+p_{i+1}) h_2\left(\frac{p_i}{p_i+p_{i+1}}\right) \\
	& \le \sum_{i=1}^{L-1} (p_i+p_{i+1}) < 2 \sum_{i=1}^{L} p_i = 2.\\
\end{aligned}
\end{equation}
Hence,
\begin{equation}
\lim_{N \rightarrow \infty} \frac{\Delta_A}{N} = 0.
\end{equation}
\end{IEEEproof}

\subsubsection{Proof of \corref{scale_users_max}}
\label{sec:proof_scale_users_max}
\begin{IEEEproof}
By \thmref{converse_max}, we get no savings if $\min \mathcal{X} > 0$.
By \thmref{max_with_zero}, if $\min \mathcal{X} = 0$, we have
\begin{equation}
\begin{aligned}
\frac{\Delta_M}{N} &= NH(X) - \sum_{n=1}^{N} H_G(X_n) \\
				&= N\left(p_1+p_2\right)h_2 \left(\frac{p_1}{p_1+p_2}\right)
\end{aligned}
\end{equation}
and
\begin{equation}
\lim_{N \rightarrow \infty} \frac{\Delta_M}{N} = \left(p_1+p_2\right)h_2 \left(\frac{p_1}{p_1+p_2}\right).
\end{equation}
\end{IEEEproof}

\subsubsection{Proof of \thmref{argmax-hom-sq}}
\label{sec:argmax-hom-sq-proof}
\begin{IEEEproof}
	The optimal Bayes estimator will select one of the users that reports being in the highest interval.
	\begin{equation}
		n_j \triangleq \sum_{i = 1}^{N} \mathds{1}_{\ell_{j-1} \leq X_i \leq \ell_j}.
	\end{equation}
	We then have
	\begin{equation}
		\begin{aligned}
			\expected{X_{\hat{Z}_A}}
			&= \sum_{j = 1}^{K} \left[E_j \sum\limits_{\substack{\sum_{k}^j n_k = N \\ n_j > 0}} \binom{N}{n_1, \ldots, n_j}p_1^{n_1} \cdots p_j^{n_j}\right]\\
			&= \sum_{j = 1}^{K} \left[E_j \left(\left(\sum_{k=1}^jp_k\right)^N-\left(\sum_{k=1}^{j-1}p_k\right)^N\right)\right]\\
			&= \sum_{j = 1}^{K} \left[E_j \left(F_j^N - F_{j-1}^N\right)\right].
		\end{aligned}
	\end{equation}
	The last step follows from observing
	\begin{equation}
		\sum_{\sum_{k}^{j} n_k = N} \binom{N}{n_1, \ldots, n_j}p_1^{n_1} \cdots p_j^{n_j} = \sum\limits_{\substack{\sum_{k}^j n_k = N \\ n_j > 0}} \binom{N}{n_1, \ldots, n_j}p_1^{n_1} \cdots p_j^{n_j} + \sum\limits_{\substack{\sum_{k}^j n_k = N \\ n_j = 0}} \binom{N}{n_1, \ldots, n_j}p_1^{n_1} \cdots p_j^{n_j};
	\end{equation}
	rearranging and applying the multinomial theorem yeilds
	\begin{equation}
		\sum\limits_{\substack{\sum_{k}^j n_k = N \\ n_j > 0}} \binom{N}{n_1, \ldots, n_j}p_1^{n_1} \cdots p_j^{n_j} = \left(\sum_{k=1}^jp_k\right)^N-\left(\sum_{k=1}^{j-1}p_k\right)^N.
	\end{equation}
\end{IEEEproof}

\subsubsection{Proof of \lemref{argmax-hom-sq-derivative}}
\label{sec:argmax-hom-sq-derivative-proof}
\begin{IEEEproof}
	We re-write \eq{estimator-ev} as
	\begin{equation}
		\begin{aligned}
			\expected{X_{\hat{Z}_A}} &= \sum_{j = 1}^{K} \left[E_j\left(F_j^N - F_{j-1}^N\right)\right]\\
						 &= F_{K}^NE_{K} - \sum_{j=1}^{K-1}F_j^N(E_{j+1} - E_{j}) - F_{0}^NE_{1}\\
						 &= E_{K} - \sum_{j=1}^{K-1}F_j^N(E_{j+1} - E_{j})
		\end{aligned}
	\end{equation}
	and take derivatives
	\begin{equation}
		\frac{\partial \expected{X_{\hat{Z}_A}}}{\partial \ell_{k}} = \frac{\partial}{\partial \ell_{k}}E_{K} - \sum_{j=1}^{K-1}\frac{\partial}{\partial \ell_{k}}F_j^N(E_{j+1} - E_{j}).
	\end{equation}
	If \(k \neq K - 1\), the above becomes
	\begin{equation}
		\begin{aligned}
			\frac{\partial \expected{X_{\hat{Z}_A}}}{\partial \ell_{k}}
			&= -\sum_{j=1}^{K-1}\frac{\partial}{\partial \ell_{k}}F_j^N(E_{j+1} - E_{j})\\
			&= -\frac{\partial}{\partial \ell_{k}}F_{k-1}^N(E_{k} - E_{k-1}) - \frac{\partial}{\partial \ell_{k}}F_{k}^N(E_{k+1} - E_{k}) - \frac{\partial}{\partial \ell_{k}}F_{k+1}^N(E_{k+2} - E_{k+1})\\
			&= -F_{k-1}^N\frac{\partial E_{k}}{\partial \ell_{k}} - NF_{k}^{N-1}f_k(E_{k+1} - E_{k}) - F_{k}^N(\frac{\partial E_{k+1}}{\partial \ell_{k}} - \frac{\partial E_{k}}{\partial \ell_{k}}) + F_{k+1}^N\frac{\partial E_{k+1}}{\partial \ell_{k}}\\
			&= f_k\left[-F_{k-1}^N\frac{\ell_k - E_k}{p_k} - NF_k^{N-1}(E_{k+1} - E_{k}) - F_k^N\left(\frac{E_{k+1} - \ell_k}{p_{k+1}} - \frac{\ell_k - E_k}{p_k}\right) + F_{k+1}^N\frac{E_{k+1} - \ell_k}{p_{k+1}}\right]\\
		\end{aligned}
	\end{equation}
	If \(k = K - 1\), the above becomes
	\begin{equation}
		\begin{aligned}
			\frac{\partial \expected{X_{\hat{Z}_A}}}{\partial \ell_{K-1}} &= \frac{\partial}{\partial \ell_{K-1}}E_{K} - \frac{\partial}{\partial \ell_{K-1}}F_{K-2}^N(E_{K-1} - E_{K-2}) - \frac{\partial}{\partial \ell_{K-1}}F_{K-1}^N(E_{K} - E_{K-1})\\
										      &=  -F_{K-2}^N \frac{\partial E_{K-1}}{\partial \ell_{K-1}} - NF_{K-1}^{N-1}f_{K-1}(E_{K} - E_{K-1}) - F_{K-1}^N(\frac{\partial E_{K}}{\partial \ell_{K-1}} - \frac{\partial E_{K-1}}{\partial \ell_{K-1}}) + F_K^N\frac{\partial E_{K}}{\partial \ell_{K-1}}
		\end{aligned}
	\end{equation}
	The above follows from recognizing that \(F_K = 1\) and we see that the expression for \(k \neq K - 1\) holds for \(k = K - 1\).
\end{IEEEproof}

\subsubsection{Proof of \corref{argmax-hom-sq-derivative}}
\label{sec:argmax-hom-sq-derivative-cor-proof}
\begin{IEEEproof}
	\begin{equation}
		\begin{aligned}
			\frac{\partial \expected{X_{\hat{Z}_A}}}{\partial \ell_{k}}
			&= f_k\left[\frac{(F_{k+1}^2 - F_k^2)(E_{k+1} - \ell_k)}{p_{k+1}} + \frac{(F_k^2 - F_{k-1}^2)(\ell_k - E_k)}{p_k} - 2F_k(E_{k+1} - E_{k})\right]\\
			&= f_k\left[\frac{(F_{k+1}^2 - F_k^2)(E_{k+1} - \ell_k)}{F_{k+1} - F_k} + \frac{(F_k^2 - F_{k-1}^2)(\ell_k - E_k)}{F_k - F_{k-1}} - 2F_k(E_{k+1} - E_{k})\right]\\
			&= f_k\left[(F_{k+1} + F_k)(E_{k+1} - \ell_k) + (F_k + F_{k-1})(\ell_k - E_k) - 2F_k(E_{k+1} - E_{k})\right]\\
			&= f_k\left[(F_{k+1} -F_k)E_{k+1} + (F_k - F_{k-1})E_k - (F_{k+1} - F_{k-1})\ell_k\right]\\
			&= f_k\left[\int_{\ell_{k}}^{\ell_{k+1}} \! x f(x) \, \mathrm{d}x + \int_{\ell_{k-1}}^{\ell_{k}} \! x f(x) \, \mathrm{d}x - \ell_k\int_{\ell_{k-1}}^{\ell_{k+1}} \! f(x) \, \mathrm{d}x\right]
		\end{aligned}
	\end{equation}
\end{IEEEproof}

\subsubsection{Proof of \thmref{argmax-het-sq}}
\label{sec:argmax-het-sq-proof}
\begin{IEEEproof}
	For an \(\argmax\) quantizer, an average distortion is
	\begin{equation}
		\label{eq:argmax-step0}
		\begin{aligned}
			\expected{D\left((X_1, X_2), \hat{z}\right)} %
			&= \expected{\expected{D((X_1, X_2), \hat{z}) \mid U_1, U_2}}\\
			&= \sum_{k_1, k_2} \expected{Z_M - X_{\hat{z}} \mid U_1 = k_1, U_2 = k_2} \prob{U_1 = k_1, U_2 = k_2}\\
			&= \sum_{(k_1, k_2) \in \Zbf_1} \expected{X_1 - X_{\hat{z}} \mid X_1 \in \Lca_{1, k_1}, X_2 \in \Lca_{2, k_2}} \prob{X_1 \in \Lca_{1, k_1}, X_2 \in \Lca_{2, k_2}}\\
			&+ \sum_{(k_1, k_2) \in \Zbf_2} \expected{X_2 - X_{\hat{z}} \mid X_1 \in \Lca_{1, k_1}, X_2 \in \Lca_{2, k_2}} \prob{X_1 \in \Lca_{1, k_1}, X_2 \in \Lca_{2, k_2}}\\
			&+ \sum_{(k_1, k_2) \in \Zbf_0} \expected{X_{Z_A} - X_{\hat{z}}\mid X_1 \in \Lca_{1, k_1}, X_2 \in \Lca_{2, k_2}} \prob{X_1 \in \Lca_{1, k_1}, X_2 \in \Lca_{2, k_2}}
		\end{aligned}
	\end{equation}
	where \(\Zbf_1 = \{(k_1, k_2) : \ell_{2, k_2} \leq \ell_{1, k_1 - 1}\}\), %
	\(\Zbf_2 = \{(k_1, k_2) : \ell_{1, k_1} \leq \ell_{2, k_2 - 1}\}\), and %
	\(\Zbf_0 = \{(k_1, k_2) : \max(\ell_{1, k_1 - 1}, \ell_{2, k_2 - 1}) \leq \min(\ell_{1, k_1}, \ell_{2, k_2}) \leq \max(\ell_{1, k_1}, \ell_{2, k_2})\}\).
	Observe that
	\begin{equation}
		\label{eq:argmax-step1}
		\begin{aligned}
			\hat{z}(U_1, U_2)
			&= \argmin_{z} \expected{d((X_1, X_2), z) \mid U_1 = k_1, U_2 = k_2}\\
			&= \argmin_{z} \expected{Z_M - X_{z} \mid X_1 \in \Lca_{1, k_1}, X_2 \in \Lca_{2, k_2}}\\
			&= \argmax_{z} \expected{X_{z} \mid X_1 \in \Lca_{1, k_1}, X_2 \in \Lca_{2, k_2}}.
		\end{aligned}
	\end{equation}
	When \((U_1, U_2) \in \Zbf_1\), \(Z_A = 1\) and a distortion of \(0\) can be attained with \(\hat{z} = 1\).
	When \((U_1, U_2) \in \Zbf_2\), \(Z_A = 2\) and a distortion of \(0\) can be attained with \(\hat{z} = 2\).
	For region \(\Zbf_0\), \(Z_M\) may be equal to either \(X_1\) or \(X_2\) and we see from \eq{argmax-step1} that the Bayes estimator is
	\begin{equation}
		\hat{z}(U_1, U_2) =\begin{dcases}
			1 & \text{if } \expected{X_1 \mid X_1 \in \Lca_{1, U_1}} \geq \expected{X_2 \mid X_2 \in \Lca_{2, U_2}} \\
			2 & \text{otherwise}.
		\end{dcases}
	\end{equation}
	To derive the expression for the average distortion in the region \(\Zbf_0\), we partition into the two sets \(\Zbf_{01}\) and \(\Zbf_{02}\) and break the last summation in \eq{argmax-step0} into two parts and substitute the appropriate conditional density functions and the expression for the Bayes estimator.
\end{IEEEproof}

\subsubsection{Proof of \thmref{max-het-sq}}
\label{sec:max-het-sq-proof}
\begin{IEEEproof}
	For a \(\max\) quantizer, the average distortion can be expressed as follows
	\begin{equation}
		\label{eq:max-proof-step0}
		\begin{aligned}
			\expected{d((X_1, X_2), \hat{z})} %
			&= \expected{\expected{d(((X_1, X_2), \hat{z}(U_1, U_2)) \mid U_1, U_2}}\\
			&= \sum_{k_1, k_2 } \expected{d(((X_1, X_2), \hat{z}) \mid U_1 = k_1, U_2 = k_2 } \prob{ U_1 = k_1, U_2 = k_2 }\\
			&= \sum_{(k_1, k_2) \in \Zbf_1} \expected{d(X_1, \hat{z}) \mid X_1 \in \Lca_{1, k_1}, X_2 \in \Lca_{2, k_2}} \prob{X_1 \in \Lca_{1, k_1}, X_2 \in \Lca_{2, k_2}}\\
			&+ \sum_{(k_1, k_2) \in \Zbf_2} \expected{d(X_2, \hat{z}) \mid X_1 \in \Lca_{1, k_1}, X_2 \in \Lca_{2, k_2}} \prob{X_1 \in \Lca_{1, k_1}, X_2 \in \Lca_{2, k_2}}\\
			&+ \sum_{(k_1, k_2) \in \Zbf_0} \expected{d(Z_M, \hat{z}) \mid X_1 \in \Lca_{1, k_1}, X_2 \in \Lca_{2, k_2}} \prob{X_1 \in \Lca_{1, k_1}, X_2 \in \Lca_{2, k_2}}
		\end{aligned}
	\end{equation}
	In order to find a quantizer minimizing the average distortion, we need to evaluate a minimum distortion term as follows,
	\begin{equation}
		\begin{aligned}
			\hat{z}(U_1, U_2) %
			&= \argmin_{z} \expected{d(Z_M, z) \mid U_1 = k_1, U_2 = k_2}\\
			&= \argmin_{z} \expected{Z_M - \hat{z} \mathds{1}_{Z_M \geq z} \mid X_1 \in \Lca_{1, k_1}, X_2 \in \Lca_{2, k_2} }\\
			&= \argmax_{z} z\prob{Z_M \geq z \mid X_1 \in \Lca_{1, k_1}, X_2 \in \Lca_{2, k_2}}
		\end{aligned}
	\end{equation}
	When \((U_1, U_2) \in \Zbf_1\), then \(Z_M = X_1\) and we have
	\begin{equation}
		\hat{z}(U_1, U_2) = \argmax_{z} z\prob{X_1 \geq z \mid X_1 \in \Lca_{1, k_1}}
	\end{equation}
	In order to find an optimum estimation \(\hat{z}_1^*\) minimizing the average distortion in region \(\Zbf_1\), the necessary and sufficient condition is to find \(z\) maximizing
	\begin{equation}
		\label{eq:max-proof-z1}
		z\prob{X_1 \geq z \mid X_1 \in \Lca_{1, k_1}} = \begin{dcases}
			z & z \leq \ell_{1, k_1 - 1}\\
			z\frac{F_X(\ell_{1, k_1}) - F_X(z)}{F_X(\ell_{1, k_1}) - F_X(\ell_{1, k_1 - 1})} & z \in \Lca_{1, k_1}\\
			0 & z \geq \ell_{1, k_1}
			\end{dcases}
	\end{equation}
	Since the maximum of each region can be included on the boundary from \(\Lca_{1, k_1}\), it is suffices to evaluate as follows,
	\begin{equation}
		\hat{z}_1^* = \argmax_{z \in \Lca_{1, k_1}} z\frac{F_X(\ell_{1, k_1}) - F_X(z)}{F_X(\ell_{1, k_1} ) - F_X(\ell_{1, k_1 - 1})}
	\end{equation}
	If it is possible to take the first and second derivative of \eq{max-proof-z1} with respect to \(z\), the estimation \(\hat{z}_1^*\) can be determined by
	\begin{equation}
		\hat{z}_1^* = \begin{dcases}
			\sol \left\{ z : F_X(\ell_{1, k_1}) = F_X(z) + zf_X(z), 2f_X(z) + zf'_X(z) \geq 0\right\} & z \in \Lca_{1, k_1}\\
			\ell_{1, k_1 - 1} & \text{otherwise}
		\end{dcases}
	\end{equation}
	In order to find an average distortion in the region \(\Zbf_1\), we first give \(\expected{X_1 \mid X_1 \in \Lca_{1, k_1}}\) as follows
	\begin{equation}
		\expected{X_1 \mid X_1 \in \Lca_{1, k_1}} = \int_{\ell_{1, k_1 - 1}}^{\ell_{1, k_1}} x\frac{f_X(x )}{F_X(\ell_{1, k_1}) - F_X(\ell_{1, k_1 - 1})} \, \mathrm{d}x.
	\end{equation}
	The average conditional minimum distortion when \((U_1, U_2) \in \Zbf_1\) is
	\begin{multline}
		\expected{X_1 \mid X_1 \in \Lca_{1, k_1} } - \hat{z}_1^*\prob{X_1 \geq \hat{z}_1^* \mid X_1 \in \Lca_{1, k_1}} =\\
		\int_{\ell_{1, k_1 - 1}}^{\ell_{1, k_1}} x\frac{f_X(x)}{F_X(\ell_{1, k_1}) - F_X(\ell_{1, k_1 - 1})} \, \mathrm{d}x - \hat{z}_1^* \frac{F_X(\ell_{1, k_1}) - F_X(\hat{z}_1^*)}{F_X(\ell_{1, k_1}) - F_X(\ell_{1, k_1 - 1})}
	\end{multline}
	Therefore, the average distortion in \(\Zbf_1\) is,
	\begin{multline}
		\sum_{(k_1, k_2) \in \Zbf_1} \expected{d(X_1, \hat{z}) \mid X_1 \in \Lca_{1, k_1}, X_2 \in \Lca_{2, k_2}}\prob{X_1 \in \Lca_{1, k_1}, X_2 \in \Lca_{2, k_2}} =\\
		\sum_{(k_1, k_2) \in \Zbf_1} \left[\int_{\ell_{1, k_1 - 1}}^{\ell_{1, k_1}} xf_X(x) \, \mathrm{d}x - \hat{z}_1^*\left(F_X(\ell_{1, k_1}) - F_X(\hat{z}_1^*)\right)\right] \left(F_X(\ell_{2, k_2}) - F_X(\ell_{2, k_2 - 1})\right)
	\end{multline}

	When \((U_1, U_2) \in \Zbf_2\), then \(Z_M = X_2\) and by symmetry we have that the estimation \(\hat{z}_2^*\) minimizing the average distortion in region \(\Zbf_2\)
	\begin{equation}
		\hat{z}_2^* = \argmax_{z} z\prob{X_2 \geq z \mid X_2 \in \Lca_{2, k_2}}.
	\end{equation}
	If is is possible to take the first and second derivatives, then \(\hat{z}_2^*\) can be determined by
	\begin{equation}
		\hat{z}_2 ^* = \begin{dcases}
			\sol \left\{z \mid F_X(\ell_{2, k_2}) = F_X(z) + zf_X(z), 2f_X(z) + zf'_X(z) \geq 0\right\} & z \in \Lca_{2, k_2}\\
			\ell_{2, k_2 - 1} & \text{otherwise}
		\end{dcases}
	\end{equation}
	The average conditional minimum distortion when \((U_1, U_2) \in \Zbf_2\) is
	\begin{multline}
		\expected{X_2 \mid X_2 \in \Lca_{2, k_2}} - \hat{z}_2 ^* \prob{X_2 \geq \hat{z}_2^* \mid X_2 \in \Lca_{2, k_2}} = \\
		\int_{\ell_{2, k_2 - 1}}^{\ell_{2, k_2}} x\frac{f_X(x)}{F_X(\ell_{2, k_2}) - F_X(\ell_{2, k_2 - 1})} \, \mathrm{d}x - \hat{z}_2^*\frac{F_X(\ell_{2, k_2}) - F_X(\hat{z}_2^*)}{F_X(\ell_{2, k_2}) - F_X(\ell_{2, k_2 - 1})}
		\end{multline}
	Therefore, the average distortion in region \(\Zbf_2\) is
	\begin{multline}
		\sum_{(k_1, k_2) \in \Zbf_2} \expected{d(X_2, \hat{z}) \mid X_1 \in \Lca_{1, k_1}, X_2 \in \Lca_{2, k_2}}\prob{X_1 \in \Lca_{1, k_1}, X_2 \in \Lca_{2, k_2}} = \\
		\sum_{(k_1, k_2) \in \Zbf_2} \left[\int_{\ell_{2, k_2 - 1}}^{\ell_{2, k_2}} xf_X(x) \, \mathrm{d}x - \hat{z}_2^*\left(F_X(\ell_{2, k_2}) - F_X(\hat{z}_2^*)\right)\right]\left(F_X(\ell_{1, k_1}) - F_X(\ell_{1, k_1 - 1})\right)
	\end{multline}

	When \((U_1, U_2) \in \Zbf_0\), the maximum value \(Z_M\) may be equal to either \(X_1\) or \(X_2)\) and
	\begin{equation}
		\hat{z}_0^* = \argmax_{z} z\prob{Z_M \geq z \mid X_1 \in \Lca_{1, k_1}, X_2 \in \Lca_{2, k_2}}.
	\end{equation}

	For the case of \((U_1, U_2) \in \Zbf_{01} \subseteq \Zbf_0\) (i.e., \(\max(\ell_{2, k_2 - 1}, \ell_{1, k_1 - 1}) < \ell_{2, k_2} \leq \ell_{1, k_1}\)), the CDF of \(Z_M\) given \(X_1 \in \Lca_{1, k_1}\) and \(X_2 \in \Lca_{2, k_2}\) is
	\begin{equation}
		\begin{aligned}
			F_{Z_M \mid X_1 \in \Lca_{1, k_1}, X_2 \in \Lca_{2, k_2}}(z) %
			&= F_{X \mid X \in \Lca_{1, k_1}}(z)F_{X \mid X \in \Lca_{2, k_2}}(z)\\
			&= \begin{dcases}
				0 & z \leq \max(\ell_{1, k_1 - 1}, \ell_{2, k_2 - 1})\\
				F_{X \mid X \in \Lca_{1, k_1}}(z)F_{X \mid X \in \Lca_{2, k_2}}(z) & \max(\ell_{1, k_1 - 1}, \ell_{2, k_2 - 1}) \leq z \leq \ell_{2, k_2}\\
				F_{X \mid X \in \Lca_{1, k_1}}(z) & \ell_{2, k_2} \leq z \leq \ell_{1, k_1}\\
				1 & z \geq \ell_{1, k_1}
			\end{dcases}
		\end{aligned}
	\end{equation}
	then, \(z \prob{Z_M \geq z \mid X_1 \in \Lca_{1, k_1}, X_2 \in \Lca_{2, k_2}}\) is
	\begin{equation}
		z\prob{Z_M \geq z \mid X_1 \in \Lca_{1, k_1}, X_2 \in \Lca_{2, k_2}} = \begin{dcases}
			z & z \leq \max(\ell_{1, k_1 - 1}, \ell_{2, k_2 - 1})\\
			z\left[1 - F_{X \mid X \in \Lca_{1, k_1}}(z)F_{X \mid X \in \Lca_{2, k_2}}(z)\right] & \max(\ell_{1, k_1 - 1}, \ell_{2, k_2 - 1}) \leq z \leq \ell_{2, k_2}\\
			z\left[1 - F_{X \mid X \in \Lca_{1, k_1}}(z)\right] & \ell_{2, k_2} \leq z \leq \ell_{1, k_1}\\
			0 & z \geq \ell_{1, k_1}
		\end{dcases}
	\end{equation}
	Since the maximum of each region can be also included on the boundary from \(\max (\ell_{1, k_1 - 1}, \ell_{2, k_2 - 1} ) \leq \hat{z} \leq \ell_{1, k_1}\), it is suffices to evaluate as follows,
	\begin{equation}
		\hat{z}_{01}^* = \argmax_{z} \begin{dcases}
			w_{11}(z) & \max(\ell_{1, k_1 - 1}, \ell_{2, k_2 - 1}) \leq z \leq \ell_{2, k_2}\\
			w_{12}(z) & \ell_{2, k_2} \leq z \leq \ell_{1, k_1}.
		\end{dcases}
	\end{equation}
	where
	\begin{equation}
		\label{eq:W1-proof}
		\begin{aligned}
			w_{11}(z) &= z\left(1 - F_{X \mid X \in \Lca_{1, k_1}}(z)F_{X \mid X \in \Lca_{2, k_2}}(z)\right)\\
			w_{12}(z) &= z\left(1 - F_{X \mid X \in \Lca_{1, k_1}}(z)\right)
		\end{aligned}
	\end{equation}
	If we can take first and second derivatives, then for \(\max(\ell_{1, k_1 - 1}, \ell_{2, k_2 - 1}) \leq z \leq \ell_{2, k_2}\), a condition such that \(w'_{11}(\hat{z}) = 0\) is
	\begin{multline}
		(zf_X(z) + F_X(z) - F_X(\ell_{1, k_1}))F_X(\ell_{2, k_2 - 1}) + (zf_X(z) + F_X(z) - F_X(\ell_{2, k_2}))F_X(\ell_{1, k_1 - 1}) = \\
		+ 2zF(x)f(x) + F_X^2(z) - F_X(\ell_{1, k_1})F_X(\ell_{2, k_2})
	\end{multline}
	and \(w''_{11}(z) \leq 0\) is
	\begin{equation}
		\begin{aligned}
			\left(2f_X(z) + zf'_X(z)\right)\left(F_X(\ell_{2, k_2 - 1}) - F_X(\ell_{1, k_1 - 1}) - 2F_X(z)\right) \leq 2zf_X^2(z)
		\end{aligned}
	\end{equation}
	A maximizer \(\hat{z}_{11}^*\) for \(w_{11}(z)\) is given by
	\begin{equation}
		\label{eq:Z11-proof}
		\hat{z}_{11}^* = \begin{dcases}
			\sol \left\{z : w'_{11}(z) = 0, w''_{11}(z) \leq 0\right\} & \max(\ell_{1, k_1 - 1}, \ell_{2, k_2 - 1}) \leq z \leq \ell_{2, k_2}\\
			\max(\ell_{1, k_1 - 1}, \ell_{2, k_2 - 1}) & \text{otherwise}
		\end{dcases}
	\end{equation}
	For \(\ell_{2, k_2} \leq z \leq \ell_{1, k_1}\), based on the first and second derivative of the function \(w_{12}(z)\) with respect to \(z\), a maximizer \(\hat{z}_{12}^*\) for \(w_{12}(z)\) is given by
	\begin{equation}
		\label{eq:Z12-proof}
		\hat{z}_{12}^* = \begin{dcases}
			\sol \left\{z : F_X(\ell_{1, k_1}) = F_X(z) + zf_X(z), 2f_X(z) + zf'_X(z) \geq 0\right\} & \ell_{2, k_2} \leq z \leq \ell_{1, k_1}\\
			\ell_{2, k_2} & \text{otherwise}
		\end{dcases}
	\end{equation}
	The estimation \(\hat{z}_{01}^*\) minimizing the distortion in region \(\Zbf_{01}\) is
	\begin{equation}
		\hat{z}_{01}^* = \begin{dcases}
			\hat{z}_{11}^* & w_{11}(\hat{z}_{11}^*) \geq w_{12}(\hat{z}_{12}^*)\\
			\hat{z}_{12}^* & \text{otherwise}.
		\end{dcases}
	\end{equation}

	For the case \((U_1, U_2) \in \Zbf_{02} \subseteq \Zbf_0\) (i.e., \(\max (\ell_{1, k_1 - 1}, \ell_{2, k_2 - 1}) < \ell_{1, k_1} \leq \ell_{2, k_2}\)) we have that
	\begin{equation}
		z\prob{Z_M \geq z \mid X_1 \in \Lca_{1, k_1}, X_2 \in \Lca_{2, k_2}} = \begin{dcases}
			z & z\leq \max(\ell_{1, k_1 - 1}, \ell_{2, k_2 - 1})\\
			z\left[1 - F_{X \mid X \in \Lca_{1, k_1}}(z)F_{X \mid X \in \Lca_{2, k_2}}(z)\right] & \max(\ell_{1, k_1 - 1}, \ell_{2, k_2 - 1}) \leq z \leq \ell_{1, k_1}\\
			z\left[1 - F_{X \mid X \in \Lca_{2, k_2}}(z)\right] & \ell_{1, k_1} \leq z \leq \ell_{2, k_2}\\
			0 & z \geq \ell_{2, k_2}
		\end{dcases}
	\end{equation}
	Since the maximum of each region can be also included on the boundary from \(\max(\ell_{1, k_1 - 1}, \ell_{2, k_2 - 1}) \leq \hat{z} \leq \ell_{2, k_2}\), it is suffices to evaluate as follows,
	\begin{equation}
		\hat{z}_{02}^* = \argmax_{z} \begin{dcases}
			w_{21}(z) & \max(\ell_{1, k_1 - 1}, \ell_{2, k_2 - 1}) \leq z \leq \ell_{1, k_1}\\
			w_{22}(z) & \ell_{1, k_1} \leq z \leq \ell_{2, k_2}
		\end{dcases}
	\end{equation}
	where
	\begin{equation}
		\label{eq:W2proof}
		\begin{aligned}
			w_{21}(z) &= z\left(1 - F_{X \mid X \in \Lca_{1, k_1}}(z)F_{X \mid X \in \Lca_{2, k_2}}(z)\right)\\
			w_{22}(z) &= z\left(1 - F_{X \mid X \in \Lca_{2, k_2}}(z)\right]
		\end{aligned}
	\end{equation}
	Following a similar argument as above, a maximizer \(\hat{z}_{21}^*\) of \(w_{21}(z)\) is given by
	\begin{equation}
		\label{eq:Z21-proof}
		\hat{z}_{21}^* = \begin{dcases}
			\sol \left\{z : w'_{21}(z) = 0, w''_{21}(z) \leq 0\right\} & \max(\ell_{1, k_1 - 1}, \ell_{2, k_2 - 1}) \leq z \leq \ell_{1, k_1}\\
			\max(\ell_{1, k_1 - 1}, \ell_{2, k_2 - 1}) & \text{otherwise}
		\end{dcases}
	\end{equation}
	and a maximizer \(\hat{z}_{22}^*\) of \(w_{22}(z)\) is given by
	\begin{equation}
		\label{eq:Z22-proof}
		\hat{z}_{22}^* = \begin{dcases}
			\sol \left\{z : F_X(\ell_{2, k_2}) = F_X(z) + zf_X(z), 2f_X(z) + zf'_X(z) \geq 0\right\} & \ell_{1, k_1} \leq z \leq \ell_{2, k_2}\\
			\ell_{1, k_1} & \text{otherwise}.
		\end{dcases}
	\end{equation}
	The estimation \(\hat{z}_{02}^*\) minimizing the distortion in region \(\Zbf_{02}\) is
	\begin{equation}
		\hat{z}_{02}^* = \begin{dcases}
			\hat{z}_{21}^* & w_{21}(\hat{z}_{21}^*) \geq w_{22}(\hat{z}_{22}^*)\\
			\hat{z}_{22}^* & \text{otherwise}.
		\end{dcases}
	\end{equation}

	An expression for the distortion as a function of the quantizer parameters \(\bm{\ell}\) and the distribution \(F(\cdot)\) is found by substituting in the Bayes estimator for in the different regions into \eq{max-proof-step0}.
\end{IEEEproof}

\subsubsection{Proof of \thmref{pair-het-sq}}
\label{sec:pair-het-sq-proof}
\begin{IEEEproof}
	For a \(\max\) and \(\argmax\) quantizer, the average distortion is
	\begin{equation}
		\label{eq:pair-proof-step0}
		\begin{aligned}
			\expected{d_{M, A}((X_1, X_2 ), (\hat{z}, \hat{i}))} %
			&= \expected{\expected{d_{M,A}((X_1, X_2 ), (\hat{z}, \hat{i})) \mid U_1, U_2}}\\
			&=\sum_{k_1, k_2} \expected{ Z_M - \hat{z} \mathds{1}_{\hat{z} \leq X_{\hat{i}}} \mid U_1 = k_1, U_2 = k_2 } \prob{ U_1 = k_1, U_2 = k_2 }\\
			&=\sum_{(k_1, k_2 ) \in \Zbf_1} \expected{ X_1 - \hat{z} \mathds{1}_{\hat{z} \leq X_{\hat{i}}} \mid X_1 \in \Lca_{1, k_1 }, X_2 \in \Lca_{2, k_2} } \prob{ X_1 \in \Lca_{1, k_1}, X_2 \in \Lca_{2, k_2} }\\
			&+\sum_{(k_1, k_2 ) \in \Zbf_2} \expected{ X_2 - \hat{z} \mathds{1}_{\hat{z} \leq X_{\hat{i}}}\mid X_1 \in \Lca_{1, k_1 }, X_2 \in \Lca_{2, k_2} } \prob{ X_1 \in \Lca_{1, k_1}, X_2 \in \Lca_{2, k_2} }\\
			&+\sum_{(k_1, k_2 ) \in \Zbf_0} \expected{ Z_M - \hat{z} \mathds{1}_{\hat{z} \leq X_{\hat{i}}}\mid X_1 \in \Lca_{1, k_1 }, X_2 \in \Lca_{2, k_2} } \prob{ X_1 \in \Lca_{1, k_1}, X_2 \in \Lca_{2, k_2} }
		\end{aligned}
	\end{equation}
	Similar to the case of \(\max\) quantizer, to find a quantizer minimizing an average distortion, we need to evaluate a minimum distortion term as follows,
	\begin{equation}
		\begin{aligned}
			\argmin_{z, i} & \expected{D_{M,A}((Z_M, Z_A), (z, i)) \mid X_1 \in \Lca_{1, k_1}, X_2 \in \Lca_{2, k_2}}\\
				       &= \argmin_{z, i} \expected{Z_M - z\mathds{1}_{X_i \geq z \mid X_1 \in \Lca_{1, k_1}, X_2 \in \Lca_{2, k_2}}}\\
				       &= \argmax_{z, i} \left\{z\prob{X_i \geq z \mid X_1 \in \Lca_{1, k_1}, X_2 \in \Lca_{2, k_2}}\right\}
		\end{aligned}
	\end{equation}
	
	When \((U_1, U_2) \in \Zbf_1\), then \(Z_M = X_1\) and \(Z_A = 1\) and we take as our estimate \(\hat{i}^* = 1\) which gives
	\begin{equation}
		\hat{z}_1^* = \argmax_{z} z\prob{X_1 \geq z \mid X_1 \in \Lca_{1, k_1}}
	\end{equation}
	In order to find an optimum estimation \(\hat{z}_1^*\) minimizing the average distortion in region \(\Zbf_1\), the necessary and sufficient condition is to find \(z\) maximizing
	\begin{equation}
		z\prob{ X_1 \geq z \mid X_1 \in \Lca_{1, k_1} } = \begin{dcases}
			z & z \leq \ell_{1, k_1 - 1}\\
			z\frac{F_X(\ell_{1, k_1}) - F_X(z)}{F_X(\ell_{1, k_1}) - F_X(\ell_{1, k_1 - 1})} & z \in \Lca_{1, k_1}\\
			0 & z \geq \ell_{1, k_1}
		\end{dcases}
	\end{equation}
	Since the maximum of each region can be included on the boundary from \(\Lca_{1, k_1}\), it is suffices to evaluate as follows,
	\begin{equation}
		\hat{z}_1^* = \argmax_{z \in \Lca_{1, k_1}} z\frac{F_X(\ell_{1, k_1}) - F_X(z)}{F_X(\ell_{1, k_1}) - F_X(\ell_{1, k_1 - 1})}
	\end{equation}
	Based on the first and second derivative with respect to \(z\), the estimation \(\hat{z}_1^*\) can be determined by
	\begin{equation}
		\hat{z}_1^* = \begin{dcases}
			\sol \left\{z : F_{X}(\ell_{1, k_1}) = F_{X}(z) + zf_{X}(z), \; 2f_{X}(z) + zf'_{X}(z) \geq 0 \right\} & z \in \Lca_{1, k_1}\\
			\ell_{1, k_1 - 1} & \text{otherwise}
		\end{dcases}
	\end{equation}
	In order to find an average distortion in the region \(\Zbf_1\), we first give \(\expected{X_1 \mid X_1 \in \Lca_{1, k_1}}\) as follows
	\begin{equation}
		\expected{X_1 \mid X_1 \in \Lca_{1, k_1}} = \int_{\ell_{1, k_1 - 1}}^{\ell_{1, k_1}} x\frac{f_X(x)}{F_X(\ell_{1, k_1}) - F_X(\ell_{1, k_1 - 1})} \, \mathrm{d}x.
	\end{equation}
	The average minimum distortion when \((U_1, U_2) \in \Zbf_1\) is
	\begin{multline}
		\expected{X_1 \mid X_1 \in \Lca_{1, k_1}} - \hat{z}_1^* \prob{X_1 \geq \hat{z}_1^* \mid X_1 \in \Lca_{1, k_1}} = \\
		\int_{\ell_{1, k_1 - 1}}^{\ell_{1, k_1}} x\frac{f_X(x)}{F_X(\ell_{1, k_1}) - F_X(\ell_{1, k_1 - 1})} \, \mathrm{d}x - \hat{z}_1^*\frac{F_X(\ell_{1, k_1}) - F_X(\hat{z}_1^*)}{F_X(\ell_{1, k_1}) - F_X(\ell_{1, k_1 - 1})}
	\end{multline}

	When \((U_1, U_2) \in \Zbf_2\), then \(Z_M = X_2\) and \(Z_A = 2\) and we take as our estimate \(\hat{i}^* = 2\) which gives
	\begin{equation}
		\begin{aligned}
			\hat{z}_2 ^* = \argmax_z z\prob{X_2 \geq z \mid X_2 \in \Lca_{2, k_2}}
		\end{aligned}
	\end{equation}
	Similar to the region \(\Zbf_1\), the estimation \(\hat{z}_2^*\) can be determined by the solution as follows,
	\begin{equation}
		\hat{z}_2^* = \begin{dcases}
			\sol \left\{z : F_X(\ell_{2, k_2}) = F_X(z) + zf_X(z), 2f_X(z) + zf'_X(z) \geq 0\right\} & z \in \Lca_{2, k_2}\\
			\ell_{2, k_2 - 1} & \text{otherwise}
		\end{dcases}
	\end{equation}
	The average minimum distortion when \((U_1, U_2) \in \Zbf_2\) is
	\begin{multline}
		\expected{X_2 \mid X_2 \in \Lca_{2, k_2}} - \hat{z}_2^*\prob{X_2 \geq \hat{z}_2^* \mid X_2 \in \Lca_{2, k_2}} = \\
		\int_{\ell_{2, k_2 - 1}}^{\ell_{2, k_2}} x\frac{f_X(x)}{F_X(\ell_{2, k_2}) - F_X(\ell_{2, k_2 - 1})} \, \mathrm{d}x - \hat{z}_2^*\frac{F_X(\ell_{2, k_2}) - F_X(\hat{z}_2^*)}{F_X(\ell_{2, k_2}) - F_X(\ell_{2, k_2 - 1})}
	\end{multline}
	
	When \((U_1, U_2) \in \Zbf_0\), then \(Z_M\) could be either \(X_1\) or \(X_2\) and it suffices to compare 
	\begin{equation}
		z\prob{Z_M \geq z \mid X_1 \in \Lca_{1, k_1}, X_2 \in \Lca_{2, k_2}}
	\end{equation}
	under \(Z_M = X_1\) and \(Z_M = X_2\).
	We need to find \(\hat{z}^*\) and \(\hat{i}^*\) as follows,
	\begin{equation}
		\argmax_{z, i} z\frac{F_X(\ell_{i, k_i}) - F_X(z)}{F_X(\ell_{i, k_i}) - F_X(\ell_{i, k_i - 1})}
	\end{equation}
	
	An expression for the distortion as a function of the quantizer parameters \(\bm{\ell}\) and the distribution \(F(\cdot)\) is found by substituting in the Bayes estimator for in the different regions into \eq{pair-proof-step0}.
\end{IEEEproof}

\end{document}